\documentclass[twocolumn,superscriptaddress]{revtex4-1}
\usepackage[english]{babel}
\usepackage[T1]{fontenc}
\usepackage{amsmath}
\usepackage[colorlinks]{hyperref}
\hypersetup{
plainpages=true,
breaklinks=true,
hypertexnames=false,
 pageanchor=true,
colorlinks=true,linkcolor={blue},
citecolor={magenta},
urlcolor={blue},
anchorcolor={black}}
\usepackage{tikz}
\usepackage{lipsum}
\usepackage{pifont}
\usepackage{booktabs}
\usepackage{multirow}
\usepackage{adjustbox}

\def\bra#1{\mathinner{\langle{#1}|}} 
\def\ket#1{\mathinner{\left|{#1}\right\rangle}}

\def\code#1{\texttt{#1}}
\newcommand{\Eq}[1]{Eq.~(\ref{#1})}

\usepackage{MnSymbol}
\usepackage{pythonhighlight}

\definecolor{orangepiqs}{RGB}{255,153,51}
\definecolor{greenpiqs}{RGB}{87,185,90}
\definecolor{bluepiqs}{RGB}{57,92,167}
\definecolor{redpiqs}{RGB}{176,53,57}
\definecolor{purplepiqs}{RGB}{106,0,162}
\definecolor{cyanpiqs}{RGB}{0,162,165}

\newcommand{\mydisk}[1]{\tikz{\draw[black,fill=#1] (0,0) circle (.8ex);}}
\newcommand{\mydisksmall}[1]{\tikz{\filldraw[black,fill=#1] (0,0) circle (.5ex);}}
\newcommand{\mysquare}[1]{\tikz{\filldraw[draw=black,fill=#1] (0,0)
rectangle (0.2cm,0.2cm);}}
\newcommand{\mytriangle}[1]{\tikz{\filldraw[draw=black,fill=#1] 
(0,0) --(0.2cm,0) -- (0.1cm,0.2cm) -- (0,0);}}
\newcommand{\mydtriangle}[1]{\tikz{\filldraw[draw=black,fill=#1] (0.1cm,0) --
(0.2cm,0.2cm) -- (0,0.2cm)--(0.1cm,0);}}
\newcommand{\mydiamond}[1]{\tikz{\filldraw[line width=0.1mm,draw=black,fill=#1] (0.1cm,0) --
(0.2cm,0.1cm) -- (0.1cm,0.2cm) --(0,0.1cm)--(0.1cm,0);}}
\newcommand{\myedisk}[1]{\tikz{\draw[line width=0.4mm,#1] (0,0) circle (.8ex);}}
\newcommand{\myedisksmall}[1]{\tikz{\draw[line width=0.4mm,draw=#1] (0,0) circle (.5ex);}}
\newcommand{\myesquare}[1]{\tikz{\draw[line width=0.4mm,draw=#1] (0,0)
rectangle (0.2cm,0.2cm);}}

\newcommand{\myediamond}[1]{\tikz{\draw[line width=0.4mm,draw=#1] (0.1cm,0) --
(0.2cm,0.1cm) -- (0.1cm,0.2cm) --(0,0.1cm)--(0.1cm,0);}}

\makeatletter
\begin{document}
\title{Open quantum systems with local and collective incoherent processes: \\ Efficient numerical simulation using permutational invariance} 
\date{\today}
\author{Nathan Shammah}
\email{nathan.shammah@gmail.com}
\affiliation{Theoretical Quantum Physics Laboratory, RIKEN Cluster for Pioneering Research, Wako-shi, Saitama 351-0198, Japan}
\author{Shahnawaz Ahmed}
\affiliation{Theoretical Quantum Physics Laboratory, RIKEN Cluster for Pioneering Research, Wako-shi, Saitama 351-0198, Japan}
\affiliation{BITS Pilani Goa Campus, Sancoale, Goa 403726, India}

\author{Neill Lambert}
\affiliation{Theoretical Quantum Physics Laboratory, RIKEN Cluster for Pioneering Research, Wako-shi, Saitama 351-0198, Japan}

\author{Simone De Liberato}
\affiliation{School of Physics and Astronomy, University of Southampton, Southampton, SO17 1BJ, United Kingdom}

\author{Franco Nori}
\affiliation{Theoretical Quantum Physics Laboratory, RIKEN Cluster for Pioneering Research, Wako-shi, Saitama 351-0198, Japan}
\affiliation{Department of Physics, University of Michigan, Ann Arbor, Michigan 48109-1040, USA}
%%%
\begin{abstract}
{The permutational invariance of identical two-level systems allows for an exponential reduction in the computational resources required to study the Lindblad dynamics of coupled spin-boson ensembles evolving under the effect of both local and collective noise. 
Here we take advantage of this speedup to study several important physical phenomena in the presence of local incoherent processes, in which each degree of freedom couples to its own reservoir. 
Assessing the robustness of collective effects against local dissipation is paramount to predict their presence in different physical implementations.
We have developed an open-source library in Python, the Permutational-Invariant Quantum Solver (PIQS), which we use to study a variety of phenomena in driven-dissipative open quantum systems. 
We consider both local and collective incoherent processes in the weak, strong, and ultrastrong-coupling regimes. 
Using PIQS, we reproduced a series of known physical results concerning collective quantum effects and extended their study to the local driven-dissipative scenario. 
Our work addresses the robustness of various collective phenomena, e.g., spin squeezing, superradiance, quantum phase transitions, against local dissipation processes.
}
\end{abstract}
\maketitle
%%%%%
\section{Introduction}
\label{intro}
%%%%%
The simplest quantum model of an atom interacting with light is that of a two-level system (TLS) interacting with a single or multi-mode electromagnetic environment. The intrinsic nonlinearity of a TLS allows this model to capture a wide breadth of quantum optics phenomena, from the emission of anti-bunched light under strong driving, to ultrastrong coupling with artificial atoms. Nowadays the quantum properties of \emph{single} TLSs can be investigated on many different physical platforms beyond atoms, including trapped ions and NV centers, quantum dots and superconducting circuits \cite{Buluta11,Gu17}.

The steady improvement of experimental quantum technologies allows one to engineer the dynamics of \emph{many} TLSs. 
From a fundamental point of view, collective ensembles of TLSs can display nontrivial quantum correlations \cite{Dicke54} and be used to simulate more complex systems. Since a TLS is the physical embodiment of a qubit, its coherent control is sought after for quantum computing and information applications \cite{Georgescu14}.
% Figure 1
\begin{figure}[ht!]
\includegraphics[width = 8cm]{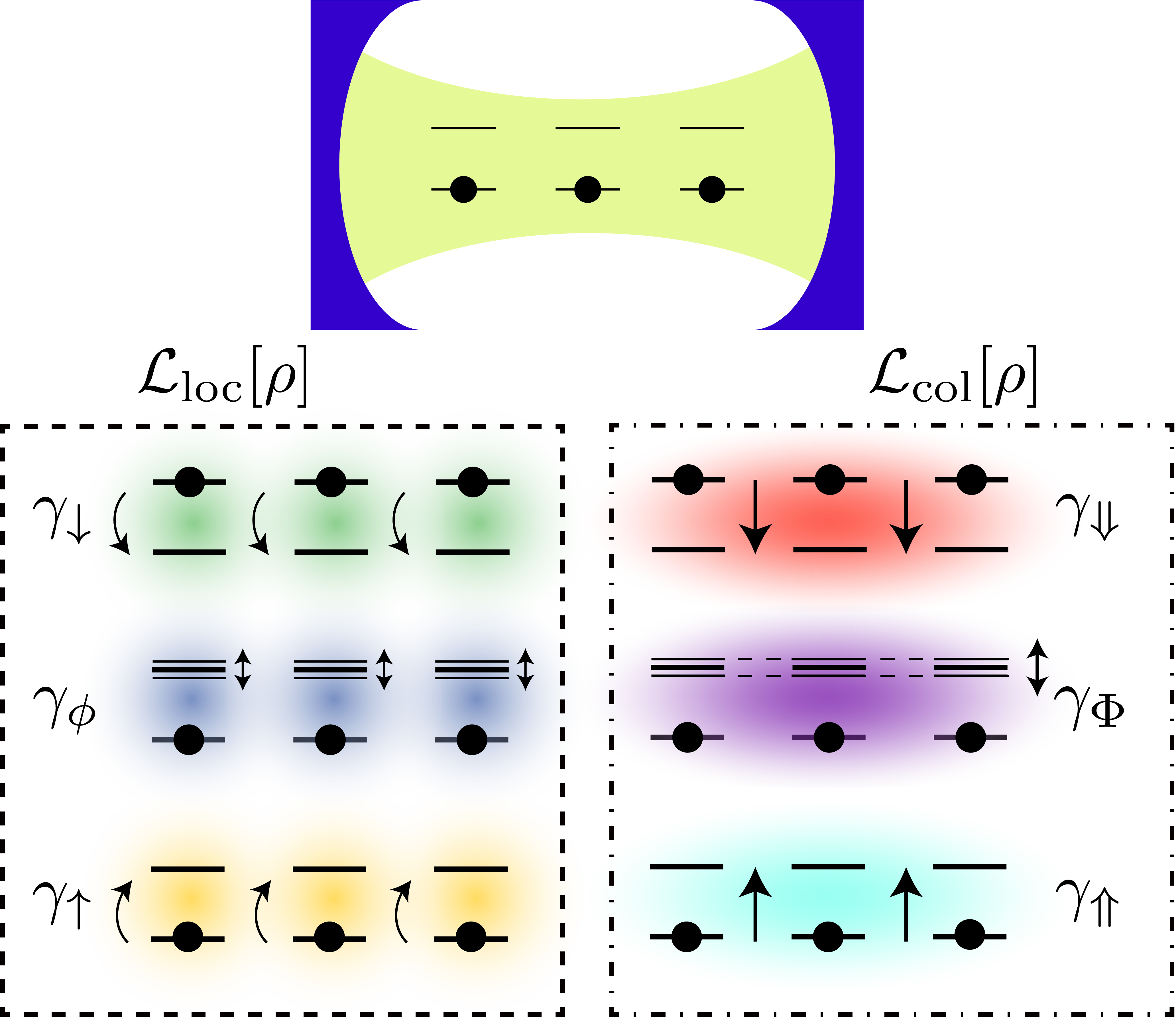}
\caption{\label{figxa} 
{\bf Open quantum system dynamics.} 
An ensemble of identical two-level systems is collectively coupled to a bosonic cavity through a coherent dynamics. 
The action of  dissipative processes on the two-level system dynamics is quantified by different rates for homogeneous local (dashed box) and collective processes (dot-dashed box), given by $\mathcal{L}_\mathrm{loc}[\rho]$ and $\mathcal{L}_\mathrm{col}[\rho]$, respectively: local and collective emission, set by $\gamma_{\downarrow}$ and $\gamma_{\Downarrow}$, local and collective dephasing, set by $\gamma_{\phi}$ and $\gamma_{\Phi}$, and local and collective pumping, set by $\gamma_{\uparrow}$ and $\gamma_{\Uparrow}$, respectively.}
\end{figure}

A major challenge in preserving the coherence of many-body quantum systems is that the system is perturbed by uncontrolled interactions with the environment, which can disrupt such fragile quantum coherence. Local impurities, defects, as well as phononic and photonic environments, prompt single TLSs to excite or de-excite incoherently or undergo dephasing, see Figure~\ref{figxa}. 
This breaks the time reversibility of the unitary dynamics that distinguishes isolated quantum systems. 
From a theoretical point of view, one can treat this problem of an \emph{open quantum system} with a range of powerful tools developed over the last few decades \cite{Haroche,CarmichaelI,CarmichaelII, Breuer02,Gardiner04}. 

The interplay of collective and local dissipative interactions in TLSs has been the subject of increased interest, as it has been shown to lead to the emergence of robust quantum effects. 
The prototypical example of collective behaviors in closed systems is the $\sqrt{N}$ enhancement of light-matter coupling. In the open setting, where only collective dissipation is considered, the cooperative enhancement in the collective decay leads to superradiant light emission \cite{Dicke54}.

Recently it has been shown that by incoherently pumping an ensemble of TLSs that can radiate superradiantly in the bad-cavity limit, one obtains an interesting coexistence of local and collective effects which lead to robust steady-state superradiance \cite{Meiser10a,Meiser10} and quantum synchronization \cite{Giorgi13,Xu14,Zhu15,Shankar17}. Steady-state superradiance results in the line-narrowing of emitted light, a feature proposed to improve the precision of atomic clocks by orders of magnitude \cite{Meiser09,Bohnet12,Norcia16, Norcia17,Tieri17}. Recent theoretical studies even propose to embrace the unavoidable nature of dissipative interactions and turn it into a resource, for example engineering a reservoir to steer the evolution of the open quantum system \cite{Diehl08,Verstraete09,Giorgi12,Manzano13,Peng14,Metelmann15,Reiter16,Qin18,Leroux18}.

However, when local and collective processes coexist in a driven-dissipative ensemble of TLSs, the complexity of the system makes the direct solution of the dynamics intractable analytically. This is an obstacle for reliable theoretical predictions that could guide experimental work. 
So far, different alternative strategies have been developed to tackle this problem: if the dissipative environment is fixed, one can design the unitary contribution to steer the dynamics to stable states \cite{Wang13,Sauer13}; the spectrum of the Lindbladian can be analyzed to find decoherence-free subspaces and the existence of multiple steady states \cite{Lidar98,You05,Paulisch16,AlbertPhD,Albert14,Albert16}; the corner-space renormalization method can be used to factorize a large lattice into sub-lattices \cite{Finazzi15}; the Keldysh path-integral formalism can be used to map the problem in terms of Majorana fermions, and other condensed-matter methods allow the treatment of both periodic driving and dissipation \cite{DallaTorre13,DallaTorre16,Sieberer16,Shchadilova18,Kirton18r}.

Besides special cases in which the dynamics can be solved analytically \cite{FossFeig13,FossFeig17}, one generally resorts to a bosonic approximation, e.g., with a Holstein-Primakoff transformation \cite{Holstein40,Klein91,Lambert04,Kavokin,Combescot08,Hayn11,Kessler12,Laurent15,Cortese17a,Kirton18r}, which is valid only in the dilute-excitation regime. 
Alternatively, a semiclassical approximation in which higher moments of TLS operators are factorized can be performed, truncating the hierarchy of equations and closing the system \cite{CarmichaelII,Meiser10,Scully15a,Shammah17,Kirton17b}. 
The cumulant expansion method is indeed an effective mean-field theory that improves for large $N$ but it gives access to limited information about the system's state (only up to second moments) \cite{Kirton17b,Kirton18r}.

Moreover, numerically simulating an open quantum system is typically more demanding than simulating its closed counterpart. Numerical methods can be employed \cite{CarmichaelI}, but, in the most general case, the dimension of the Hilbert space grows exponentially as $2^N$ with the linear increase in the number $N$ of TLSs. In the open quantum setting, even if one assumes only Markovian interactions with the environment, this means that the Liouvillian space grows as $4^N$. This challenge makes the numerical solution of the system intractable through a straightforward approach for large $N$.

A possible solution emerges if the system under study allows one to consider all of the TLSs as identical and identically prepared at an initial time. Within the validity of these assumptions, the open quantum dynamics across equivalent sectors of the Liouvillian space effectively maps in the same way due to the permutational symmetry of the system. The identical behavior and mapping lead to a drastic reduction of the resources required to describe the open system evolution \cite{Sarkar87b,Sarkar87,Martini01,Chase08,Baragiola10,Lee11,Hartmann16,Moroder12, Novo13,Xu13,Richter15,Bolanos15,Zhang15a,Wilson16,Xu,Gegg,Gegg16,Gegg17a,Gegg17b,Shankar17,Damanet16b,Kirton17,Kirton17b,Shammah17,Gong16,Zhang18a,Garttner17,Garttner18}. 

Hence standard numerical methods can exactly solve the dynamics of $N$ TLSs evolving under the action of homogenous local Lindblad dissipation terms without the need to deal with an exponentially large Liouvillian space. 
The treatment of identical TLSs evolving under such homogenous and local collective processes allows the scaling of the Liouvillian space to be ${O}(N^4)$, with actually only $O(N^3)$ non-zero matrix elements in the density matrix of the system. For a special class of problems, which comprises many interesting models, the scaling can be further reduced to ${O}(N^2)$.  
%%%
\subsection{This work}
%%%
In this paper we introduce PIQS, an open source computational library that exploits the permutational symmetry of identical TLSs. 
We use PIQS to solve different physical models that can be described with \emph{local and collective} Lindblad superoperators. Our implementation, in the form of a Python package, is a low-effort easy-to-use tool which can be set-up in a few minutes and applied to a wide class of problems. At the same time PIQS is optimized for performance and robustly tested. PIQS is natively integrated with QuTiP, the quantum toolbox in Python, in order to allow the user to take full advantage of the features provided by QuTiP \cite{Johansson12,Johansson13}.  

Using PIQS, in the last section of this paper we numerically investigate a number of model systems and effects, including superradiant light emission, steady-state superradiance, spin squeezing, limit cycles, and dissipatively-coupled ensembles of TLSs. 
We initially reproduce various known results, successfully testing the validity and efficacy of our library. 
By including \emph{local} dissipation terms, we then demonstrate that some of these effects become unstable, establishing important limits on the class of physical implementations in which those different effects could potentially be observed. Moreover, we investigate how the TLS \emph{nonlinearity} affects dissipative systems in the ultrastrong-coupling regime.

The article is structured as follows. 
In Section~\ref{theo} the Lindblad master equation representing the physical model is introduced. 
We show how permutational invariance is applied to simplify the description of the density matrices, the overall dynamics and the rate-equation dynamics for populations. 
We also provide an overview of existing works that exploit permutational invariance, pointing out connections among apparently unrelated works. We discuss how the simplification can be used to study multiple ensembles of TLSs in different cavities. 

In Section~\ref{piqs}, we describe the structure of the computational library, with information on the usage of the code, development and description of the various parts and optimizations that allow the study of many spin ensembles. 
Furthermore we provide information regarding the time and space complexity of the problems and assess the performance of the code.

In Section~\ref{results}, we demonstrate how PIQS can be applied to study the influence of local dissipation on collective phenomena. 
As representative examples we investigate how the time evolution of different quantum states can enhance superradiant light emission \cite{Shammah17} and spin squeezing \cite{Chase08}. 
Local emission processes are found detrimental for steady-state superradiant light emission \cite{Meiser10a}, precluding its straightforward application to thermodynamics systems governed by detailed balance. 
We also study phase transitions in such noisy open dynamics, extending previous work \cite{Kirton17,Kirton17b}. 
Moreover, we analyze a coherently driven system undergoing collective incoherent dissipation, as well as local dissipation processes. 
The collective spin oscillations of such systems have recently been interpreted as the signature of a quantum phase transition and named boundary time crystals \cite{Iemini17}. We generalize the study of such limit cycles in the presence of local dissipation.
Being able to study multiple spin ensembles and their coupling to one or multiple bosonic cavities, we also investigate dissipatively-coupled ensembles of TLSs, and the transient exchange of collective excitation in a noisy environment \cite{Hama16}. 

Finally, we apply permutational invariance to the ultrastrong-coupling regime. 
Python notebooks containing the code for each physical model treated here, and other investigations, can be found online \cite{Piqs}. Appendix~\ref{app1} details the explicit derivation of the matrix elements of the Lindbladian.

%%%%
\subsubsection*{Original results}
\label{newres}
%%%%
Before moving to the next sections, let us provide a brief summary of the original results obtained in this investigation, as discussed in Section~\ref{results}:
\begin{itemize}
{\item In Section~\ref{sre}, we take advantage of PIQS to discriminate different time evolutions for different initial states evolving under the same superradiant dynamics. We illustrate how very different time evolutions can arise for states with same second moments, such as for the Greenberger-Horne-Zeilinger (GHZ) state and the superradiant Dicke state $\ket{\frac{N}{2},0}$. We find that when the system is initialized in the $\ket{0,0}$ Dicke state, the introduction of local dephasing is favorable for light emission. Moreover, we verify that under the non-optimal superradiant decay, the symmetric and antisymmetric coherent spin states emit light in a similar fashion.}
{\item In Section~\ref{ssr}, we find that, in the bad-cavity limit, steady-state superradiance does not display a threshold pump rate if detailed balance governs local losses and local pumping, at any temperature.}
{\item In Section~\ref{lmg}, we investigate two-axis spin squeezing beyond the Dicke symmetric ladder, finding that the state with longest spin-squeezed time evolution does not belong to the symmetric Dicke ladder for the local de-excitation channel. We consider the trade-off between spin squeezing and spin squeezing time, for all non-symmetrical collective states, addressing the interplay of local and collective dissipation and the emergence of spin squeezing with the system size growth.}
{\item In Section~\ref{qpt}, we assess the effect of collective loss and pumping to the stability of the superradiant phase in the presence of local dephasing and collective incoherent processes. We then turn to the study of a driven-dissipative system that has been shown to sustain time crystallization, and we probe how the introduction of local and collective pure dephasing affects the visibility of the collective spin oscillations related to limit cycles.}
{\item Generalizing our approach to multiple ensembles of TLSs, in Section~\ref{negtemp}, we show that negative-temperature effects are not robust to local dephasing, local losses or even collective losses of each of the TLS ensembles. Conversely, we propose to exploit such dissipative dynamics for transient spin-excitation schemes in arrays of coupled ensembles.}
{\item Finally, in Section~\ref{sec:usc}, we extend the use of permutational invariance to the ultrastrong-coupling regime. We illustrate how the correct master equation with dressed local jump operators, can be analytically derived easily from the weak-coupling model, exploiting permutational symmetry. We then address the time evolution of the collective TLS population inversion and the steady-state photon emission spectrum in the range $N\gg 1$ for the open Dicke model in the ultrastrong-coupling regime with dressed light-matter dissipation terms.}
\end{itemize}

%%%%
\section{Theory}
\label{theo}
%%%%
Here, we study the dynamics of a collection of $N$ identical TLSs, described by a Lindblad master equation,
\begin{eqnarray}
\label{master}
\dot{\rho} &=& 
- \frac{i}{\hbar}\lbrack H,\rho \rbrack
 +\frac{\gamma_{\Downarrow}}{2}\mathcal{L}_{J_{-}}[\rho] +\frac{\gamma_{\Phi}}{2}\mathcal{L}_{J_{z}}[\rho]  +\frac{\gamma_{\Uparrow}}{2}\mathcal{L}_{J_{+}}[\rho] \nonumber\\&&
 +\sum_{n=1}^{N}\left(\frac{\gamma_{\downarrow}}{2}\mathcal{L}_{J_{-,n}}[\rho] +\frac{\gamma_{\phi}}{2}\mathcal{L}_{J_{z,n}}[\rho]
+\frac{\gamma_{\uparrow}}{2}\mathcal{L}_{J_{+,n}}[\rho]\right),
\end{eqnarray}
where $\rho$ is the density matrix of the full system and $H$ is the TLS ensemble Hamiltonian. Here, $[J_{x,n},J_{y,m}]=i\delta_{m,n}J_{z,n}$, $[J_{+,n},J_{-,m}]=2\delta_{m,n}J_{z,n}$ and $J_{\pm,n}=J_{x,n}\pm iJ_{y,n}$. We will use the spin operators $J_{\alpha,n}=\frac{1}{2}\sigma_{\alpha,n}$ for $\alpha=\{x,y,z\}$ and $J_{\pm,n}=\sigma_{\pm,n}$. Note that $H$ should be invariant under TLS permutation, i.e., it can be constructed using any combination of the collective operators $J_{\alpha}$ only and describes all-to-all spin interactions (one way of looking at this is to picture the spins as nodes in a complete graph, in which each edge, describing the coupling of the $i$-th and $j$-th spin is simply associated to the same global constant, i.e. there is no notion of lattice dimensionality in this fully-connected network \cite{Rotondo18}).

The Lindblad superoperators are defined by $\mathcal{L}_{A}[\rho]=2A\rho A^\dagger - A^\dagger A \rho -\rho A^\dagger A$
and the $\gamma_{i}$ terms are coefficients characterizing emission, dephasing and pumping, corresponding to local and collective operators acting on the $n$-th TLS, also summarized in Table~\ref{tablej}.
% Table I v1
%\mysquare{greenpiqs} %em
%\mydisk{bluepiqs}  %deph
%\mydiamond{orangepiqs} %pump 
%\mydisk{redpiqs} %c.em
%\tikz\draw[black,fill=purplepiqs] (0,0) circle (.8ex); %c.deph
%\mydisk{cyanpiqs} %c.pump
\begin{table*}[ht!]
\centering		
\renewcommand{\arraystretch}{2} 
\renewcommand{\tabcolsep}{12pt}
\begin{tabular}{|l |c |c |l|}
\hline \hline
\bf{Process}&\bf{Jump Operator} &\bf{Rate} & \bf{Lindbladian $\mathcal{L}[\rho]$}\\
\hline \hline			
 \mysquare{greenpiqs}  
Local Emission &$J_{-,n}=\sigma_{-,n}$ 
& $\color{black}{\gamma_{\downarrow}}$ 
&$\sum_n^N2J_{-,n}\rho J_{+,n} - \{\left(\frac{N}{2}+J_z\right),\rho\}$\\
\hline
\mydisk{bluepiqs} 
Local Dephasing &$J_{z,n}=\frac{1}{2}\sigma_{z,n}$ 
&$\color{black}{\gamma_{\phi}}$ 
&
$\sum_n^N2J_{z,n}\rho J_{z,n} - \frac{N}{2}\rho$\\
\hline
\mydiamond{orangepiqs} 
Local Pumping &$J_{+,n}=\sigma_{+,n}$ 
& $\color{black}{\gamma_\uparrow}$ 
&$\sum_n^N2J_{+,n}\rho J_{-,n}- \{\left(\frac{N}{2}-J_z\right),\rho\}$\\
\hline
\myesquare{redpiqs} 
Collective Emission &$J_{-}=\sum_n^NJ_{-,n}$ 
& 
$\color{black}{\gamma_{\Downarrow}}$ 
&$\sum_{m,n}^N2J_{-,m}\rho J_{+,n}- J_{+}J_{-}\rho-\rho J_{+}J_{-}$\\
\hline
\myedisk{purplepiqs}
Collective Dephasing &$J_{z}=\sum_n^NJ_{z,n}$ 
& 
$\color{black}{\gamma_{\Phi}}$ 
&$\sum_{m,n}^N2J_{z,m}\rho J_{z,n}- J_{z}^2\rho-\rho J_{z}^2$\\
\hline
\myediamond{cyanpiqs} 
Collective Pumping &$J_{z}=\sum_n^NJ_{+,n}$ 
&
$\color{black}{\gamma_{\Uparrow}}$ 
&$\sum_{m,n}^N2J_{+,m}\rho J_{-,n}- J_{-}J_{+}\rho-\rho J_{-}J_{+}$\\
\hline
\hline
\end{tabular}
\caption{\label{tablej} Summary of the dissipative processes considered in \Eq{master} for the open quantum dynamics of an ensemble of $N$ identical TLSs. }
\end{table*}

The homogeneous \emph{local emission}, $\gamma_{\downarrow}$, usually represents radiative or non-radiative losses while the \emph{homogeneous local pumping}, $\gamma_{\uparrow}$, is the coefficient quantifying the rate of incoherent pumping. 
In the context of bosonic heat baths governed by detailed balance, we can identify $\gamma_{\downarrow}=\gamma_{0}(1+n_\text{T})$ and $\gamma_{\uparrow}=\gamma_{0}n_\text{T}$, where $n_\text{T}$ is the thermal population of the environment and $\gamma_{0}$ is a coefficient fixed for a given system. \emph{Homogeneous local dephasing}, detrimental for coherent correlations among the TLSs, is quantified by $\gamma_{\phi}$. The corresponding collective phenomena describe \emph{collective emission}, $\gamma_{\Downarrow}$, typical of superradiant decay, \emph{collective pumping}, $\gamma_{\Uparrow}$, and  \emph{collective dephasing}, $\gamma_{\Phi}$.  

It is worth recalling that \Eq{master} is derived \cite{CarmichaelI} under the assumptions that the environments are memory-less (Markov approximation), that system and environment always remain in a product state (Born approximation), and that the baths are uncorrelated. These assumptions cannot always be made, such as in the case of photosynthetic complexes \cite{Lambert13}, which are strongly coupled to structured reservoirs \cite{Iles16}.

Furthermore, in \Eq{master} the reservoir is traced out and the TLS bare basis is used to construct the system's superoperators. If the TLS are strongly coupled to another system, e.g., a cavity, one must derive a more nuanced master equation to take into account the hybridization of the system eigenstates \cite{DeLiberato09c,Ashhab10,Beaudoin11,Bamba12,Ridolfo12,Bamba14,DeLiberato14,DeLiberato17}. Fortunately, as we show in Sec. \ref{sec:usc}, it is possible to extract such an equation from the general dissipators we construct in PIQS.
(Though this procedure is not strictly necessary in systems where the coupling is only ``effectively'' strong, and where the bare-basis master equation is still valid \cite{Dimer07,Baumann10,Baden14,Kirton17,Puebla17,Braumuller17,Langford17,Lamata17,Barberena17,Lv17,Qin18,Leroux18}).

Let us stress that while the numerical methods employed hereafter allow us to \emph{exactly} solve the dynamics of \Eq{master}, the master equation is a second-order perturbative expansion of the quantum dynamics induced by the time-ordering operator, and in the form of \Eq{master} it implies performing a rotating-wave approximation.
 As such, the $\gamma_i$ coefficients of \Eq{master} must be much smaller than the coupling present in the Hamiltonian. 
Nevertheless, if only the bare-energy TLS Hamiltonian is present, $H=\hbar\omega_0J_z$, the dissipative dynamics is insensitive to the physical value of the bare frequency, $w_0$. The same argument applies to any Hamiltonian diagonal in the Dicke basis. 
In addition, in driven systems, when the real physical frequencies of the system are large, the Hamiltonian can be written in a rotating frame, where the effective Hamiltonian dynamics can be slower than dissipation, as is the case of the non-equilibrium superradiant phase transition \cite{Kirton18r}.
Thus \Eq{master} can also be used as an effective model, with the real physical frequencies different from the effective Hamiltonian parameters, 
although these vary case by case, and in general one needs to be careful in checking whether in a given case this approach can be justified.

The crucial assumption which underlines the permutational-invariant method used hereafter to solve \Eq{master} is that there the TLSs are identical and identically prepared. This means that the approach described here cannot be \emph{directly} applied to cases in which a non-negligible inhomogeneous broadening is present \cite{Lalumiere13,Lambert16} or in which each artificial atom might have a tailored coupling to the environment \cite{Tudela13,Galve17,Kockum18}. Furthermore, relative to the collective dissipation terms in \Eq{master}, the assumption made is that the TLSs couple identically to a single mode of the environment, e.g., implying that the wavelength of light is larger than the sample size of the whole TLS ensemble and dipole-dipole interactions and atomic motion are negligible, assumptions that might not always be valid in some realistic implementations \cite{Scully06,Yan13,Guerin16,Damanet16a,Damanet16c,Zhu16,Bromley16}. Moreover, the Dicke space formalism that will be applied hereafter assumes that the initial state of the system must be permutational symmetric, i.e., one is limited to the initial condition in which none of the artificial atoms are individually addressable. Dynamics of non permutation-symmetric initial states could in general be studied, as long as the calculated observables are permutational symmetric, e.g., collective spin moments.

The fact that all TLSs are identical affords us a critical simplification because, even if local processes are present, we do not need to work in the full $4^N$ Liouvillian space, due to the permutational symmetry of the system and the permutational invariance of the Lindblad superoperators, which possess a SU(4) symmetry. We can reduce the Liouvillian space to be of the order of $N^4$ instead, with actually only $O(N^3)$ non-zero matrix elements for the density matrix of a permutational-symmetric quantum system. 

The reduction in complexity can be captured by using the Dicke state basis, which will be introduced in the next section. 
The symmetric properties of identical TLSs initially prepared through the manipulation of collective spin operators only produce a sparse block-diagonal density matrix. 
Using the Dicke basis, \Eq{master} can be efficiently rewritten as a symmetrized Liouvillian superoperator $\mathcal{D}_\text{S} $,
\begin{eqnarray}
\dot{\rho} = \mathcal{D}_\text{S}[\rho],
\end{eqnarray}
allowing to simulate ensembles with a large number of TLSs, of the order of hundreds. 
This computational advantage enables the study of TLS ensembles coupled to bosons that also experience a driven-dissipative dynamics, so that \Eq{master} can be generalized to the light-matter Liouvillian $\mathcal{D}_\text{SB}[\rho]$,
\begin{eqnarray}
\label{masterbos}
\dot{\rho}&=&\mathcal{D}_\text{S}[\rho] -\frac{i}{\hbar}\lbrack H_\text{B} + H_\text{SB},\rho \rbrack
+\frac{w}{2}\mathcal{L}_{a^\dagger}[\rho]
+\frac{\kappa}{2}\mathcal{L}_{a}[\rho],\nonumber\\
\end{eqnarray}
where $\lbrack a, a^\dagger\rbrack=1$, $w$ is the bosonic pump coefficient and $\kappa$ is the bosonic cavity decay.  
Now $\rho$ is defined as a tensor product of the density matrices of the bosonic and spin subspaces. 
Here $H_\text{B}$ is a bosonic Hamiltonian, while $H_\text{SB}$ is the spin-boson interaction. 
Given that the bosonic Hilbert space has infinite dimension, numerical implementations require an effective cut-off on the photon number $n_\text{ph}$. 
The dimension of the total light-matter Liouvillian space is then multiplied by $n_\text{ph}^2$.
Note that \Eq{master} and \Eq{masterbos} can be further generalized to describe arrays or lattices of multiple TLS ensembles in multiple bosonic cavities, see Figure~\ref{figxa} for the case of a single cavity. 
In this modular architecture, adding each dissipative sub-system implies a tensor product with the Liouvillian spaces of the other sub-systems.
This means that, in the most general case, the size of the problem grows very fast. %%%
%%%
\subsection{Dicke states}
\label{pi}
%%%
Before introducing the details of the Liouvillian dynamics, let us begin by making some general considerations on the structure of the Hilbert space of an isolated system. 
The study of the collective behavior of ensembles of identical TLSs can be simplified by introducing the Dicke states \cite{Dicke54}, which are the eigenstates of the collective (pseudo-)spin operators
\begin{eqnarray}
\label{Dicke}
\mathbf{J}^2\ket{j,m}&=&j(j+1)\ket{j,m}\\
J_z\ket{j,m}&=&m\ket{j,m}
\label{Dicke2}\\
J_{\pm}\ket{j,m}&=&A^{\pm}_{j,m}\ket{j,m\pm1}=\sqrt{(j\mp m)(j\pm m + 1)}\ket{j,m\pm1},\nonumber\\
\label{Dicke3}
\end{eqnarray} 
where $j\leq N/2$ and $|m|\leq j$, with $j,m$ integer or half-integer and $j_\text{min}=0,1/2$ for $N$ as an even or odd number of TLSs, respectively. The collective spin algebra ensures that in \Eq{Dicke3}, $A^{\pm}_{j,m}$ are real semi-definite coefficients. The representation of collective processes using the Dicke states reduces the $2^N$-dimensional Hilbert space to a dimension of the $O(N^2)$ thanks to the intrinsic permutational symmetry of $J_\alpha$ operators.
The symmetric Dicke states, defined as the Dicke states with $j=\frac{N}{2}$, have an intuitive construction in terms of the uncoupled eigenstates of the single TLSs, as they are the symmetric superposition of a state with $k$ excited TLSs, 
\begin{eqnarray}
\label{nkm}
\ket{\frac{N}{2}, k - \frac{N}{2} } = \frac{1}{\sqrt{N \choose k}} \mathcal{S} \bigg[ \ket{e}^{\otimes k} \otimes \ket{g}^{\otimes (N-k)} \bigg]
\end{eqnarray}
where $\mathcal{S}$ is the symmetrization operator, $N \choose k$ is the binomial coefficient, and $\ket{g}$ and $\ket{e}$ are the ground and excited states, respectively, in the uncoupled basis for each TLS, and $(k - \frac{N}{2})=m$. 

Note that the action of the collective operators in Eqs.~(\ref{Dicke}-\ref{Dicke3}) restricts the Hilbert space for $N$ TLSs to $(N + 1)$ symmetric states. 
The Dicke states with $j<\frac{N}{2}$ can be constructed iteratively  \cite{Mandel}, as done in the standard construction of Clebsch-Gordan coefficients. The state $\ket{\frac{N}{2}-1,\frac{N}{2}-1}$ is the only other state orthogonal to $\ket{\frac{N}{2},\frac{N}{2}-1}$ with $(N-1)$ excitations; all of the other states of the kind $\ket{\frac{N}{2}-1,m}$ in that Dicke ladder are found just by applying the ladder operator $J_{-}$. 

The states in the other Dicke ladders can be found by iterating the orthogonalization process for every state $\ket{j,j-1}$.
Each Dicke state $\ket{j,m}$ has a degeneracy  
\begin{eqnarray}
\label{eq8d}
d^j_N&=& (2j + 1)\frac{N! }{(\frac{N}{2}+  j + 1)!(\frac{N}{2} -  j )!},
\end{eqnarray}
with respect to an irreducible representation in the uncoupled TLS basis. 
It is this degeneracy at the core of the formalism that allows the exponential reduction in resources employed in PIQS. 
Note that non-symmetric Dicke states with $j<\frac{N}{2}$ have been addressed in multiple spin systems \cite{Bradac16} and have recently been obtained in a single Rydberg atom with high excitation number \cite{Facon16}. 
%%
%%
%%%
\subsection{Permutational symmetry}
\label{ld}
%%%
Let us now approach the Liouvillian dynamics. 
The challenge introduced by the local operators $J_{\alpha,n}$ present in \Eq{master} is that in general one does not know how they act on the Dicke states. 

For $j=\frac{N}{2}$, the problem is trivial, as any Dicke state $\ket{\frac{N}{2},m}$ of \Eq{nkm} is a symmetric superposition of $k=(\frac{N}{2}-m)$ excited TLSs in the uncoupled basis. 

For $j<\frac{N}{2}$ though, there exist $d^N_j$ degenerate non-symmetric superpositions of TLSs in the uncoupled basis and the action of $J_{\alpha,n}$ requires writing the state explicitly in the uncoupled basis. 
In order to map the action of local processes onto the Dicke states one needs to define a more general Dicke state, $\ket{j,m,\alpha_j}$ \cite{Dicke54}, where the additional symmetry quantum number $\alpha_{j}$ removes the degeneracy $d^j_N$. Using this basis, a general density matrix can be written as
\begin{eqnarray}
\label{eq0}
\rho&=&\sum_{\substack{j, m, \alpha_{j}\\j', m', \alpha_{j'}}}p_{jm\alpha_{j},j'm'\alpha_{j'}}\ket{j,m,\alpha_{j}}\bra{j',m',\alpha_{j'}},
\end{eqnarray} 
where ${p_{jm\alpha_{j},j'm'\alpha_{j'}}} =\langle j, m, \alpha_{j}|\rho |j',m',\alpha_{j'}\rangle$, while $p_{jm\alpha_{j}}\equiv p_{jm\alpha_{j},jm\alpha_{j}}$, is the probability density of a given diagonal matrix element, and there is always a mapping to a given microscopic representation in the uncoupled TLS basis. The usefulness of \Eq{eq0} is limited, as finding all of the $2^N$ representations is tedious and effective strategies employing them are known only for the subspace with $m=(\mp\frac{N}{2}\pm1)$, relevant for single-photon excitation processes \cite{Svidzinsky08,Svidzinsky10,Svidzinsky12,Chen12,Vetter16,Han16}.

\begin{figure*}[ht!]
\centering
	\includegraphics[width=14cm]{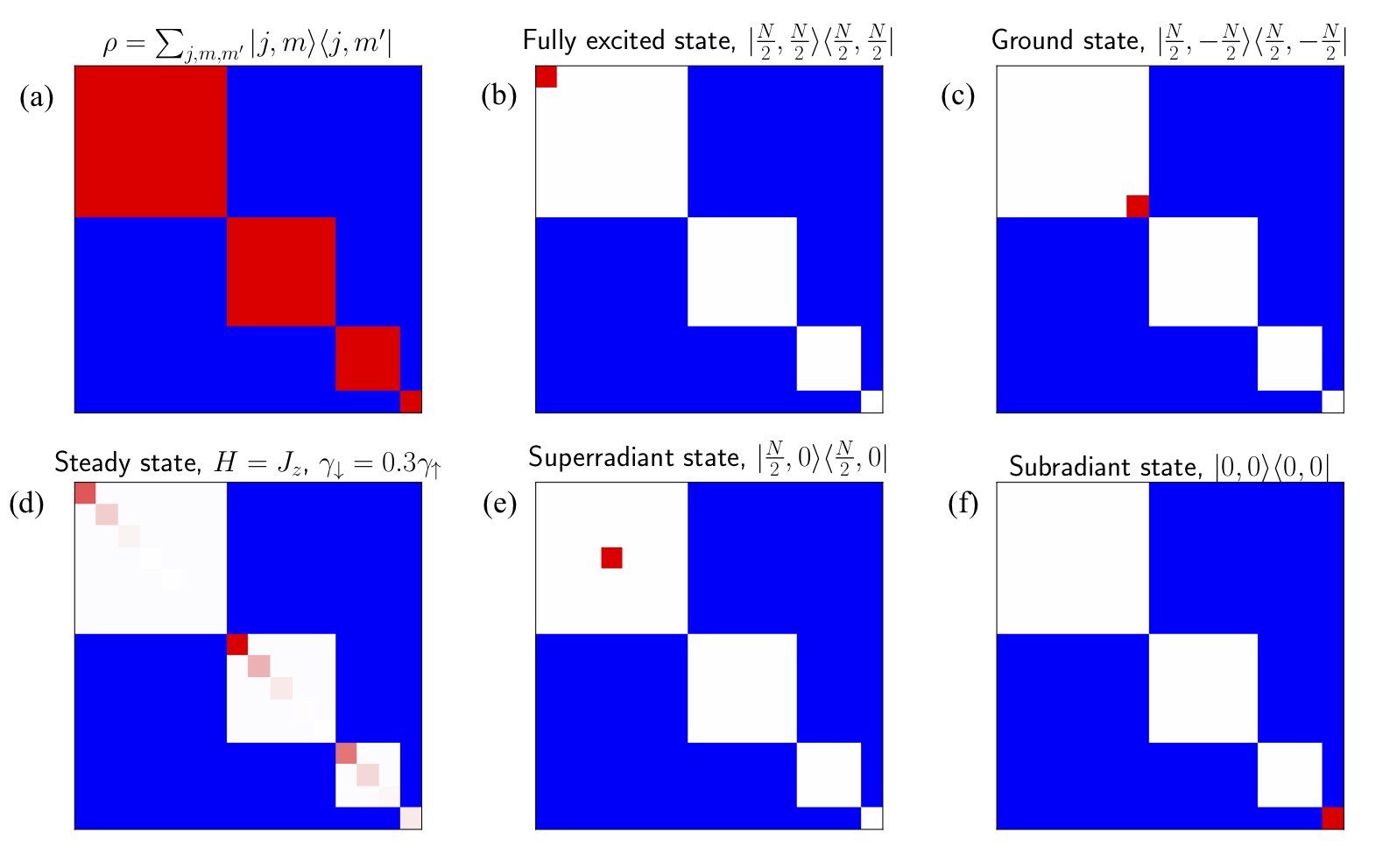}
	\caption{\label{figxds} {\bf Dicke space structure.}
(a) Block-diagonal structure of the density matrix in the Dicke basis, $\ket{j,m}\bra{j,m'}$, for $N=6$. Each block has a different value of $j$, decreasing from the top-left one for which $j = \frac{N}{2}$. The size of each block is set by the multiplicity of $m$. 
Red equals 1, white equals 0. 
The blue region outside of the diagonal blocks (darker shading) is shown to highlight sectors that cannot be populated.
(b) The density matrix of the Dicke state $\ket{\frac{N}{2},\frac{N}{2}}$, which is the fully excited state (top left matrix element). 
(c) The Dicke state $\ket{\frac{N}{2},-\frac{N}{2}}$ is the ground state if the Hamiltonian $H=\hbar\omega_0J_z$ (bottom right matrix element in the first block). 
(d) Qualitative representation (saturated palette) of the steady state for $H=\hbar\omega_0J_z$, and $\gamma_\downarrow =0.3\gamma_\uparrow$, $\gamma_\uparrow=\omega_0$. 
Due to the choice of the Hamiltonian, only the matrix elements $p_{jmm}$ on the main diagonal can be populated, and the choice of the relative strength of the dissipation determines the fact that the matrix elements $p_{jjj}$ are the most populated ones.  
(e) The density matrix of the maximally-symmetric Dicke state $\ket{\frac{N}{2},0}$ (central matrix element in the first block).
(f) The density matrix of the subradiant Dicke state $\ket{0,0}$.
All these states are diagonal in the Dicke space, $p_{jmm'}=0$ for $m\neq m'$ (bottom right matrix element).
}
\end{figure*}

While \Eq{master} does not limit its action to the $(N+1)$ symmetric states of the coupled Dicke basis $\ket{\frac{N}{2},m}$, it does preserve the permutational symmetry of the system \cite{Sarkar87b,Sarkar87,Martini01,Chase08, Baragiola10,Lee11,Hartmann16,Moroder12,Novo13,Xu13,Damanet16b,Shammah17} because the Lindblad superoperators are invariant under the SU(4) transformation. 
This observation is \emph{crucial} to enable the simplification that is at the core of PIQS.
The density matrix of $N$ identical TLSs that are initially prepared through the action of collective spin operators is permutational symmetric \cite{Messiah99}, 
\begin{eqnarray}
\label{eq1}
\rho&=&\sum_{\pi\in S_N} P_\pi\rho P_\pi^\dagger,
\end{eqnarray}
where $P_\pi$ is a given permutation operator and the sum is over any possible permutation $\pi$. According to \Eq{eq1} we can thus project $\rho$ on the basis $\ket{j,m}\bra{j',m'}$ of the Dicke states. Furthermore \Eq{master} does not create coherences (matrix elements with $m\neq m'$) between Dicke states with $j\neq j'$, so that we can write any density matrix in the \emph{Dicke basis} simply as 
\begin{eqnarray}
\label{eq1b}
\rho&=&\sum_{j, m, m'}p_{jmm'}\ket{j,m}\bra{j,m'},
\end{eqnarray} 
with $p_{jmm'}=\bra{j,m}\rho\ket{j,m'}$. 
Note that \Eq{eq1b} is a block diagonal matrix, as shown in Figure~\ref{figxds}(a) for $N=6$. 
Red represents $p_{jmm'}=1$ and blue 0, as it corresponds to sectors that cannot be accessed by the permutational symmetric density matrix written as in \Eq{eq1b}.

For any permutationally-symmetric density matrix, each block with fixed $j$ contains the matrix elements corresponding to the degenerate irreducible representations $\ket{j,m}\bra{j,m'}$.
The value of $j$ decreases from $j=\frac{N}{2}$ (top-left block in the states of Figure~\ref{figxds}) to $j_\text{min}$ (bottom-right block) 
and the size of each block is given by the spin multiplicity $(2j+1)$.
Each block $j$ has a degeneracy $d_j^N$, given by \Eq{eq8d}, and $d_j^N$ grows with decreasing $j$ (except for the only case $d_{j_\text{min}+1}^N>d_{j_\text{min}}^N$).
This means that, while the corresponding blocks become smaller and smaller with decreasing $j$, moving from left to right in the matrix of Figure~\ref{figxds}, the number of degenerate states represented by each block actually increases exponentially. 
The total number of elements in the density matrix is thus $\sum_{j=j_\text{min}}^{N/2}(2j+1)^2=\frac{1}{6}(N+1)(N+2)(N+3)=O(N^3)$. 

In order to better illustrate the power of this representation, in panels (b)-(f) of Figure~\ref{figxds} we show some density matrices that are diagonal in the Dicke basis, with the white parts representing elements with $p_{jmm'}=0$. 
The Dicke state $\ket{\frac{N}{2},\frac{N}{2}}$, Figure~\ref{figxds}(b), and $\ket{\frac{N}{2},-\frac{N}{2}}$, Figure~\ref{figxds}(c), represent the states with all spins up and all spins down, respectively, which, for $H=\hbar\omega_0J_z$, correspond to the fully-excited and ground states, respectively. The maximally symmetric superradiant state $\ket{\frac{N}{2},0}$, shown in Figure~\ref{figxds}(e) also contains a non-zero matrix element in its largest block, with $j=\frac{N}{2}$, while the subradiant state $\ket{0,0}$, Figure~\ref{figxds}(f), contains a non-zero matrix element in the degenerate block with $j=0$.
The steady state of \Eq{master} with  $H=\hbar\omega_0J_z$, $\gamma_\downarrow=0.3\gamma_\uparrow$, and setting $\gamma_\downarrow=\omega_0$, is a classical mixture of Dicke states and (like any steady or thermal state of a Hamiltonian diagonal in the Dicke basis) is characterized by matrix elements only on the main diagonal.

\begin{figure*}[ht!]
\centering
	\includegraphics[width=16cm]{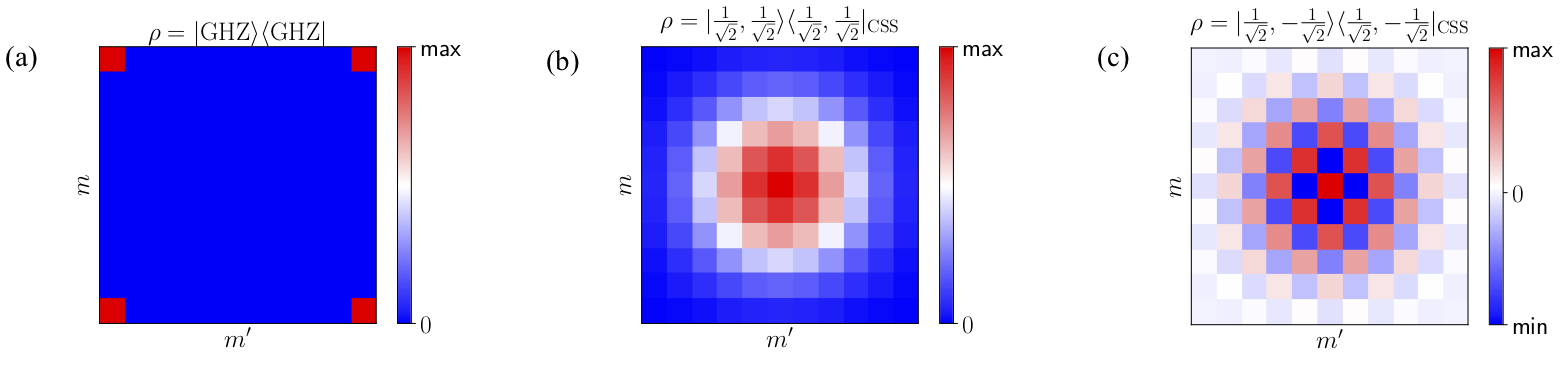}
	\caption{\label{figxdm} {\bf Non-diagonal density matrices.}
Representation of quantum states in the Dicke basis $\ket{j,m}\bra{j,m'}$. We show for $N=10$ only the block with $j=\frac{N}{2}$, which corresponds to the largest one in Figure~\ref{figxds}(a), since it is the only block with non-zero matrix elements. 
(a) The GHZ state (four red matrix elements at the corners of the first block), (b) the symmetric coherent spin state (CSS) with $a=b=\frac{1}{\sqrt{2}}$ in \Eq{cssab1}, (c) and the antisymmetric CSS, with $a=-b=\frac{1}{\sqrt{2}}$ in \Eq{cssab1}.
}
\end{figure*}

In Figure~\ref{figxdm}, we instead show the upper left block, with $j=\frac{N}{2}$, of density matrices that are not diagonal in the Dicke basis. These are density matrices in which the only non-zero terms occupy the $j=\frac{N}{2}$ block. 
In Figure~\ref{figxdm}(a), we show the GHZ state, which in the uncoupled and Dicke basis, respectively, is written as 
\begin{eqnarray}
\label{ghz}
\ket{\mathrm{GHZ}}&=&\frac{1}{\sqrt{2}}\left(\ket{e}^{\otimes N}+\ket{g}^{\otimes N}\right)=\frac{1}{\sqrt{2}}\left(\ket{\frac{N}{2},\frac{N}{2}}+\ket{\frac{N}{2},-\frac{N}{2}}\right).\nonumber\\
\end{eqnarray}

In Figure~\ref{figxdm}(b) and Figure~\ref{figxdm}(c), we show the symmetric and antisymmetric coherent spin states (CSS), respectively, 
which are a specific case of the general CSS state \cite{Ma11,Pezze16}
\begin{eqnarray}
\label{cssab1}
\ket{a,b}_\text{CSS}&=&\bigotimes_{n=1}^N\left(a\ket{e}+b\ket{g}\right)\\
&=& \sum_{m= -N/2}^{N/2} \sqrt{d_m}\alpha^{\frac{N}{2}+m}\beta^{\frac{N}{2}-m}\ket{\frac{N}{2},m},
\label{cssab}
\end{eqnarray}
with $a=\frac{1}{\sqrt{2}}$ and $b=\pm\frac{1}{\sqrt{2}}$, respectively. 
In \Eq{cssab}, $d_m={N \choose {\frac{N}{2}+m}}$.
The CSS states are also commonly written in the polar basis, 
\begin{eqnarray}
\label{csst}
\ket{\theta,\varphi}_\text{CSS}&=&\bigotimes_{n=1}^N\left(\cos\frac{\theta}{2}\ket{e}+e^{i\varphi}\sin\frac{\theta}{2}\ket{g}\right).
\end{eqnarray}
%%%
% Figure 2
%Figure 3
The diagonal matrix elements  $p_{jmm}$ correspond to populations of Dicke states $\ket{j,m}$. 
%%%%
The total number of Dicke states $n_\text{DS}$ in the triangular Dicke space is 
\begin{eqnarray}
\label{eq9}
n_\text{DS}&=&\sum_{j=j_\text{min}}^{N/2}(2j+1)=\left(\frac{N}{2}+1\right)^2-\frac{1}{4}\text{mod}_2(N),
\end{eqnarray} 
where the modulo term in the formula above takes care of ensembles with odd number of systems, for which $j_\text{min}=\frac{1}{2}$. 

%%%%
\subsection{Permutational-invariant dynamics}
%%%%
Equation~(\ref{master}) can be rewritten by flattening the density matrix into a vector, $\rho\rightarrow \vec{\rho}$, and applying to it a matrix representation of the superoperator that contains the full Liouvillian as a $n_\text{DS}^2\cdot n_\text{DS}^2$ matrix $D$, with $n_\text{DS}$ given by \Eq{eq9}, which is the sum of a unitary, $\tilde{H}$, and a dissipative part, $\tilde{L}$,
\begin{eqnarray}
\label{eq14a}
\frac{d}{dt}\vec{\rho}&=&D \vec{\rho}=\left(i\tilde{H}+\tilde{L}\right)\vec{\rho}.
\end{eqnarray}

The non-zero matrix elements from \Eq{master} are obtained by projecting onto the Dicke states $\bra{\bar{j},\bar{m}}\dot{\rho}\ket{\bar{j}',\bar{m}'}$. 
For simplicity of notation hereafter we set $\bar{j},\bar{j}'\rightarrow j,j'$ and $\bar{m}, \bar{m}'\rightarrow m,m'$.
Due to the permutational symmetry of the Lindblad superoperators, all terms with ${j}\neq {j}'$ are zero. 

\begin{figure*}[ht!]
\centering
\includegraphics[width=17cm]{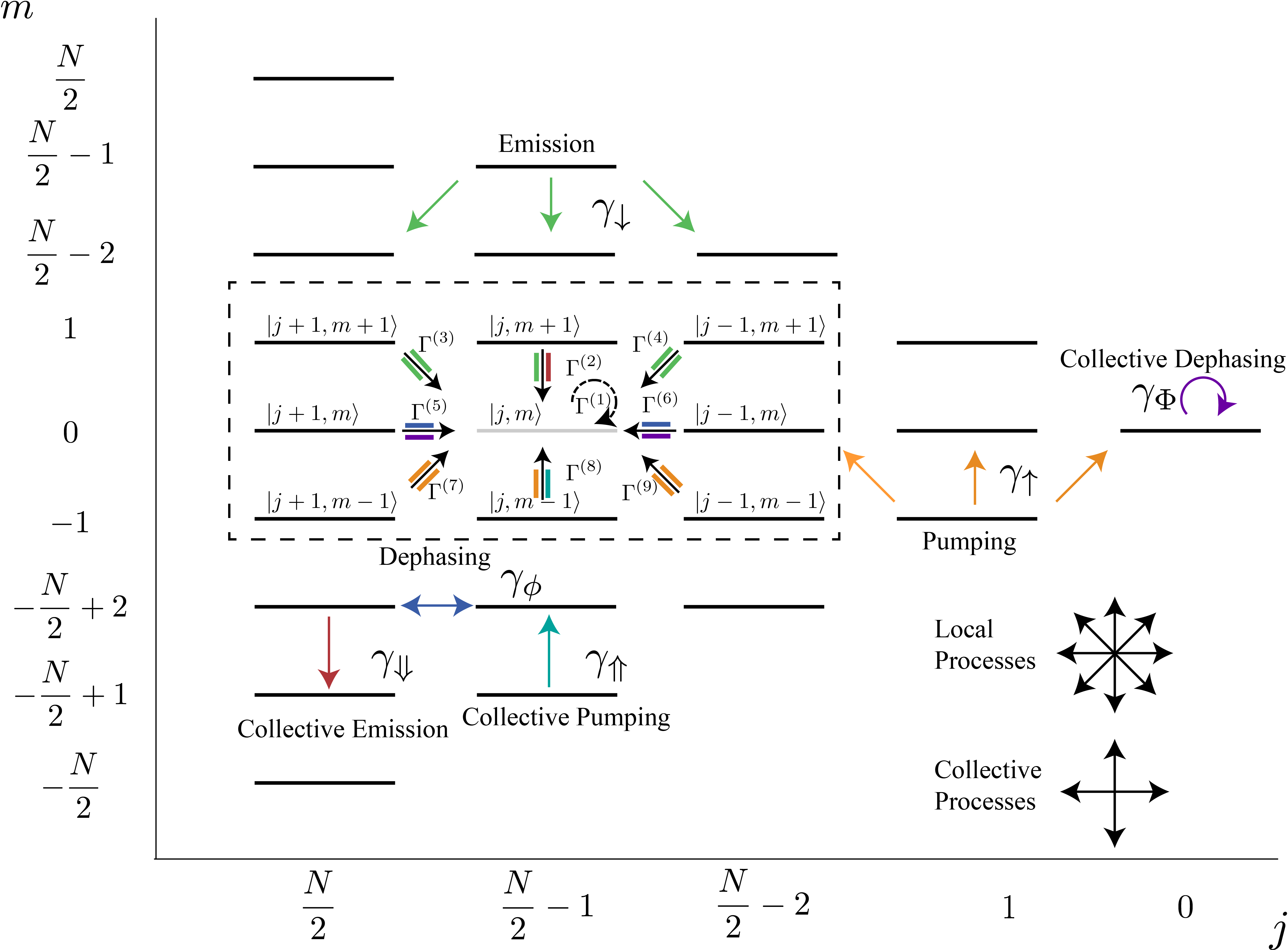}
\caption{\label{dickespace} Dicke space for $N=8$ showing the Dicke states $\ket{j,m}$ arranged in Dicke ladders, each with a given degeneracy $d_j^N$.
Sketch of the dynamics experienced by a Dicke state, expressed in terms of the rates $\Gamma^{(i)}$, where we omit the subscripts  for clarity $\Gamma^{(i)}=\Gamma^{(i)}_{j,m,m'}$. 
The effect of different driven-dissipative collective and incoherent processes is shown with colored arrows. 
We show the Dicke space \cite{Shammah17} and the action of each rate onto a given Dicke state in the center of the Dicke space.
All processes contribute to the rate $\Gamma^{(1)}_{j,m,m'}$. 
Local pure dephasing, local emission and local pumping connect neighbouring Dicke ladders.
Collective emission and collective pumping connect vertically subsequent rungs in the same Dicke ladder, while collective pure dephasing destroys correlations among different Dicke states.
In the dashed rectangle, the contribution to the time evolution of the state probability is expressed in terms of incoming probability from the other eight neighboring states according to the rates $\Gamma^{(i)}$, with $i>1$, and population depletion at a rate $\Gamma^{(1)}$ (dashed black arrow). 
}
\end{figure*}

We adopt the notation introduced in Ref.~\cite{Damanet16b} to treat collective emission and local emission, and generalize it to include incoherent pumping \cite{Meiser10a}, and local dephasing \cite{Shammah17,Bradac16}, as well as adding the corresponding collective processes. 
We can then write the projection of the symmetrized Lindbladian matrix $\tilde{L}$ as
\begin{widetext}
\begin{eqnarray}
\label{eq14}
\frac{d}{dt}p_{jmm'}&=&-\Gamma^{(1)}_{j,m,m'}p_{jmm'}
+\Gamma^{(2)}_{j,m+1,m'+1}p_{jm+1m'+1}+\Gamma^{(3)}_{j+1,m+1,m'+1}p_{j+1m+1m'+1} +\Gamma^{(4)}_{j-1,m+1,m'+1}p_{j-1m+1m'+1}\nonumber \\ && +\Gamma^{(5)}_{j+1,m,m'}p_{j+1mm'}+\Gamma^{(6)}_{j-1,m,m'}p_{j-1mm'} 
+\Gamma^{(7)}_{j+1,m-1,m'-1}p_{j+1m-1m'-1}+\Gamma^{(8)}_{j,m-1,m'-1}p_{jm-1m'-1}\nonumber\\&&
+ \Gamma^{(9)}_{j-1,m-1,m'-1}p_{j-1m-1m'-1},
\end{eqnarray}
\end{widetext}
where the explicit expressions for $\Gamma^{(i)}_{j,m,m'}$ are given in Appendix~\ref{app1}. 
An important feature of \Eq{eq14} is that the dynamics does not mix populations ($m=m'$) with coherences ($m\neq m'$), since any displacement in $m$ is matched by the same displacement in $m'$. 
There are only nine non-zero terms determining a given density-matrix-element evolution \cite{Chase08}, whose meaning can be immediately understood by setting $m=m'$, for clarity, and by inspecting the Dicke space \cite{Shammah17} in Figure~\ref{dickespace}.  

The dynamics of the probability density relative to a given Dicke state depends on the interaction with its eight nearest-neighboring Dicke states. 
% Figure 4
%%% 
By considering the effect on a single Dicke state in the Dicke space picture investigated in \cite{Shammah17}, it then becomes clear what the different contributions to \Eq{eq14} are: the upper state with same cooperative number, $\Delta j=0$, decays at a rate $\Gamma^{(2)}_{j,m+1,m'+1}$ that is determined by the collective and local emission processes, proportional to $ \gamma_{\Downarrow}$ and $\gamma_{\downarrow}$.

While the terms in $ \gamma_{\Downarrow}$ cannot change $j$, the local emission, which is $\propto\gamma_{\downarrow}$, can prompt transitions from states with $\Delta j=\pm 1$, which are accounted for by the terms in $\Gamma^{(3)}_{j+1,m+1,m'+1}$ and $\Gamma^{(4)}_{j-1,m+1,m'+1}$. 
As these processes do not conserve the cooperative number, they mix the populations of neighboring Dicke ladder. 

Local and collective dephasing, $\gamma_{\phi}$ and $\gamma_{\Phi}$, respectively, are energy-conserving processes, since $\Delta m=0$,  
and they determine the rates $\Gamma^{(5)}_{j+1,m,m'}$ and $\Gamma^{(6)}_{j-1,m,m'}$. 
Finally, similarly to collective emission, collective pumping $\propto\gamma_{\Uparrow}$ contributes only to $\Gamma^{(8)}_{j,m-1,m'-1}$, with $\Delta j =0$, while incoherent pumping, $\propto\gamma_{\uparrow}$, also prompts transitions with $\Delta j= \pm 1$, accounted for by $\Gamma^{(7)}_{j+1,m-1,m'-1}$ and $\Gamma^{(8)}_{j,m-1,m'-1}$. Note that in the more favorable case of a diagonal problem, for which the Hamiltonian is diagonal in $\mathbf{J}^2$, ${J}_z$, the scaling of the system's size is effectively only of $O(N^2)$.

%%%

%%%
\subsection{Survey of previous results}
\label{lit}
%%%
The permutational invariance of \Eq{master} has been independently explored by different authors in many recent and less recent works. 
In this Section we provide a global literature review on the topic of permutational-invariant master equations, which we hope will be useful to place this timely topic in its due context. 
Until very recently \cite{Gegg17a,Gegg}, the literature on this topic has been extremely fragmented.

In Table \ref{table1} we have summarized the results contained in the relevant publications, highlighting the features studied, and the dynamics generating them, showing both the Hamiltonian and the dissipative terms studied, using the rates for the spin mechanisms, \Eq{master}, and bosonic dissipation, \Eq{masterbos}.

The use of permutational invariance for the treatment of \Eq{master}, has, to the best of our knowledge, first been derived in Refs.~\cite{Sarkar87b,Sarkar87}, where it was applied to study optical bistability in systems in which both collective and local emission processes, proportional to $\gamma_\Downarrow$ and $\gamma_\downarrow$, respectively, were present. 

Recent research interest in the Dicke model, mainly stimulated by Refs.~\cite{Dimer07} and \cite{Baumann10}, focuses on how collective couplings can induce quantum phase transitions. At the same time, methods describing the interplay of collective and local effects date back to the early years of quantum optics. Its open system approach was applied to laser theory, for which local pumping and dissipation are fundamental. One such example is the phase-space representation of Ref. \cite{Haken70}, which relies on the permutational symmetry of identical two-level systems.

More generally, even in the presence of local dissipation, it has been shown that it is sufficient to derive the $(N+1)(N+2)(N+3)/6$ normal-ordered moments of collective operators to provide all statistical information for an ensemble of $N$ identical TLSs \cite{CarmichaelI}. One way to look at this is that after simultaneously absorbing $N$ quanta, the system saturates, so this limits the amount of correlations that can be present in the ensemble of TLSs. 

Such kind of operator-based approaches are equivalent to the density-matrix formalism used hereafter and derived in Ref.~\cite{Martini01}, where it was applied to the numerical study of anti-bunching in the emitted spectrum of a collection of $N$ coherently driven TLSs \cite{Carmichael91}.
 
The core of the method applied in this paper, and detailed in Appendix~\ref{app1}, was developed in Ref.~\cite{Chase08} for general local processes. 
 There, the action onto the Dicke states of the Lindblad superoperators relative to local emission, dephasing and pumping processes was analyzed in a general framework. 
 In Refs.~\cite{Chase08,Baragiola10} the robustness of spin squeezing against local and collective depolarization channels was tested. The collective depolarization channel can be expressed in terms of the Lindblad superoperators of \Eq{master},
 \begin{eqnarray}
 \label{depol}
\mathcal{L}_{J_{x}}[\rho]
+\mathcal{L}_{J_{y}}[\rho]
+\mathcal{L}_{J_{z}}[\rho]
&=&
\frac{1}{2}\left(
\mathcal{L}_{J_{-}}[\rho]+\mathcal{L}_{J_{+}}[\rho]
\right)+\mathcal{L}_{J_{z}}[\rho],\nonumber\\
\end{eqnarray}
and similarly one can do this for the local depolarization channel. Furthermore, in Ref.~\cite{Baragiola09b} the effect of different continuous measurement protocols under stochastic processes have been addressed.
 
The same approach, which leverages the Dicke space formalism, has been employed for the study of the superradiant and subradiant light emission of a collection of identical particles whose motional degrees of freedom let a bosonic or fermionic statistics emerge \cite{Damanet16b}. Moreover, the interplay between superradiant and subradiant light emission in the presence of dephasing and nonradiative emission mechanisms has been investigated employing the Dicke triangle to visualize the processes in the high-excitation and dilute excitation regimes \cite{Shammah17,Shammah}. Hereafter we will follow the formalism adopted in those works.

Independently from Ref.~\cite{Chase08}, in Ref.~\cite{Hartmann16} it has been shown explicitly that the Lindblad evolution preserves SU(4) symmetry, deriving a set of superoperators accounting for the local processes. 
The superoperator approach has been further extended in Refs.~\cite{Xu13,Xu}, mapping it onto Dicke states. In Ref.~\cite{Xu13} nonlinear effects in noisy open quantum systems in cavity QED have been investigated, addressing the study of lasing and steady-state superradiance and illustrating deviations from semiclassical approximations. A Ramsey spectroscopy scheme, robust against decoherence, has been proposed in Ref.~\cite{Xu15}. Recently, in Ref.~\cite{Gong16}, it has been proposed to exploit the local incoherent pumping to obtain a peculiar resonance fluorescence spectrum from Rydberg polaritons: the central peak of the Mollow triplet displays both line narrowing, typical of steady-state superradiance, and a non-classical photon statistics with anti-bunching, typical of the resonance fluorescence of a single TLS. 
In Ref.~\cite{Gong16}, the U(1) symmetry of the model has been used to reduce the computational resources required for the TLS basis from O($N^3$) to only O($N^2$).
In Ref.~\cite{Tieri17} the crossover from the bad-cavity limit characterizing steady-state superradiance to the good-cavity limit of lasing has been explored using several methods, including Monte-Carlo simulations in the reduced Liouvillian space using SU(4) generators \cite{Tieri17}. Superradiant lasing has also been studied through Monte-Carlo simulations of $N>10^5$ TLSs which, beyond local excitation, de-excitation and pure dephasing processes, also included the possibility of TLS loss from, and feeding into, the ensemble \cite{Zhang18b}. 
Stemming from steady-state superradiance investigations in cavity QED, dissipatively-induced spin synchronization in an ion trap has been addressed in Ref.~\cite{Shankar17}, displaying potential applications in metrology.

% Table II v3
\begin{table*}[htb!]
\centering		
\renewcommand{\arraystretch}{2} 
\renewcommand{\tabcolsep}{4pt}
\begin{adjustbox}{max width=\textwidth}
\begin{tabular}{|l|c|ccc|ccc|cc|}
\toprule
\multicolumn{10}{|c|}{
$
\Large \dot{\rho} = - \frac{i}{\hbar}\lbrack H,\rho \rbrack
+\stackrel{{\mathcal{L}}_\text{col}[\rho]}{\overbrace{
\underset{\myesquare{redpiqs} }{\frac{\gamma_{\Downarrow}}{2}\mathcal{L}_{J_{-}}[\rho] }
+
\underset{\myedisksmall{purplepiqs} }{\frac{\gamma_{\Phi}}{2}\mathcal{L}_{J_{z}}[\rho]}
+
\underset{\myediamond{cyanpiqs} }{\frac{\gamma_{\Uparrow}}{2}\mathcal{L}_{J_{+}}[\rho]}}}
+\stackrel{{\mathcal{L}}_\text{loc}[\rho]}{\overbrace{
\sum_{n=1}^{N}\left(
\underset{
\mysquare{greenpiqs} 
}{\frac{\gamma_{\downarrow}}{2}\mathcal{L}_{J_{-,n}}[\rho]} 
+\underset{\mydisksmall{bluepiqs}}{\frac{\gamma_{\phi}}{2}\mathcal{L}_{J_{z,n}}[\rho]}
+
\underset{\mydiamond{orangepiqs} }{\frac{\gamma_{\uparrow}}{2}\mathcal{L}_{J_{+,n}}[\rho]}
\right)}}
+\stackrel{{\mathcal{L}}_\text{cav}[\rho]}{\overbrace{
\underset{\mydtriangle{red} }{\frac{\kappa}{2}\mathcal{L}_{a}[\rho]}
+
\underset{\mytriangle{black} }{\frac{w}{2}\mathcal{L}_{a^\dagger}[\rho]}
}}$}\\ 
\multicolumn{10}{|c|}{}
\\
\hline
\multicolumn{1}{|c|}{\bf{Features}}& 
\bf{Hamiltonian} $\boldsymbol{H}$ &\multicolumn{3}{c|}{\bf{Collective TLS} $\boldsymbol{\mathcal{L}}_\text{col}[\rho]$}&\multicolumn{3}{c|}{\bf{Local TLS}  $\boldsymbol{\mathcal{L}}_\text{loc}[\rho]$}&\multicolumn{2}{c|}{\bf{Cavity}  $\boldsymbol{\mathcal{L}}_\text{cav}[\rho]$}\\
&  TLS/TLS-cavity 
&\multicolumn{1}{c|}{
Emission} 
&\multicolumn{1}{c|}{Dephasing} 
&\multicolumn{1}{c|}{Pump}
&\multicolumn{1}{c|}{Emission}
&\multicolumn{1}{c|}{Dephasing}
&\multicolumn{1}{c|}{Pump}
&\multicolumn{1}{c|}{Emission}
&\multicolumn{1}{c|}{Pump}\\
&   
&\multicolumn{1}{c|}{
 $\gamma_{\Downarrow}$} 
&\multicolumn{1}{c|}{ $\gamma_{\Phi}$} 
&\multicolumn{1}{c|}{$\gamma_{\Uparrow}$}
&\multicolumn{1}{c|}{$\gamma_{\downarrow}$}
&\multicolumn{1}{c|}{$\gamma_{\phi}$}
&\multicolumn{1}{c|}{$\gamma_{\uparrow}$}
&\multicolumn{1}{c|}{$\kappa$}
&\multicolumn{1}{c|}{$w$}\\

%\myesquare{redpiqs} %c.em
%\myedisk{purplepiqs} %c.deph
%\myediamond{cyanpiqs} %c.pump
%\mysquare{greenpiqs} %em
%\mydisk{bluepiqs}  %deph
%\mydiamond{orangepiqs} %pump 
%
\hline \hline		
Dicke space formalism 
\cite{Chase08,Baragiola10}
&  General
&   \myesquare{redpiqs} 
&\myedisk{purplepiqs} 
&\myediamond{cyanpiqs}
&\mysquare{greenpiqs}
& \mydisk{bluepiqs}
&\mydiamond{orangepiqs} 
& 
&  
\\
\hline		
Superoperators \cite{Hartmann16} 
&  $\mathrm{General}$
&   
& 
& 
&\mysquare{greenpiqs}
& \mydisk{bluepiqs}
&\mydiamond{orangepiqs}
& $\color{red}{}$
&  \\
\hline	
Superoperators \& Dicke space \cite{Xu13}
&  General
&   \myesquare{redpiqs} 
&\myedisk{purplepiqs} 
&\myediamond{cyanpiqs}
&\mysquare{greenpiqs}
& \mydisk{bluepiqs}
&\mydiamond{orangepiqs}
& \mydtriangle{red}
& \mytriangle{black}\\	
\hline	
Multi-level systems \cite{Bolanos15,Gegg16}
&  $\mathrm{General}$
&   \myesquare{redpiqs} 
& 
&\myediamond{cyanpiqs}
&\mysquare{greenpiqs}
& \mydisk{bluepiqs}
&\mydiamond{orangepiqs}
& \mydtriangle{red}
& \mytriangle{black}\\	
\hline
\hline	
Optical bistability  \cite{Sarkar87b,Sarkar87}
&  $\hbar\omega_xJ_x$
&   \myesquare{redpiqs} 
& 
& 
&\mysquare{greenpiqs}
& $\color{bluepiqs}{}$
& 
& $\color{red}{}$
&  
\\
\hline	
Collective quantum jumps \cite{Lee11}
&  $\hbar\omega_xJ_x-\hbar\omega_{zz}J_z^2$
&   
& 
& 
&\mysquare{greenpiqs}
& $\color{bluepiqs}{}$
& 
& $\color{red}{}$
&  
\\
\hline
Bistability in the XY model \cite{Wilson16}
&  $\hbar \omega_{x}J_x-\hbar \omega_{0}J_z-\hbar \omega_{xy}\left(J_x^2+J_y^2\right)$
&   
& 
& 
&\mysquare{greenpiqs}
& $\color{bluepiqs}{}$
& 
& $\color{red}{}$
&  \\	
\hline
Two-axis spin squeezing*  \cite{Chase08,Baragiola10}
&  $-i\hbar\Lambda(J_+^2-J_-^2)$
& \myesquare{redpiqs}
&\myedisk{purplepiqs}
&\myediamond{cyanpiqs}
&\mysquare{greenpiqs}
& \mydisk{bluepiqs}
&\mydiamond{orangepiqs}
& \mydtriangle{red}
&  
\\
\hline
Entanglement witness \cite{Garttner18}
&  $-\hbar \omega_{xx} J_x^2-\hbar \omega_0 J_z$
&   
& 
& 
&\mysquare{greenpiqs}
& \mydisk{bluepiqs}
&\mydiamond{orangepiqs}
& $\color{red}{}$
& \\	
\hline
\hline	
Steady-state superradiance \cite{Xu13,Tieri17}
&  $\hbar\omega_0J_z$
&   \myesquare{redpiqs} 
& 
& 
&$\color{greenpiqs}{}$
& $\color{bluepiqs}{}$
&\mydiamond{orangepiqs}
& \mydtriangle{red}
& \mytriangle{black}\\	
\hline
Ramsey spectroscopy \cite{Xu15}
&  $\hbar\omega_0J_z$
&   \myesquare{redpiqs} 
& 
& 
&$\color{greenpiqs}{}$
& \mydisk{bluepiqs}
&\mydiamond{orangepiqs}
& $\color{red}{}$
&  \\	
\hline
Superradiant emission \cite{Damanet16b}
&  $\hbar\omega_0 J_z$
&   \myesquare{redpiqs} 
& 
& 
&\mysquare{greenpiqs}
& $\color{bluepiqs}{}$
& 
& $\color{red}{}$
&  \\	
\hline		
Superfluorescence/subradiance \cite{Shammah17}
&  $\hbar\omega_0 J_z$
&   \myesquare{redpiqs} 
& 
& 
&\mysquare{greenpiqs}
& \mydisk{bluepiqs}
& 
& $\color{red}{}$
&  \\
\hline
Spin synchronization \cite{Shankar17}
&  $\hbar\omega_0 J_z$
&   \myesquare{redpiqs} 
& 
&\myediamond{cyanpiqs}
&$\color{greenpiqs}{}$
& $\color{bluepiqs}{}$
&\mydiamond{orangepiqs}
& $\color{red}{}$
& \\
\hline
Superradiant lasing$^\dagger$ \cite{Zhang18b}
&  $\hbar\omega_0 J_z$
&   \myesquare{redpiqs} 
& 
& 
&\mysquare{greenpiqs}
& \mydisk{bluepiqs}
&\mydiamond{orangepiqs}
& $\color{red}{}$
&  \\
\hline
\hline
Non-classical light \cite{Martini01}
&  $\hbar \omega_x (a+a^\dagger )+ \hbar g(aJ_++a^\dagger J_-)$
&   
& 
& 
&\mysquare{greenpiqs}
& $\color{bluepiqs}{}$
& 
& \mydtriangle{red}
&  
\\
\hline	
State engineering \cite{Wood16}
&  $\hbar g(J_+a+J_-a^\dagger)$
&   \myesquare{redpiqs} 
& 
&\myediamond{cyanpiqs}
&$\color{greenpiqs}{}$
& \mydisk{bluepiqs}
& 
& $\color{red}{}$
&  
\\
\hline	
Lasing \cite{Xu13,Tieri17}
& $\hbar g(J_+a+J_-a^\dagger)$
&   \myesquare{redpiqs} 
& 
& 
&$\color{greenpiqs}{}$
& $\color{bluepiqs}{}$
&\mydiamond{orangepiqs}
& \mydtriangle{red}
& \mytriangle{black}\\	
\hline	
Photon anti-bunching \cite{Gong16}			
&  $\hbar g(J_+a+J_-a^\dagger)$
&   
& 
& 
&\mysquare{greenpiqs}
& \mydisk{bluepiqs}
&\mydiamond{orangepiqs}
& \mydtriangle{red}
& \\
\hline
Super/subradiance  \cite{Gegg16,Gegg17a}
&  $\hbar g(J_+a+J_-a^\dagger)$
&   
& 
& 
&\mysquare{greenpiqs}
& $\color{bluepiqs}{}$
& 
& \mydtriangle{red}
&  \\	
\hline
\hline	
Spaser \cite{Richter15,Zhang15a,Zhang18a}
&  $\hbar g J_x(a+a^\dagger)$
&   \myesquare{redpiqs} 
& 
&\myediamond{cyanpiqs}
&\mysquare{greenpiqs}
& \mydisk{bluepiqs}
&\mydiamond{orangepiqs}
& \mydtriangle{red}
& \mytriangle{black}\\	
\hline	
Superradiant PT \cite{Kirton17}
&  $\hbar g J_x(a+a^\dagger)$
&   
& 
& 
&\mysquare{greenpiqs}
& \mydisk{bluepiqs}
& 
& \mydtriangle{red}
& \mytriangle{black}\\	
\hline
PT, Lasing, Chaos \cite{Kirton17b}
&  $\hbar g (J_+a+J_-a^\dagger)+ \hbar g' (J_-a+J_+a^\dagger)$
&   
& 
& 
&\mysquare{greenpiqs}
& \mydisk{bluepiqs}
&\mydiamond{orangepiqs}
& \mydtriangle{red}
& \mytriangle{black}\\				
\hline
Super/subradiant PT, squeezing \cite{Gegg17a}
&  $\hbar g J_x(a+a^\dagger)+\hbar \omega_x J_x$
&   
& 
& 
&\mysquare{greenpiqs}
& \mydisk{bluepiqs}
& 
& \mydtriangle{red}
& \\
\bottomrule				
\end{tabular}
\end{adjustbox}
\caption{\label{table1} Features studied in driven-dissipative open quantum systems comprising several TLSs in works in which permutational-invariant methods were applied. 
The works are grouped according to the general theory developed or according to the Hamiltonian studied, with $\omega_0$, $\omega_x$, $\omega_{xx}$, $\omega_{xy}$, $\Lambda$, $g$, and $g'$ frequency parameters. 
For the master equation $ \dot{\rho}=i\lbrack H,\rho\rbrack+\mathcal{L[\rho]}$ we show the relative interaction Hamiltonian, the rates relative to collective TLS processes, homogeneous local TLS processes, and cavity rates. 
PT stands for Phase Transition, and spaser stands for Surface Plasmon Amplification by Stimulated Emission of Radiation.
$^*$In Ref.~\cite{Chase08,Baragiola10} the collective and local depolarization channel is considered, fixing $\frac{1}{2}\gamma_\Downarrow=\frac{1}{2}\gamma_\Uparrow=\gamma_\Phi$ and $\frac{1}{2}\gamma_\downarrow=\frac{1}{2}\gamma_\uparrow=\gamma_\phi$. $^{\dagger}$ In Ref.~\cite{Zhang18b} TLS addition and subtraction is treated exploiting permutational symmetry.
}	
\end{table*}

\emph{Multi}-level systems have been studied in Refs.~\cite{Richter15,Bolanos15,Zhang15a,Gegg16,Gegg,Sutherland17,Zhang18a,Warnakula18}, always in the context of cavity QED. 
In Ref.~\cite{Richter15}, it has been shown that once local dephasing processes are accounted for, the mechanism that triggers the coherent surface plasmon amplification by stimulated emission of radiation (spaser) requires higher pump rates than those achieved in previous experiments, considerably limiting the possibilities under which a spaser could operate in a realistic nano-device.
This method has since been employed for the study of superradiant light emission and subradiance of an ensemble of TLSs \cite{Gegg,Gegg17b,Gegg17a}. In particular, in Ref.~\cite{Gegg} a comprehensive outlook on the use of permutational invariance for two and multi-level-system master equations is given, with group theoretic considerations and the use of graph theory to algorithmically further reduce the complexity of the master equations. 

In Ref.~\cite{Psiquasp} PsiQuaSP, an open-source computational library written in C, was introduced, allowing the study of \emph{multi}-level systems interacting with bosonic fields. 
PsiQuaSP exploits state of-the-art numerical libraries in C/C++, such as PETSc, to perform efficient matrix multiplication and time integration of differential equations and allows the user to either use a set of pre-built Lindbladians or to define an ad-hoc generic Liouvillian operator.
 
In Refs.~\cite{Shammah17,Gegg17a} it has been shown that dark Dicke states $\ket{j,-j}$ can be engineered through an interplay of collective and local losses and local dephasing. Reference~\cite{Gegg17a} contains also a study of spin squeezing for entanglement estimation. 
With regards to state engineering, in Ref.~\cite{Wood16}, a method to obtain a pure state through cavity cooling to the ground state has been proposed, in which local dephasing is used as a resource. 

In the context of phase transitions of driven-dissipative open quantum systems, recent studies \cite{Wilson16,Gegg17a,Kirton17,Kirton17b} have characterized the effect of local noise on the occurrence of the steady-state phases. 
Collective quantum jumps and bistability has been studied in Ref.~\cite{Lee11}.
Ref.~\cite{Wilson16} studied bistable phases in an array of bosonic cavities, effectively mapping the system onto the driven-dissipative XY model. There, permutational invariance was employed to investigate the closing of the Liouvillian gap in the presence of local losses, $\gamma_\downarrow$.
In Ref.~\cite{Kirton17}, cavity losses, dephasing, and local emission processes were considered, showing that their interplay induces a nontrivial suppression and restoration of the superradiant phase.  
Always in Ref.~\cite{Kirton17}, the exact diagonalization method was employed without using the Dicke-state formalism. 
The relevant code is developed in Python and available online as the \emph{Permutations} library \cite{Kirton}. 
In Ref.~\cite{Kirton17b}, the effect of incoherent and coherent pumping was addressed, studying the phase diagram in relation to the superradiant, normal, and lasing phases and addressing the emergence of chaos. Ref.~\cite{Sutherland17} addressed superradiance in the presence of inhomogeneous dipole-dipole interactions. In Ref.~\cite{Garttner17}, permutation-symmetric simulations of \Eq{master} have been used to support the experimental study of many-body correlations in a quantum simulator with more than 100 trapped ions, while in Ref.~\cite{Garttner18} the use of out-of-time-order correlations has been extended to probe entanglement.

Permutational invariance has been applied in several quantum information studies, independently from the dynamics of \Eq{master}. 
While an exhaustive survey of these contributions is beyond the scope of this section, let us mention that the Dicke basis has been used to perform efficient state tomography and to characterize multi-partite entanglement \cite{Wesenberg02,DAriano03,Toth10,Martin10,Moroder12,Novo13,Wolfe14,Gao14,Schwemmer14}. 
The permutational symmetry of an ensemble of identical qubits has also been applied in quantum information processing strategies, including for efficient quantum algorithms and optimal information compression \cite{Wesenberg02,Bacon06,Yang16,Yang17,Mozrzymas18}. 
Finally, we point out that while the computational reduction here employed arises from the SU(4) symmetry of the Lindblad operators for a collection of spins, and through the properties of Lie algebras this method has been extended to $m$-level systems \cite{Gegg16,Gegg17b,Gegg17a,Gegg}, studies of bosonic lattices have also exploited permutational symmetry of fully-connected models, so far to describe closed-system dynamics only \cite{Sciolla11,Sciolla13}.

%%%
%%%
\section{Permutational-Invariant Quantum Solver (PIQS)}
\label{piqs}
%%%
The Permutational-Invariant Quantum Solver (PIQS) is an object-oriented framework to study ensembles of $N$ identical TLSs evolving under both \emph{collective and local} dissipation processes \cite{Piqs}. 
PIQS is a robustly tested library developed in Python, designed to be memory efficient, fast, and easy to use. 
The memory efficiency comes from the employment of sparse matrices. 
The speed is due to the use of low-level C++ code behind the scenes. 
The ease-of-use is enabled by Python.  
By being tightly integrated with QuTiP, the open-source quantum toolbox in Python \cite{Johansson12,Johansson13},
PIQS is a modular tool that extends the reach of investigations involving the dissipative dynamics of large ensembles of TLSs.

Moreover, we provide a collection of functions to explore the properties of the Dicke space and define the algebra of spin operators and density matrices of important quantum states.
An extensive documentation and several tutorials based on Jupyter notebooks facilitate the interactive use of the tool \cite{Piqs,Kluyver16}.
PIQS is released as an open-source project to promote open science, joining the growing pool of open-source software libraries that are being developed for the simulation of open quantum systems \cite{Tan99,Johansson12,Johansson13,Psiquasp,Kirton,Kramer17}.
%%%
\subsection{A TLS ensemble as a \texttt{Dicke} object}
%%%
In PIQS, an ensemble of TLSs is represented as an instance of the $\code{Dicke}$ class. 
The basic setup of the ensemble just requires the number $N$ of TLSs.
The Liouvillian can be constructed simply by defining the rate coefficients, $\gamma_{i}$, according to \Eq{master} and a given
Hamiltonian. 
All of the dissipative rates of \Eq{master} can be specified as attributes of a \code{Dicke} object using the keywords \code{emission}, \code{dephasing}, \code{pumping}, \code{collective\_emission}, \code{collective\_dephasing}, and \code{collective\_pumping}. 
For example, one can build the matrix for the symmetrized Liouvillian superoperator, $\mathcal{D}_\text{S}$, simply by defining the \code{emission} rate in units of inverse time and calling the function \code{liouvillian()} of the \code{Dicke} class,
\begin{python}
from piqs import Dicke

N = 10
ensemble = Dicke(N)
ensemble.emission = 0.2
D = ensemble.liouvillian()
\end{python}

The Liouvillian is constructed as an object of the $\code{Qobj}$ class, which is the class for quantum objects in QuTiP. This integration allows the user to study the dynamical properties of the system using the many functions present in QuTiP. 
Any initial density matrix can evolve in time according to the equation $\dot{\rho}=\mathcal{D}_\text{S}[\rho]$, or its Liouvillian can be used to define a more complex light-matter system, including bosonic degrees of freedom and multiple TLS ensembles. In any case, the Liouvillian superoperator is represented by a matrix $D$ in the Liouvillian space as shown in \Eq{eq14a}.
%%%
\subsection{Initial states and operators}
%%%
PIQS provides the spin algebra of collective operators and important quantum states to let the user quickly set up a model and explore its physics. 
For example, the functions $\code{jspin(N, "+")}$, $\code{jspin(N, "-")}$ provide the collective jump operators $J_{\pm}$. 
The operators of the collective algebra can be called both in the Dicke basis (set as default and specified by the argument \code{basis="dicke"}) or in the uncoupled basis (\code{basis="uncoupled"}) of $2^N$ TLSs.
The Hamiltonian for a TLS ensemble specified as a \code{Dicke} object can be built by the user in the Dicke basis with this collective algebra.

We provide functions to quickly define density matrices corresponding to quantum states of interest, such as for the coherent spin states, which can be called with \code{css(N, a, b)} corresponding to \Eq{cssab}, while the symmetric CSS is called simply with \code{css(N)}. The states can also be initialized in the uncoupled basis, \Eq{cssab1}; the ``cartesian'' coordinates $a$ and $b$ are set as default and can be invoked with \code{coordinates="cartesian"}, while the polar coordinates $\theta$, $\varphi$ of \Eq{csst} can be specified setting \code{coordinates="polar"}.
The GHZ state of \Eq{ghz} is called with $\code{ghz(N)}$, while the fully excited state, and the ground state, are invoked with \code{excited(N)} and \code{ground(N)}, respectively.
These density matrices, just like the collective spin operators, are given both in the coupled Dicke basis as well as in the uncoupled $2^{N}$ full Hilbert space. Moreover, the user can initialize in the Dicke basis any Dicke state,
$\texttt{dicke(N, j, m)}$. 

 $\ $

 $\ $
 
An example of these instances is shown hereafter,
\begin{python}
from piqs import (ghz, dicke, jspin)

rho1, rho2 = ghz(N), dicke(N,N/2,0)
[jx, jy, jz] = jspin(N)
H = jz + 0.5*jx
\end{python}

While the $\code{Dicke}$ object defines an ensemble of  $N$ TLSs and its properties, all operators, density matrices, and superoperators are given as instances of QuTiP's quantum object class ($\code{Qobj}$). 
\subsection{The Liouvillian as a flexible quantum object (\texttt{Qobj})}
The $\texttt{lindbladian()}$ and $\texttt{liouvillian()}$ functions can be used to construct the matrix form of the respective superoperators as a sparse matrix wrapped as an instance of QuTiP's $\code{Qobj}$ class. This gives a remarkable flexibility to study a combination of TLSs and bosonic cavities, in the weak, strong, and ultrastrong-coupling regimes.
\subsubsection*{Multiple TLS ensembles in open bosonic cavities}
The \texttt{super\_tensor} function from QuTiP allows the user to construct two instances of TLS ensembles with PIQS and then combine them together. 
It is easy to construct more complex connectivities, e.g., placing the ensemble(s) in a single or in multiple bosonic cavities. 
To this end, the possibility of defining tensored Liouvillian spaces becomes crucial, as it allows us to place \emph{a posteriori} the Liouvillian matrices describing the individual parts, which have a superoperator form, in the tensor Liouvillian space.
In QuTiP's modular environment it becomes natural to extend the reach of investigations, for example placing the TLS ensemble into a photonic leaky cavity, 
\begin{python}
from piqs import num_dicke_states
from qutip import *
import numpy as np

# identity superoperators
nds = num_dicke_states(N)
nph = 20
a = destroy(nph)
itls = to_super(qeye(nds))
iph = to_super(qeye(nph))
# photonic Liouvillian
D_ph = liouvillian(a.dag()*a, [a])
# total TLS Liouvillian
D_tls = super_tensor(D_ph,itls) \
        + super_tensor(iph,D)
# light-matter interaction
H_i = tensor(a + a.dag(), jx)
D_i = -1j*spre(H_i)+1j*spost(H_i)
D_tot = D_tls + D_i
       
\end{python}
%%
% Figure 5
\begin{figure*}[ht!]
\begin{center}
\includegraphics[width=16cm]{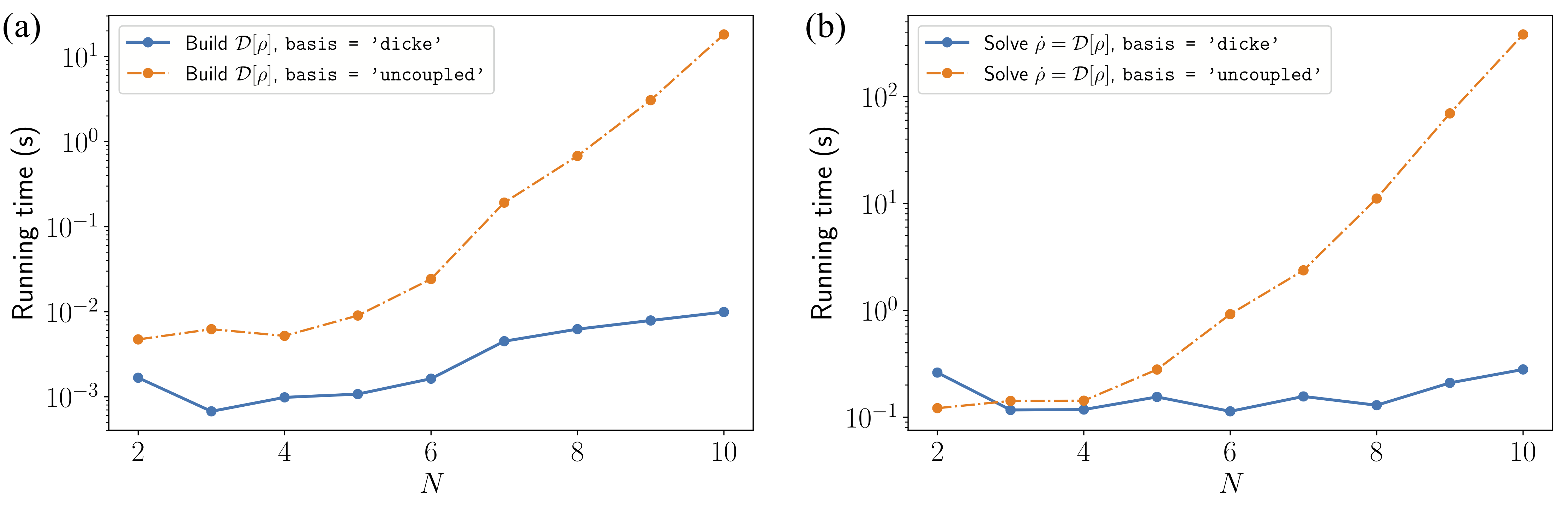}
\caption{\label{figx6} 
{\bf  PIQS performance -- small $N$.} 
Running time as a function of the number $N$ of TLSs in the system, up to $N= 10$.
(a) Semi-logarithmic plot giving the time (in seconds) required to construct the Liouvillian superoperator, with $H= \omega_0J_z$ in the Dicke basis (blue circles joined by solid lines) and in the uncoupled basis (orange circles joined by dashed lines).
We set $\omega_{0}=\omega_{x}=1$, the local processes $\gamma_{\downarrow}=\gamma_{\phi}=\gamma_{\uparrow}=0.1$, and the collective processes $\gamma_{\Downarrow}=\gamma_{\Phi}=\gamma_{\Uparrow}=0.01$.
(b) Semi-logarithmic plot giving the time (in seconds) required to solve the dynamics given by the Liouvillian in the Dicke basis (blue circles joined by solid lines) and in the uncoupled basis (orange circles joined by dashed lines). In both cases we use QuTiP's master equation \code{mesolve} for $H= \omega_0J_z$, where $\omega_{0}=1$, with local processes $\gamma_{\downarrow}=\gamma_{\phi}=\gamma_{\uparrow}=0.1$, and collective processes $\gamma_{\Downarrow}=\gamma_{\Phi}=\gamma_{\Uparrow}=0.01$. 
We use 1000 integration time steps from $t_0=0$, when the system is initially fully excited, to $t_\text{max}= 4 \log(N)/N\gamma_{\Downarrow}$. }
\end{center}
\end{figure*}
%
% Figure 6
\begin{figure*}[ht!]
\begin{center}
\includegraphics[width=9cm]{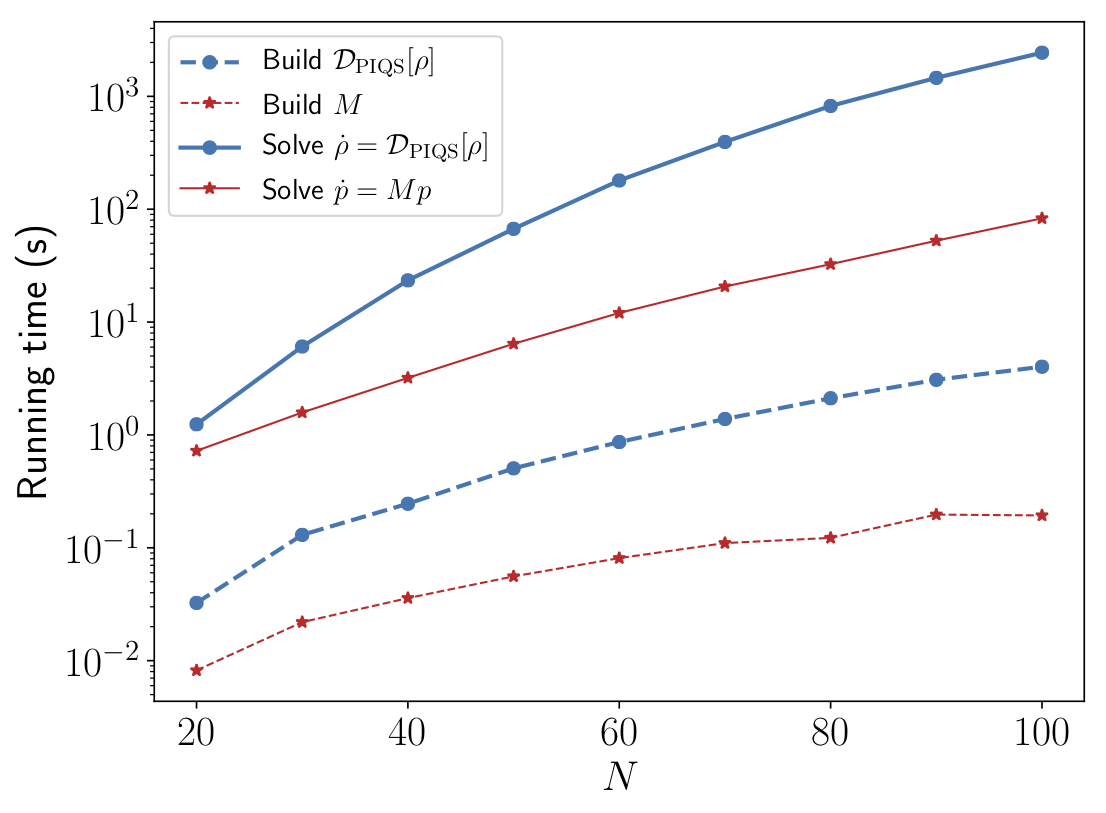}
\caption{\label{figx62} 
{\bf  PIQS performance --  large $N$.} Semi-logarithmic plots of the running time as a function of the number of TLSs $N$ in the system, up to $N= 100$.
We compare the time (in seconds) required to build the matrix governing the dynamics and to solve it.
We compare two methods that both leverage PIQS. The first one is the most flexible one, which builds the Liouvillian (blue circles joined by dashed lines) and solves the dynamics using QuTiP's \code{mesolve} (blue circles joined by solid lines). The second one, applicable in the subset of problems diagonal in the Dicke basis, uses an \emph{ad-hoc} permutationally-invariant solver, $\texttt{pisolve}$, which further decreases the system's size, building the matrix $M$ (red stars joined by dashed lines) and solving the relative rate equations (red stars joined by solid lines). 
 The dynamics is set by \Eq{master} with $H= \hbar\omega_0J_z$, where $\hbar\omega_\text{0}=1$,  the local processes are set by $\gamma_{\downarrow}=\gamma_{\phi}=\gamma_{\uparrow}=0.1$, and the collective processes are set by $\gamma_{\Downarrow}=\gamma_{\Phi}=\gamma_{\Uparrow}=0.01$. 
 We use 1000 integration time steps from $t_0=0$, when the system is initially fully excited, to $t_\text{max}= 4t_D=4 \log(N)/N\gamma_{\Downarrow}$. }
\end{center}
\end{figure*}

\subsection{Cythonized code for memory-efficient fast Lindbladian construction}
The advantage of Python comes from its easy-to-learn syntax and interactive behavior as an interpreted language.  
However, performing numerical calculations in Python is known to be much slower than compiled C/C++ code \cite{Johansson13}. 
To bypass this side effect, PIQS uses Cython routines to compute the Lindbladian, which leads to performances similar to low-level C/C++ code, all without sacrificing the advantages of Python. 
By writing the $\code{liouvillian}$ function with Cython we obtained a ten-fold increase in performance with respect to a Python version of the same function. 

Figures \ref{figx6} and \ref{figx62} show the performance of PIQS for small and large $N$, respectively, on a commercial personal computer with standard specifics (memory: 16 GB RAM at 2133 MHz; CPU: 2.3 GHz Intel Core i5). 
As a guide to the eye, the markers in Figures~\ref{figx6}, \ref{figx62} are joined by straight segments.
In Figure~\ref{figx6}(a), the plot shows the time required to construct the matrix corresponding to the Liouvillian superoperator in the $\code{dicke}$ (blue solid segments) and $\code{uncoupled}$ basis (orange dashed segments). 
We set $H= \hbar\omega_0J_z$  with $\hbar\omega_0 = 1$ and Lindbladian superoperator rates 
$\gamma_\Downarrow=\gamma_\Uparrow=\gamma_\Phi=0.01$ for collective operators and 
$\gamma_\downarrow=\gamma_\uparrow=\gamma_\phi=0.1$ for the local operators. 

Already for $N\gtrapprox10$ the calculation becomes computational intensive in the $\code{uncoupled}$ basis, as the Liouvillian space grows exponentially as $4^N$. In Figure~\ref{figx62}, a semi-logarithmic plot shows PIQS performance up to $N=100$. The dashed blue segments (circle markers) corresponds to the time required to build the matrix of the Liouvillian superoperator, showing that the Liouvillian can be built in less than 10 seconds.
\subsection{Solving the master equation with \texttt{mesolve}}
Once the Liouvillian is constructed, one can exploit QuTiP's master equation solver (\texttt{qutip.mesolve}) to solve \Eq{master} or \Eq{masterbos}. 
\begin{python}

rho_tls = ghz(N)
rho_ph = ket2dm(basis(nph,0))
rho = tensor(rho_ph, rho_tls)
t = np.linspace(0, 1, 100)

result = mesolve(D_tot, rho, t,[])
rhot = result.states
\end{python}

As larger ensembles of $N$ TLSs are considered, the real limit depends on the computational power of the machine, the scaling of the problem, known from its definition, and the effective time required to solve the dynamics of the given problem, which can vary greatly depending on the processes considered and the magnitude of the rates. 
To assess this systematically, in Figure~\ref{figx6}(b), we compare the time required to solve the master equation with QuTiP's master equation solver $\code{mesolve}$ using a Liouvillian in the $\code{dicke}$ or $\code{uncoupled}$ basis, considering every local and collective incoherent process.

We choose as the initial state $\rho_0=\ket{\frac{N}{2},\frac{N}{2}}\bra{\frac{N}{2},\frac{N}{2}}$. 
We use 1000 integration steps up to a maximum time $t_\text{max}=4t_\text{D}$, where $t_\text{D}=\text{log}(N)/N\gamma_{\Downarrow}$ is the superradiant delay time. 
We find that the performance scales similarly to the construction of the Liouvillian, Figure~\ref{figx6}(a), exploding exponentially in the \code{uncoupled} basis, while remaining fast in the \code{dicke} basis.
In Figure~\ref{figx62}, the solid blue segments joining the circles show that the exact time evolution for the collective density matrix, $\rho(t)$, can be obtained in minutes even for $N=100$ TLSs.
\subsubsection*{Diagonal solver}
When the initial state of the system is in diagonal form in the Dicke basis, such as those in Figure \ref{figxds}, and the Hamiltonian is also diagonal in the Dicke basis, the problem of \Eq{master} can be further simplified since \Eq{eq14} then couples only the diagonal matrix elements, $p_{jmm}$. 
The number of matrix elements is then only $\sum_{j=j_\text{min}}^{N/2}(2j+1)=\frac{1}{4}N(N+1)=O(N^2)$. 
This means that it is possible to define a matrix for the coefficients of the coupled linear differential equations, $M$, whose size scales only quadratically in $N$, allowing more efficient simulations, 
\begin{eqnarray}
\label{mp}
\dot{p}=Mp, 
\end{eqnarray}
where $p$ is a vector containing only the diagonal matrix elements of $\rho$, see Appendix \ref{app1} for details. 
We implemented this option in the library, so that the user can solve the dynamics with an ad-hoc permutational-invariant solver for the rate equations, $\texttt{pisolve(rho\_tls, t)}$, which generates a vector $p$ and then re-expands the results in terms of QuTiP's $\code{Results}$ class, so that the user can obtain the full density matrix $\rho(t)$ and any higher-order moment. 
Here $\texttt{rho\_tls}$ is the density matrix of the initial TLS-ensemble state and $\texttt{t}$ the integration-time list.  
An example of a program using this further space reduction to calculate the total inversion $\langle J_z\rangle (t)$ of the system defined by the $\code{Dicke}$ object $\code{ensemble}$ defined above, is given here: 
\begin{python}
# diagonal solver 
rho_tls = excited(N)
result = ensemble.pisolve(rho_tls, t)
rhot = result.states
jzt = expect(rhot, jz)

\end{python}
In Figure~\ref{figx62}, we plot the time required to build the coefficient matrix $M$ of \Eq{mp} (dashed red segments joining the stars) and to then solve the dynamics using $\code{pisolve}$  (solid red segments joining the stars), showing that for the subset of problems diagonal in the Dicke basis the system's dynamical properties can be found up to two orders of magnitude faster. 
%%%%
\section{Examples and applications}
\label{results}
%%%%
In this section we demonstrate how PIQS can be easily used to model a variety of systems in which collective effects play an important role, assessing the influence of often-neglected local dissipation. 
The interested reader can find online several interactive Python notebooks, including all of those relative to the physical models treated below \cite{Piqs}.
The driven-dissipative dynamics of models that are diagonal in the Dicke basis can be studied also with the fast solver $\code{pisolve}$. 
%%%
\subsection{Superradiant light emission}
\label{sre}
\begin{figure*}[ht!]
\begin{center}
\includegraphics[height=6.0cm]{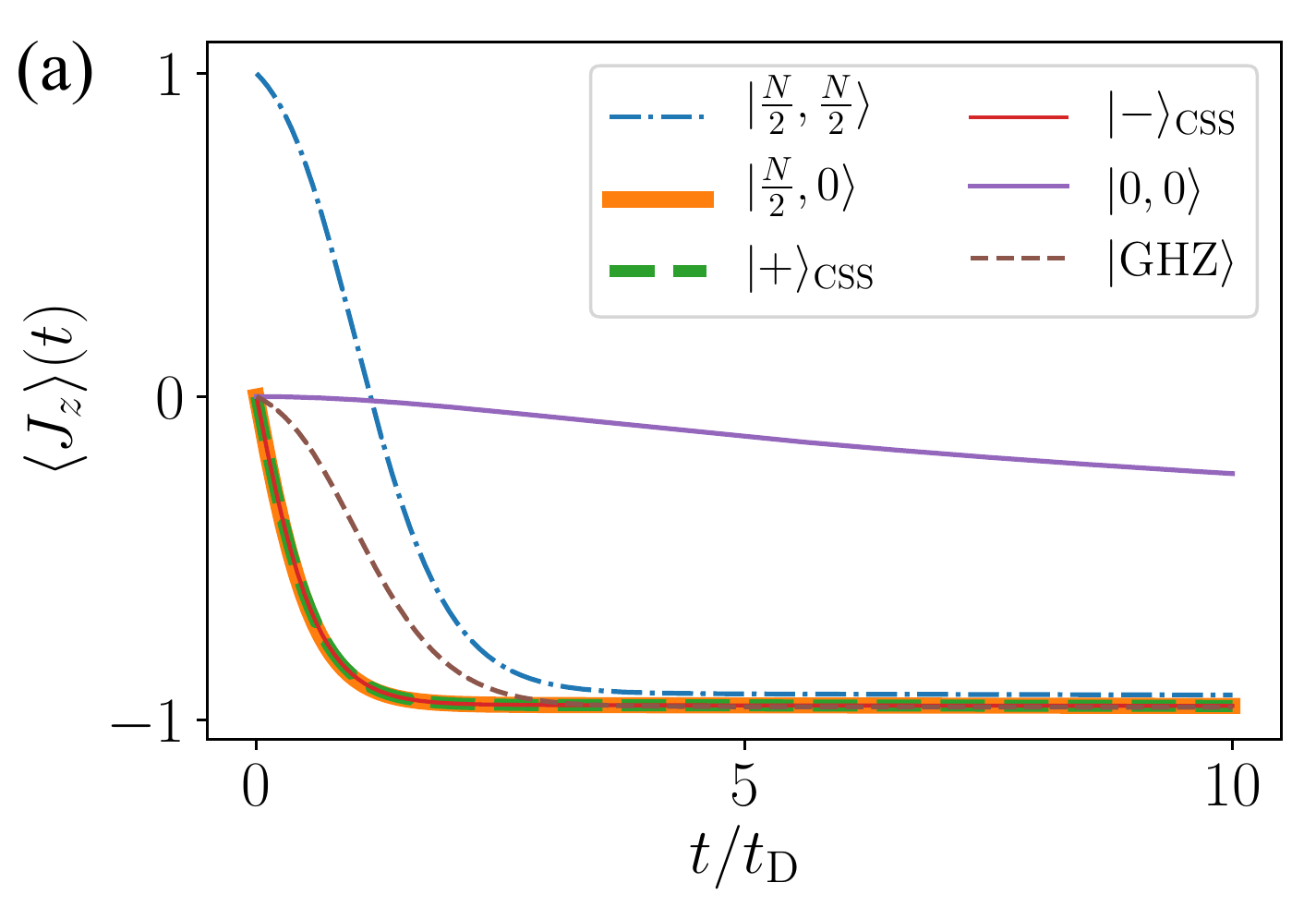}
\includegraphics[height=6.0cm]{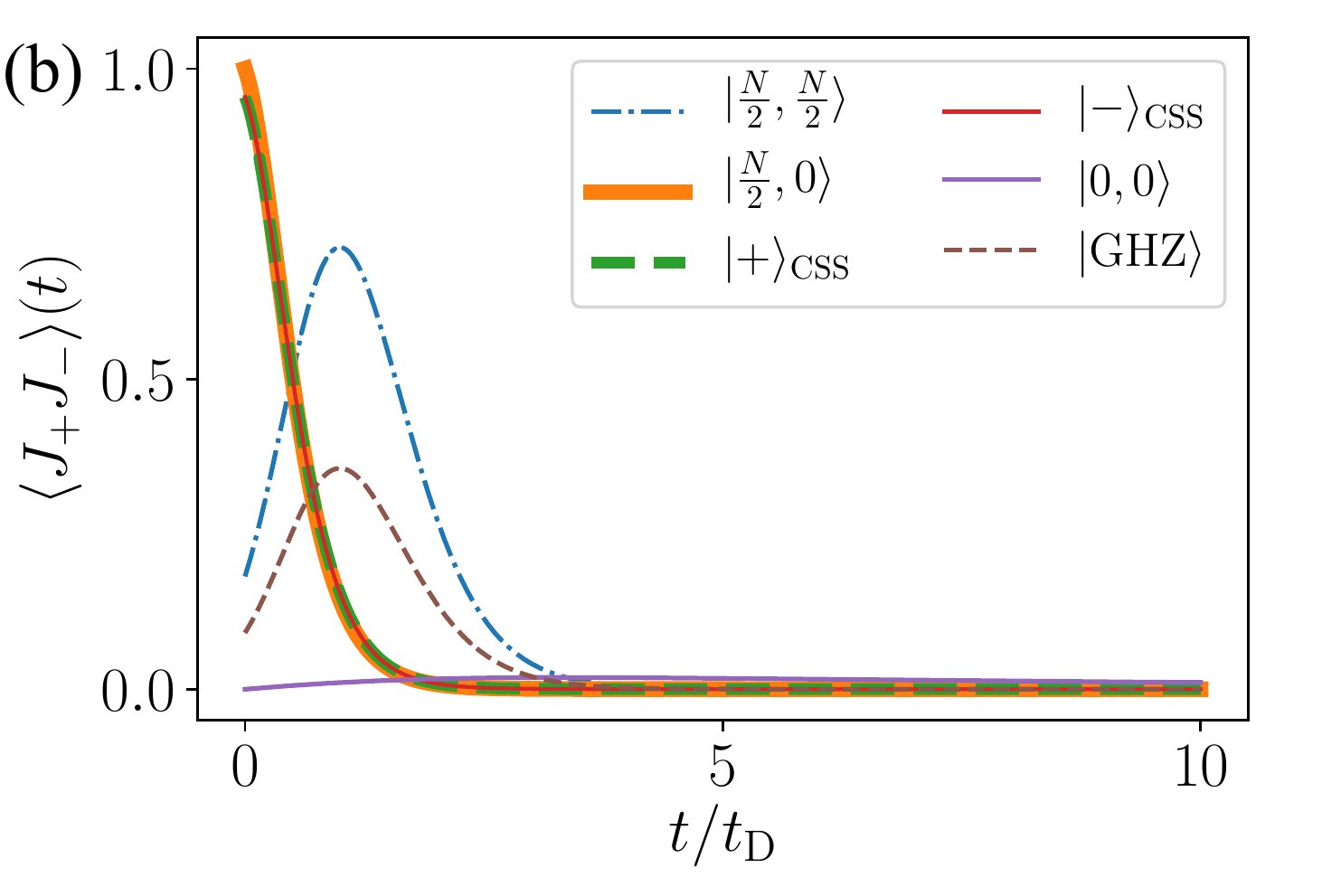}
\caption{\label{figx7} {\bf Superradiant light emission}. Superradiant decay of the total inversion, $\langle J_z\rangle(t)$, and emitted light, $\langle J_+J_-\rangle(t)$, for $N=20$ TLSs, for different initial states, with $H = \omega_0J_z$, $\omega_0\gg\gamma_{\Downarrow}$, $\gamma_{\Downarrow}/\gamma_{\phi}=1$. The time is expressed in units of the superradiant delay time, $t_\text{D}=\log N/N\gamma_{\Downarrow}$.  }
\end{center}
\end{figure*}

To begin, we study the paradigmatic example of superradiant light emission \cite{Dicke54,Bonifacio70,Lehmberg70,Bonifacio71,Bonifacio71b,Bonifacio75,Gross82,Mandel}, which can be generalized to include local phase-breaking terms, particularly relevant in large TLS ensembles and in solid-state implementations, in which sub-optimal experimental conditions spoil the simple picture of a single collective light emission coupling \cite{Lehmberg70,Skribanowitz73,Maki89,Temnov05,Scheibner07,Scully09,Scully10,Delanty11,Noe12,vanLoo13,Goban15,Damanet16a,Damanet16b,Damanet16c,Kakuyanagi16,Lambert16,Cong16,Jahnke16,Bradac16,Kim18,Angerer18} (see Ref.~\cite{Shammah17} for a more comprehensive list of references). Here we add local dephasing to the classical superradiant master equation,
\begin{eqnarray}
\label{eq2xxx}
\dot{\rho}&=&-i \omega_{0}\lbrack J_{z},\rho \rbrack
+\frac{\gamma_{\Downarrow}}{2}\mathcal{L}_{J_{-}}[\rho]
+\sum_{n=1}^{N}\frac{\gamma_{\phi}}{2}\mathcal{L}_{J_{z,n}}[\rho],
\end{eqnarray}
where $\omega_{0}$ is the bare TLS resonance, $N$ is the number of TLSs, and $\gamma_{\Downarrow}$ and $\gamma_{\phi}$ the rates of collective emission and local dephasing, respectively. The relevant parameters that affect its dynamics are $N$ and $\gamma_{\phi}/\gamma_{\Downarrow}$, which we set respectively to $20$ and $1$. In Figure~\ref{figx7}, the time evolution of the normalized total inversion, Figure~\ref{figx7}(a), and light emission, Figure~\ref{figx7}(b), are shown under the Liouvillian dynamics of \Eq{eq2xxx} for different initial states. 

Thanks to the fact that PIQS allows us to obtain the time evolution of the full collective density matrix $\rho(t)$, we can compare different initial states that have the same moments or non-classical correlations at the initial time. 
The fully excited state, $\ket{\frac{N}{2},\frac{N}{2}}$, shown by dot-dashed curves in Figure~\ref{figx7}, is the one that leads to a superfluorescent light emission after a delay time $t_\text{D}=\log (N)/(N\gamma_{\Downarrow})$ \cite{Bonifacio71} and then proceeds to a slow decay on a timescale set by $\gamma_{\phi}$ \cite{Shammah17,Bradac16,Angerer18}. 
The entangled superradiant Dicke state $\ket{\frac{N}{2},0}$ (solid orange curve), as well as the symmetric $\ket{+}_\text{CSS}$ and antisymmetric $\ket{-}_\text{CSS}$ coherent spin states, which are separable states, evolve almost identically in time, as shown by the dashed and solid curves, respectively, as predicted in Ref.~\cite{Bonifacio71b} for $\gamma_{\phi}=0$. 

The subradiant Dicke state $\ket{0,0}$, which would exhibit a frozen dynamics for $\gamma_{\phi}=0$, displays a slow decay due to the presence of dephasing, as shown in Figure~\ref{figx7}(a). 
If local incoherent losses were present, the decay of this state would be faster and no light would be emitted. 
Yet, when only dephasing is present, a small light emission occurs, as shown in Figure~\ref{figx7}(b); an intuitive analytical explanation for this dynamics can be gained by looking at the Dicke space of Figure~\ref{dickespace}: the system moves from the state $\ket{j,-j}$, with $j=0$ at $t=0$, to an inner state $\ket{j+1,-j}$ with the same excitation number, since $\Delta m=0$, and greater cooperative number, $\Delta j>0$, and falls on states with $\ket{j+1,-j-1}$, and so on, emitting a photon for each jump. 
We point out that, while superradiance has been observed experimentally since the 1970s \cite{Skribanowitz73}, the generation of collective symmetric Dicke states of TLSs has proven more elusive until recent times \cite{Linington08,Noguchi12}. Non-symmetrical Dicke states have been implemented in atomic clouds \cite{Bienaime12,Guerin16} and deterministic generation has been pioneered in superconducting circuits \cite{vanLoo13}. 

Finally, the GHZ state (dashed brown curves) displays a distinctive superradiant peak in the emitted light. 
Note that a first-order semiclassical theory or the cumulant-expansion method would not distinguish between the time evolution of the GHZ and that of the superradiant state or the (anti-)symmetric CSS, since at $t=0$, they all have identical first- and second-order moments. 
%Figure 7
%%%
\subsection{Steady-state superradiance}
\label{ssr}
%%%
The notion of a superradiant laser was introduced in Refs.~\cite{Haake93,Haake96} to describe coherent light emission of a collection of TLSs interacting with a single-mode cavity operating in the bad-cavity limit.
It was extended to an incoherently pumped steady-state light emission in Refs.~\cite{Meiser10a,Meiser10} and
verified experimentally in atomic ensembles of Rb and Sr atoms \cite{Bohnet12,Norcia16}. 
In the bad-cavity limit, the photonic degrees of freedom can be traced out and the dynamics can be described by 
\begin{eqnarray}
\label{eq2ss}
\dot{\rho}&=&-i \omega_{0}\lbrack J_{z},\rho \rbrack
+\frac{\gamma_{\Downarrow}}{2}\mathcal{L}_{J_{-}}[\rho]\nonumber\\
&&+\sum_{n=1}^{N}\left(\frac{\gamma_{\uparrow}}{2}\mathcal{L}_{J_{+,n}}[\rho]+\frac{\gamma_{\downarrow}}{2}\mathcal{L}_{J_{-,n}}[\rho]\right),
\end{eqnarray}
where here $\gamma_{\uparrow}$ is the rate of homogeneous local pumping and $\gamma_{\downarrow}$ that of local emission.
Equation (\ref{eq2ss}) corresponds to having only a rotating-wave light-matter coupling, differently from the Dicke-Hepp-Lieb phase transition \cite{Hepp73,Wang73} has been shown \cite{Meiser10a,Meiser10,Tieri17,Xu}.
For the case in which $\gamma_{\downarrow}=0$, the existence of a threshold beyond which the incoherently pumped system emits coherent light with a superradiant enhancement factor. Its dynamics is determined by $N$ and the normalised local rates $\gamma_\downarrow/\gamma_\Downarrow$ and $\gamma_\uparrow/\gamma_\Downarrow$.

\begin{figure}[ht!]
\begin{center}
\includegraphics[width=8cm]{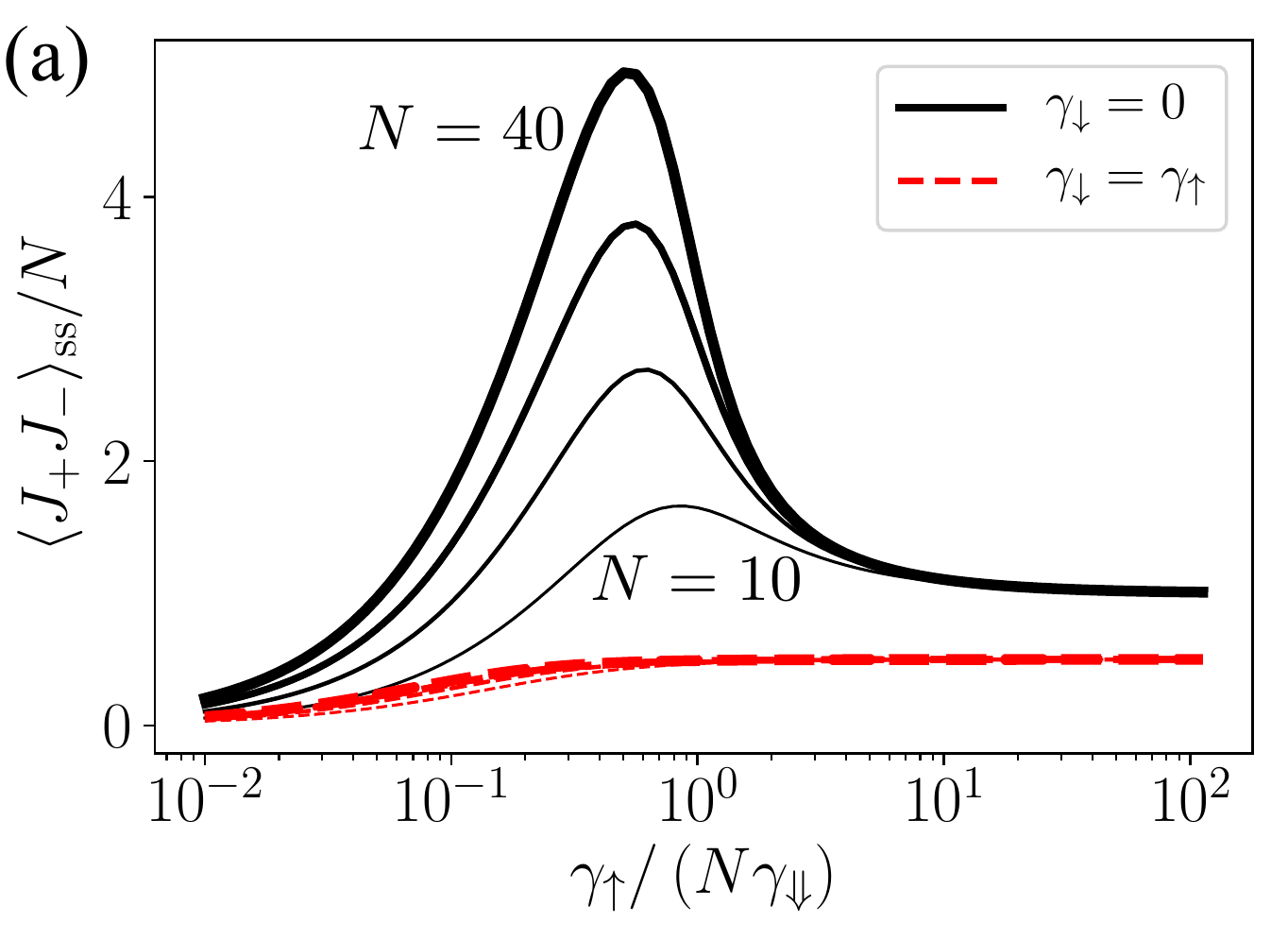}
\includegraphics[width=8cm]{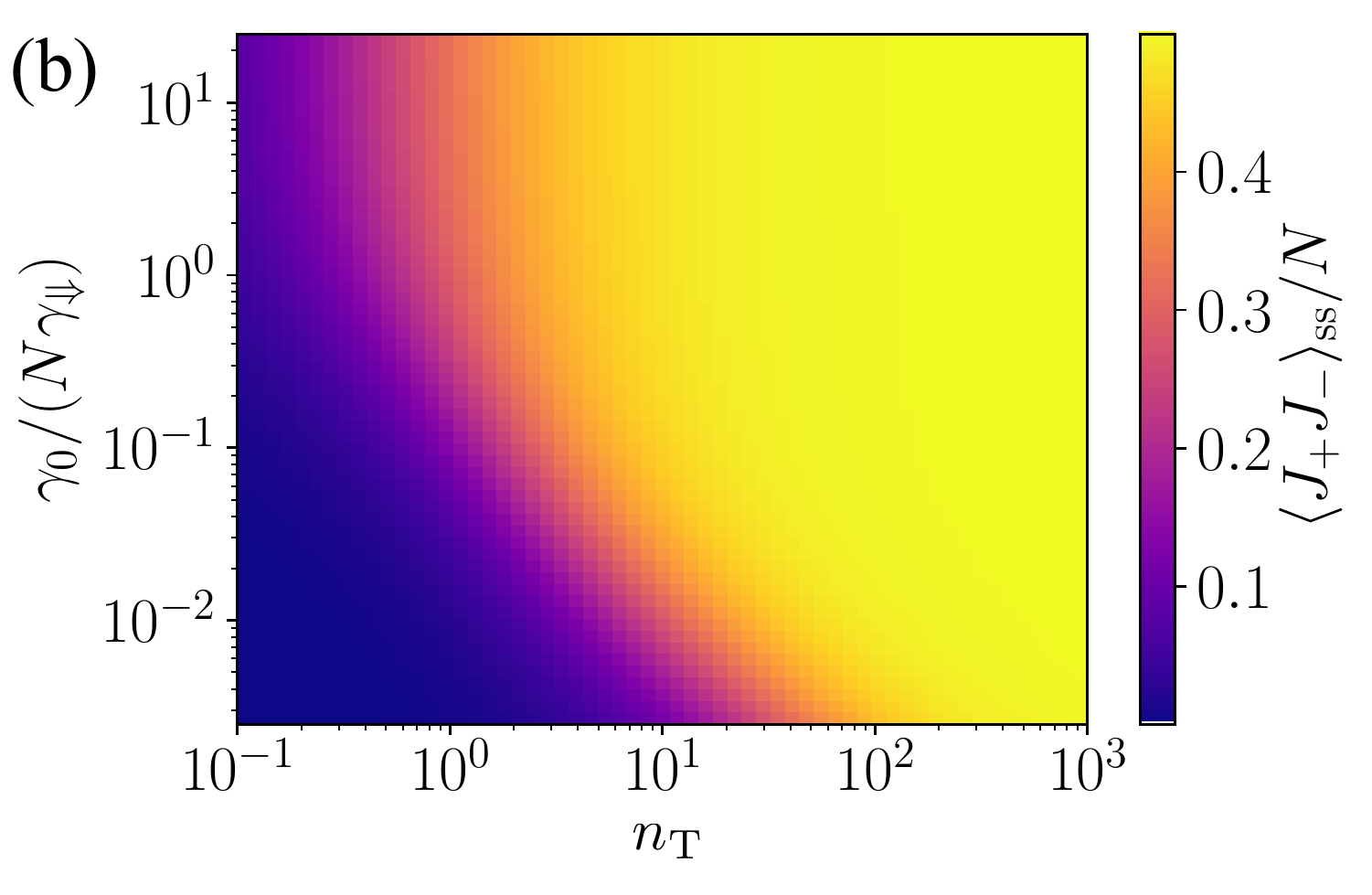}
\caption{\label{figx7b} 
{\bf Steady-state superradiance}: (a) Value of the steady-state light emission $\langle J_{+}J_{-}\rangle_\text{ss}/N$ as a function of the local pumping strength $\gamma_{\uparrow}$, which is in units of $N\gamma_{\Downarrow}$. We set $\gamma_{\Downarrow}$ and tune $\gamma_{\uparrow}$ for $N=10$ (thinnest curves), 20, 30, 40 (thickest curves). Two different cases are considered: in the first one, only collective emission is present, $\gamma_{\downarrow} = 0$ (solid black curves), see also Figure~2 of Ref.~\cite{Meiser10a}; in the second one (dashed red curves), the local emission rate ensures the detailed balance condition with respect to local pumping, $\gamma_{\downarrow}=\gamma_{0}(1+n_\text{T})$ and $\gamma_{\uparrow}=\gamma_{0}n_\text{T}$; we consider the high-temperature limit, $n_\text{T}\gg 1$, thus setting $\gamma_{\downarrow}=\gamma_{\uparrow}$. (b) We study the steady-state light emission for fixed $\gamma_\Downarrow=\omega_0$ and $N=40$, varying both $\gamma_0$ and $n_\text{T}$.}
\end{center}
\end{figure}

In Figure~\ref{figx7b}(a) we show the normalized value of the emitted light for the steady state of \Eq{eq2ss}, $\langle J_+J_-\rangle_\text{ss}$, as a function of the local pumping rate divided by $N$ for $N$=10, 20, 30, and 40, thinnest to thickest curve.
Solid black curves correspond to $\gamma_{\downarrow}=0$.
In agreement with Ref.~\cite{Meiser10a}, the value of light emission in the steady state, occurs at a point that is found around $\gamma_\uparrow=N\gamma_\Downarrow $. 
We find that the maximum value, occurring at the critical coupling, has a superradiant scaling.

The inclusion of local losses, as done in \Eq{eq2ss}, in the presence of collective phenomena, can be relevant to thermodynamics schemes, e.g., for quantum heat engines \cite{Scully01,Quan06,Quan07,DeLiberato11b,Scully11,Dorfman13,Hardal15}, and bio-inspired photon-absorption devices \cite{Creatore13,Higgins14}.
We thus consider the case in which the dissipative local interactions obey detailed balance, that is we set 
 $\frac{\gamma_{\uparrow}}{\gamma_{\downarrow}}=\frac{n_\text{T}}{1+n_\text{T}}$, where $n_\text{T}$ is the thermal occupation number. 
The red dashed lines in Figure~\ref{figx7b} show the results in the high-temperature case $n_\text{T}\gg1$, in which the system can become highly excited. We see that  thermal equilibrium in the local processes lead to the disappearance of any resonant feature, with the emission progressively dampened for larger values of $N$.
In Figure~\ref{figx7b}(b) we study how temperature affects the superradiant behaviour, fixing the collective emission rate, $\gamma_\Downarrow=\omega_0$ and the number of TLSs, $N=40$. We find that the detailed balance condition is detrimental at any temperature, i.e. for any occupation number $n_\text{T}$, and for any ratio between of the local baths figure of merit, $\gamma_0$ and the collective emission rate, $\gamma_\Downarrow$, thus showing that cooperative light emission is prevented by the detailed balance condition of the local baths.
%Figure 8
%%%
\subsection{Spin squeezing}
%%%
\label{lmg}
Spin-squeezed states can improve the sensitivity of measurements beyond the classical limit \cite{Wineland92, Kitagawa93}. They have recently been implemented on hundreds of trapped ions \cite{Bohnet16,Garttner17} and large ensembles of atoms in Bose-Einstein condensates \cite{Ma11,Pezze16,Escher11,Strobel14,Luo17,Zou18}. A typical squeezing Hamiltonian requires a second moment of the collective spin operators \cite{Baragiola10,Dooley16,Opatrny15,Hu17,Garttner18}.  

\begin{figure*}[ht!]
\centering
\includegraphics[width=18.5cm]{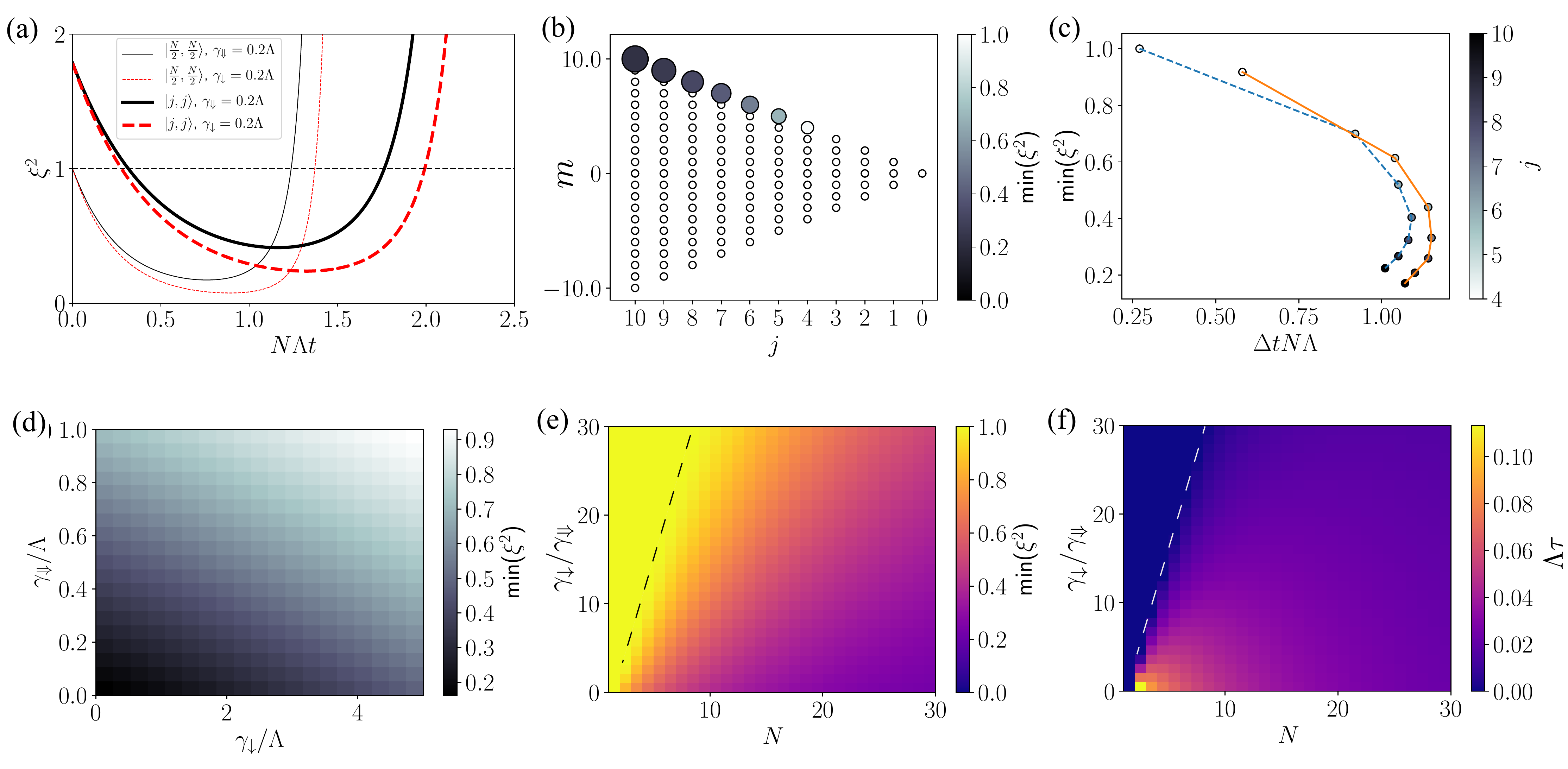}
\caption{\label{fig:squeeze} 
{\bf Spin squeezing}. (a) Time evolution of the spin squeezing parameter $\xi^2$ for two different dynamics and two different initial states: the solid curve corresponds to $\gamma_{\downarrow}=\Lambda/5$ and $\gamma_{\Downarrow}=0$, the dashed curve to $\gamma_{\downarrow}=0$ and $\gamma_{\Downarrow}=\Lambda/5$. Here we consider $N=50$ TLSs. The thin curves qualitatively reproduce Figure~3 of Ref.~\cite{Chase08} and correspond to an initially fully excited state, $\ket{\frac{N}{2},\frac{N}{2}}$. The thick curves instead correspond to the non-symmetric Dicke state with longest time under spin squeezed evolution, $\xi^2<1$, the $\ket{j,j}$ Dicke state with $j=14$. (b) Study of the minimum spin squeezing parameter $\xi^2$, for any initial Dicke state $\ket{j,m}$: We consider $\gamma_{\downarrow} = \Lambda/5$ and $\gamma_{\Downarrow}= 0$ and $N = 20$, the size of the circles and the shading of the filling give the strength of the maximum spin squeezing obtained. (c) Trade-off between the spin squeezing time $\tau$, and the minimum spin squeezing parameter, $\xi^2$, for different initial states, fixing $N = 20$ for two different dynamical conditions, either in the presence of collective dissipation only (circles joined by the solid orange segmented line) or local dissipation only (dashed blue segmented line). (d) The minimum value of the spin squeezing parameter $\xi^2$ (maximum squeezing) is explored for different values of $\gamma_{\downarrow}$ and $\gamma_{\Downarrow}$, with initial state the fully excited state for $N=20$. (e)-(f) Color plots for the minimum spin squeezing parameter value, $\xi^2$, and the time at which it is reached, $\tau$, as a function of $\gamma_\downarrow$, fixing $\gamma_\Downarrow=\Lambda/5$. The dashed lines help identify qualitatively the region with spin squeezing (parameter space to the right of the line).}
\end{figure*}

Of this important class of problems, here we demonstrate the study of the two-axis twisting Hamiltonian \cite{Kitagawa93}
\begin{eqnarray}
\label{eq:squeeze}
H&=&-i\Lambda\left(J_{+}^2-J_{-}^2\right),
\end{eqnarray}
evolving under the dissipative dynamics
\begin{eqnarray}
\label{eq:squeeze2}
\dot{\rho}&=&-\frac{i}{\hbar} \lbrack H,\rho \rbrack +\frac{\gamma_{\Downarrow}}{2}\mathcal{L}_{J_{-}}+ \frac{\gamma_{\downarrow}}{2}\sum_{n=1}^{N}\mathcal{L}_{J_{-,n}}[\rho].
\end{eqnarray}
In Ref.~\cite{Chase08}, it has been shown that the collective emission in $\gamma_{\Downarrow}$ and the homogeneous local emission in $\gamma_{\downarrow}$ affect in a different way the spin squeezing of the system, with the local emission being less detrimental than the collective emission mechanism to the attainable degree of spin squeezing. 
In Figure~\ref{fig:squeeze}(a) we plot the spin squeezing parameter $\xi^2=N\langle\Delta J_y^2\rangle/\left(\langle J_z\rangle^2+\langle J_x\rangle^2\right)$ \cite{Ma11}  for $N=50$. The horizontal black dashed line marks the boundary $\xi^2=1$ below which there is spin squeezing in the system. The black solid curves correspond to \Eq{eq:squeeze2} with $\gamma_{\downarrow}=0$ and $\gamma_{\Downarrow}=\Lambda/5$, while the red dashed curves show the homogeneous local dynamics, $\gamma_{\downarrow}=\Lambda/5$ and $\gamma_{\Downarrow}=0$. 
The thin curves correspond to a system initialized in the excited state $\ket{\frac{N}{2}, \frac{N}{2}}$ and qualitatively reproduce Figure~3 of Ref.~\cite{Chase08}.

We have then used PIQS to explore all $\ket{j,m}$ Dicke states to identify those evolving with spin squeezing \cite{Duan11,Zhang14d}.
We found that, while the excited state provides the greater degree of spin squeezing, for some non-symmetric states $\ket{j,j}$ the system displays spin squeezing with a delay and for a longer time \cite{Lucke14}. 
In Figure~\ref{fig:squeeze}(a), the thick solid curves correspond to a system initialized in the Dicke state $\ket{j,j}$, with $j<\frac{N}{2}$, setting $j=14$, which is the state which exhibits the longest time evolution as a spin squeezed state for the dynamics governed by $\gamma_\downarrow$. This is $\sim$30\% longer than for the $\ket{\frac{N}{2},\frac{N}{2}}$ state previously considered in the literature \cite{Chase08}. 

In Figure~\ref{fig:squeeze}(b) we show the spin squeezing parameter minimum value for any initial Dicke state, fixing $N=20$. It is discernible that only the first seven states $\ket{j,j}$, beginning from $j=\frac{N}{2}$ and decreasing value, display spin squeezing, for the dissipative dynamics in which only local emission is considered, $\gamma_\downarrow=\Lambda/5$. Similar qualitative results hold for the dynamics influenced by collective dissipation only, $\gamma_\Downarrow=\Lambda/5$. In Figure~\ref{fig:squeeze}(c), a study of the optimal condition for maximum spin squeezing and spin squeezing time is shown, for such seven initial states displaying spin squeezing, this time adding, to the local dynamics studied in panel (a) (markers joined by the dashed line) also a comparison to the dynamics for which $\gamma_\Downarrow=\Lambda/5$ (solid lines). Both local and dissipative dynamics have the same effect on spin squeezing, with collective emission being more detrimental. The plot shows that the relation between maximum spin squeezing and spin squeezing time is non monotonic, giving indication of different optimal conditions for cases in which either spin squeezing or the time of the spin squeezed evolution might be most relevant. 

In Figure~\ref{fig:squeeze}(d)-(f) we then turn to explore a more general mixed collective-incoherent dissipative dynamics, in which both $\gamma_\downarrow$ and $\gamma_\Downarrow$ coexist in \Eq{eq:squeeze2}. In Figure~\ref{fig:squeeze}(d) we explore the coherent-incoherent parameter space to study mixed conditions, finding that the cooperative decay is more detrimental than local decay to spin squeezing, and both have qualitatively the same detrimental effect, even when they are both present in the dynamics. 
In Figure~\ref{fig:squeeze}(e)-(f), we provide a color plot of maximum spin squeezing [panel (e)] and the time of its occurrence, $\tau$, [panel (f)], in the parameter space of local and collective dissipation as a function of the number of particles, in order to ascertain the presence of any optimal condition emerging from cooperative behaviour. We fix $\gamma_\Downarrow=\Lambda/5$ and vary both $\gamma_\downarrow$ and $N$. While the maximum spin squeezing achievable improves monotonically for greater $N$, due to the cooperative nature of the unitary spin squeezing Hamiltonian, the time of its occurrence is non trivial, and a condition for engineering longer spin-squeezing time (under non-optimal spin squeezing) can be found, which can be of interest in metrological experimental conditions in which the time to perform measurements might want to be maximized. 

An interesting direction would be to further explore the effect of local and collective dissipation terms in order to obtain robust, steady-state spin squeezing states generation \cite{DallaTorre13b}, also in the presence of a dissipative cavity, addressing the interplay with bosonic squeezing.
Since PIQS grants access to the full density matrix, it also allows us to investigate higher-order quantum correlations and other entanglement witnesses, such as the quantum Fisher information \cite{Rey08,Pezze09,Strobel14,Wang14,Zhang18d}. 

% Figure 9
%

%%%
%%%
\subsection{Phase transitions}
\label{qpt}
%%% 
%%
Driven-dissipative systems of the kind specified by \Eq{master} and \Eq{masterbos} are being intensively studied in the context of out-of-equilibrium phase transitions \cite{Morrison08a,Morrison08b,Morrison08c,Lambert09,Kessler12,DallaTorre13,Lee13,Joshi13,Zhang14b,Eisert15,DallaTorre16,Sieberer16,Maghrebi16,FossFeig17,Ohadi17,Garttner17,Shchadilova18}.
These systems challenge some of the fundamental assumptions for systems at equilibrium. 
Recent studies have highlighted the rich physics that can be inferred from the properties of the Lindbladian superoperators and their spectra \cite{Lee13,Albert14,Albert16}.
Other systems of ensembles of TLSs have shown collective phenomena such as synchronization, symmetry breaking, and long-lived state protection \cite{Komar14,Xu15,Bellomo17,Shankar17,Hannukainen17,Iemini17,Gong18,Hama16,Hama18b}.

%%%
\subsubsection*{Open Dicke model}
\label{qpt2}
%%%
The Dicke model in a driven-dissipative setting has been the object of much attention, comprehensively described in a recent review \cite{Kirton18r}. Here we will show how PIQS can be used to study a model including pumping and losses both at the local and collective level, described by the equation
\begin{eqnarray}
\label{eq4xb1}
\dot{\rho}&=&-\frac{i}{\hbar} \lbrack H,\rho \rbrack+\frac{\gamma_{\Downarrow}}{2}\mathcal{L}_{J_{-}}[\rho]+\frac{\gamma_{\Uparrow}}{2}\mathcal{L}_{J_{+}}[\rho]+\frac{\kappa}{2}\mathcal{L}_{a}[\rho]
\nonumber\\&&+\sum_{n=1}^{N}\left(\frac{\gamma_{\downarrow}}{2}\mathcal{L}_{J_{-,n}}[\rho]
+\frac{\gamma_{\phi}}{2}\mathcal{L}_{J_{z,n}}[\rho]
+\frac{\gamma_{\uparrow}}{2}\mathcal{L}_{J_{+,n}}[\rho]\right),
\nonumber\\
\end{eqnarray}
with 
\begin{eqnarray}
\label{hamdicke}
H &=\hbar \omega_{0}J_{z} + \hbar\omega_\text{cav}a^\dagger a + \hbar g J_{x}\left(a+a^{\dagger}\right).
\end{eqnarray}
% Figure 10
\begin{figure}[ht!]
\begin{center}
\includegraphics[width=8cm]{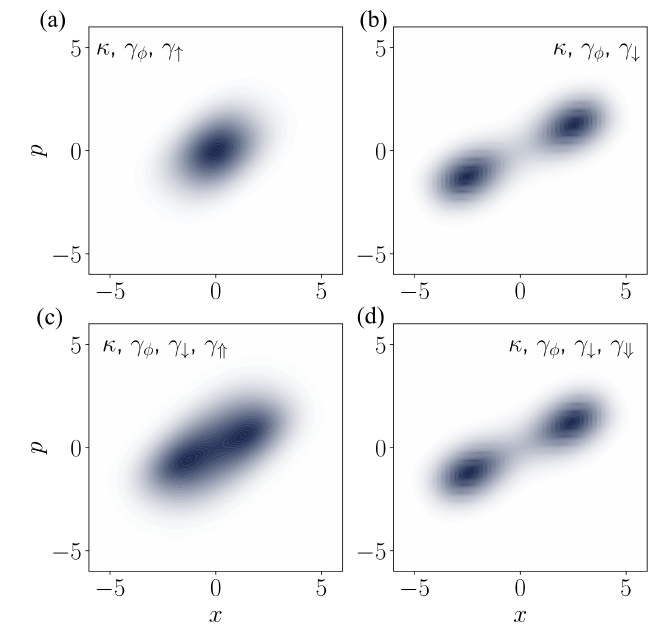}
\caption{\label{figx7d} {\bf Open Dicke model}.
Here we consider the Wigner function of the photonic part of the steady state;
$x$ and $p$ are the conjugate operators of the photonic mode operators. 
We fix the cut-off of the photonic Hilbert space at $n_\text{ph}=20$ and study the influence of local and collective processes.
For $N=10$ TLSs, $H$ is given by \Eq{eq4xb1}, with 
$\omega_\text{cav}=\omega_{0}$, $g=2\omega_{0}/\sqrt{N}$, $\kappa=\omega_{0}$ and $\gamma_\phi=0.01\omega_{0}$.
(a) Adding $\gamma_\uparrow=0.1\omega_{0}$ does not restore the superradiant phase transition. 
(b) Adding $\gamma_{\downarrow}=0.1\omega_0$ restores the superradiant phase transition, and reproduces one panel of Figure~1 of Ref.~\cite{Kirton17}.
(c) Adding both $\gamma_{\downarrow}=0.1\omega_0$ and the collective pumping $\gamma_{\Uparrow}=0.1\omega_0$ is detrimental to the superradiant phase. 
(d) adding to $\gamma_{\downarrow}=0.1\omega_0$, also  the collective emission $\gamma_{\Downarrow}=0.1\omega_0$ still allows us to resolve the superradiant phase.}
\end{center}
\end{figure}
Note that, although usually for values of the coupling constant $g$ comparable to the frequencies of the bare excitations, more refined open quantum system approaches, described in Section~\ref{sec:usc}, have to be used \cite{DeLiberato09,Beaudoin11,Ridolfo12,Bamba13,Bamba14,DeLiberato14a,DeLiberato17}, \Eq{eq4xb1} is fully justified in effective models
\cite{Dimer07,Baumann10,Kirton17,Kirton17b,Gegg17b}.
The introduction of local dissipative terms modifies the properties of the steady state of \Eq{eq4xb1}. References~\cite{DallaTorre16,Kirton17,Kirton17b} have studied the superradiant phase transition \cite{Hepp73,Wang73} in the presence of several local driven-dissipative processes. In Ref.~\cite{DallaTorre16} it has been shown that the coupling to a thermal bath affects the critical temperature of the phase transition, i.e., when the $\gamma_\downarrow$ and $\gamma_\uparrow$ are governed by detailed balance. The influence of a general Markovian bath $\sum_n\mathcal{L}_{[J_{-,n}+\lambda J_{+,n}]}(\rho)$, with $\lambda$ a dimensionless real parameter, has also been assessed in Ref.~\cite{DallaTorre16}, highlighting that for $\lambda\rightarrow1$ and hence considering the limit of $\sum_n\mathcal{L}_{[J_{x,n}]}(\rho)$, the critical point moves to ever higher light-matter couplings and there is no superradiant phase. 

Reference~\cite{Kirton17} has illustrated that the inclusion of local incoherent emission, $\gamma_\downarrow$, in the presence of local dephasing, $\gamma_\phi$, can restore the existence of the superradiant phase.
The interplay between the superradiant, normal, and lasing phases has been addressed in Ref.~\cite{Kirton17b} with regards to a general Dicke Hamiltonian with an additional degree of freedom in the coupling of the counter-rotating terms.
The occurrence of each phase has been found to depend on the ratio of coherent collective pumping in the Hamiltonian and local pumping, $\gamma_\uparrow$.
 
Here we reproduce part of these results and further extend the study to address the effect of collective pumping and emission. 
In Figure~\ref{figx7d}, we group contour plots of the Wigner function of the photonic part of the steady state of \Eq{eq4xb1}, $\rho_\text{ph}$,
\begin{eqnarray}
\label{eqwig}
W(x,p)&=&\frac{1}{\pi}\int_{-\infty}^{\infty}\bra{x-x'}\rho_\text{ph}\ket{x+x'}e^{2ipx'/\hbar}dx',
\end{eqnarray}
with $x$ and $p$ here representing the phase-space conjugate coordinates of the photonic mode operators, $a=\frac{1}{\sqrt{2}}\left(x+\frac{i}{\hbar}p\right)$. 
We set $N=10$, $\kappa=\omega_0=\omega_\text{cav}$, and assess the effect of different conditions of collective and local pumping and emission, always maintaining local dephasing, $\gamma_{\phi}=0.01\omega_0$, which is detrimental to the superradiant phase. 
We calculate the Wigner function using QuTiP's \code{wigner()} function.
In Figure~\ref{figx7d}(a), we add only incoherent pumping, $\gamma_{\uparrow}=0.1\omega_0$, which does not restore the superradiant phase and actually decreases the squeezing of the system. 
In Figure~\ref{figx7d}(b), the existence of two displaced and squeezed blobs is restored by a term in $\gamma_{\downarrow}=0.1\omega_0$, qualitatively reproducing the result obtained in Ref.~\cite{Kirton17}. 

Naively, one would expect a duality between the models of Figure~\ref{figx7d}(a)-(b), with $\kappa$, $\gamma_\phi$, $\gamma_\downarrow$ (which displays the superradiant transition) and $\kappa$, $\gamma_\phi$, $\gamma_\uparrow$ (which shows no transition). 
The reason there is no transition is because this duality also requires the sign on $\omega_0$ in \Eq{hamdicke} to be flipped. 
In the pumping case, the system is thus effectively far detuned from resonance and the superradiant phase transition does not occur.
In Figure~\ref{figx7d}(c), we find that collective pumping, with $\gamma_{\Uparrow}=0.1\omega_0$, is detrimental to the superradiant phase, while for collective losses with $\gamma_{\Downarrow}=0.1\omega_0$, shown in Figure~\ref{figx7d}(d), the fingerprint of the superradiant phase is still discernible.  
%%%
%%% Figure 11
\begin{figure*}[ht!]
\begin{center}
\includegraphics[height=12cm]{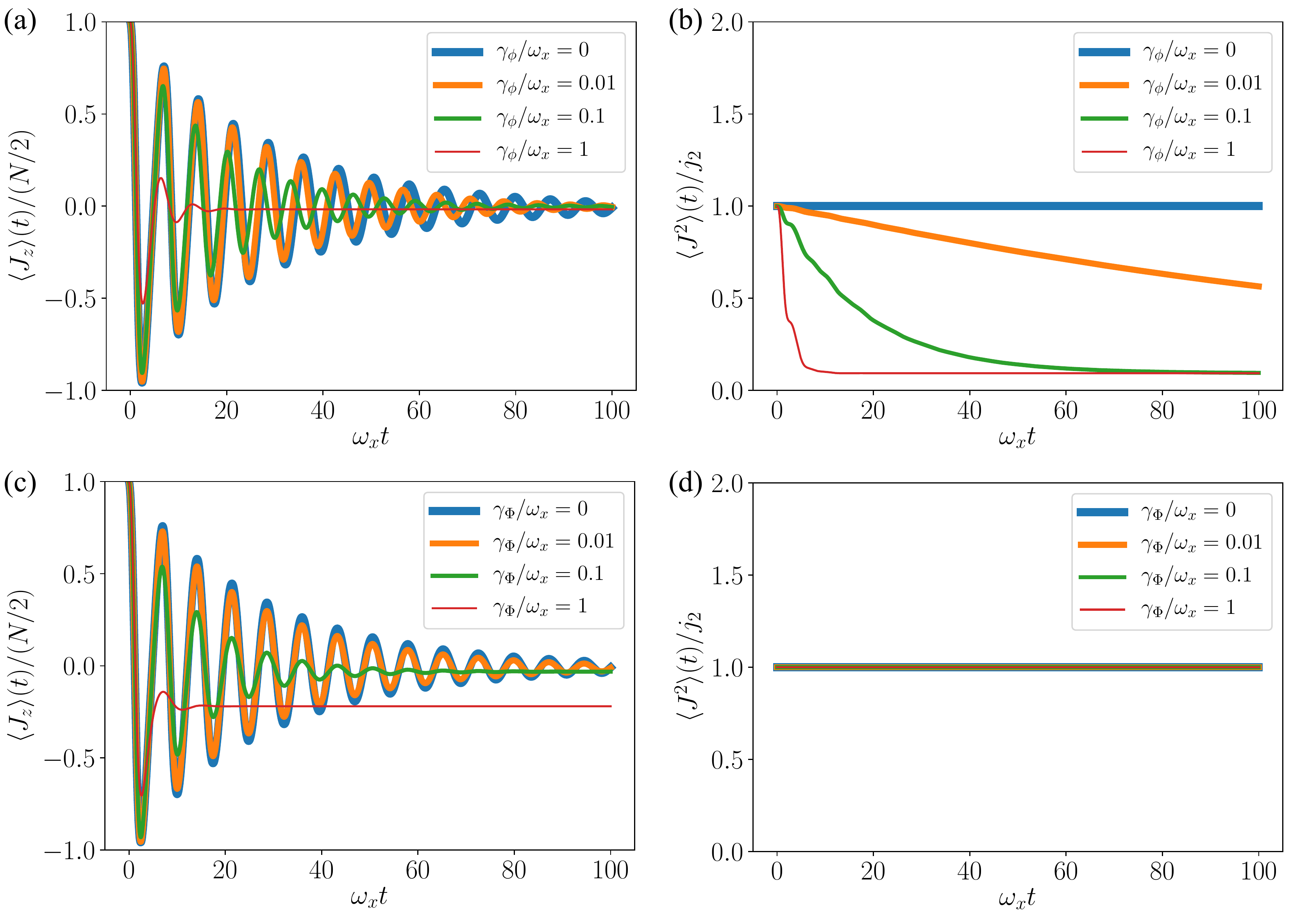}
\caption{\label{figx22} 
{\bf Boundary time crystal}. 
The dynamics of \Eq{eq4xb} is shown in the case of weak dissipation ($\omega_x\gg\frac{N}{2}\gamma_{\Downarrow} $) for $N=30$, with the system initialized in the excited state $\ket{\frac{N}{2},\frac{N}{2}}$, for different dephasing rates $\gamma_{\phi}/\omega_x=0,0.01, 0.1, 1$ (thicker to thinner curves). 
(a) The total inversion, $\langle J_{z}\rangle (t)$, normalized by $N/2$, is plotted in time. 
These results extend those obtained for $\gamma_{\phi}=0$ in Ref.~\cite{Iemini17}.
(b) The total spin length, $\langle J^2\rangle (t)$, normalized by $j_2=\frac{N}{2}(\frac{N}{2}+1)$, is shown.
In panels (c) and (d), the same quantities are plotted for different collective dephasing rates, $\gamma_{\Phi}/\omega_x=0,0.01, 0.1, 1$ (thicker to thinner curves).}
\end{center}
\end{figure*}
%%%
\subsubsection*{Limit cycles and boundary time crystals}
\label{qpt3}
%%%
%%%
The interplay of competing collective phenomena in open quantum systems can lead to limit cycles, as shown in Ref.~\cite{Iemini17} for the case of a collection of $N$ TLSs that are coherently driven at a frequency $\omega_x$, and that can also collectively decay, at a rate $\gamma_{\Downarrow}$. 

Here we consider the same model as in Refs.~\cite{Hannukainen17,Iemini17}, but generalized to include also local emission and local or collective dephasing terms,
\begin{eqnarray}
\label{eq4xb}
\dot{\rho}&=&-i \omega_{x}\lbrack J_{x},\rho \rbrack+\frac{\gamma_{\Downarrow}}{2}\mathcal{L}_{J_{-}}[\rho]+ \frac{\gamma_{\Phi}}{2}\mathcal{L}_{J_{z}}[\rho]\nonumber\\&&
+\sum_{n=1}^{N}\left(\frac{\gamma_{\downarrow}}{2}\mathcal{L}_{J_{-,n}}[\rho]
+\frac{\gamma_{\phi}}{2}\mathcal{L}_{J_{z,n}}[\rho]\right).
\end{eqnarray}
Note that, in the context of quantum optics, nonlinear effects arising from \Eq{eq4xb} have been studied also in relation to optical bistability and cooperative resonance fluorescence \cite{Bonifacio78,Drummond81,Carmichael85,Sarkar87} and more recently in connection to Hopfield neural networks \cite{Rotondo18}.
Two regimes of this model can be defined, one of strong dissipation $\omega_{x}\ll \frac{N}{2}\gamma_{\Downarrow}$, and one of weak dissipation, $\omega_{x}\gg \frac{N}{2}\gamma_{\Downarrow}$. 
In Ref.~\cite{Iemini17}, which considers $\gamma_{\downarrow}=\gamma_{\phi}=\gamma_{\Phi}=0$ and $\gamma_{\Downarrow}\propto \frac{1}{N}$, it has been shown that, in the weak dissipation regime, the total spin oscillations in $\langle J_z(t)\rangle$, set by $\gamma_{\Downarrow}\propto \frac{1}{N}$, become more clear with $N$, and the gap of the Liouvillian spectrum vanishes, a probe of PT. 
The observation of this phenomenon has been proposed in a Raman-driven cold-atom setup \cite{Dimer07} and named as a boundary time crystal \cite{Iemini17}. 

In Ref.~\cite{Iemini17}, it has been shown that the collective spin oscillations characterizing the limit cycle in the steady state are robust against nonlinear perturbations in the Hamiltonian. Using PIQS, we studied the effect of local dephasing on the collective spin oscillations. In Figure~\ref{figx22}(a), the normalized collective TLS inversion, is plotted as a function of time for $N=30$ with 
$ \omega_x/\gamma_{\Downarrow}= \frac{4}{N}$, no local emission, $\gamma_{\downarrow}=0$, and for different values of dephasing, $\gamma_{\phi}/\omega_x=0, 0.01, 0.1,1$.
We find that local dephasing affects the visibility of the collective oscillations and its detrimental effect with respect to collective processes can be traced in the decrease of the normalized total spin length, $\langle {J}^2(t)\rangle$, a measure of cooperation in the system, as shown in Figure~\ref{figx22}(b). 
Similarly, we have found that when local losses, instead of dephasing, are included, the effect is detrimental for the observation of the collective oscillations (not shown, available online) \cite{Piqs}. If collective dephasing processes, proportional to $\gamma_\Phi$, are included, we find that this collective effect does not shift the frequency of the spin oscillations, see Figure~\ref{figx22}(c). Collective dephasing leaves the total spin length unchanged, see Figure~\ref{figx22}(d). Both local and collective pure dephasing are detrimental to the visibility of the spin oscillations.
We note that a recent study has also assessed the robustness of time crystallization in a system under local noise, i.e. inhomogeneous broadening \cite{Tucker18}.

%%%

%%%
%%%
%%%
\subsection{Multiple spin ensembles}
\label{negtemp}
With PIQS it is simple to study multiple TLS ensembles coupled to a single cavity or multiple bosonic cavities.
The approach can be generalized to $k$ ensembles of TLSs, with $k>2$ and each ensemble with a given $N_k$ TLS population. 
In the case of open driven-dissipative quantum systems, \Eq{master} has been used to study a dynamical phase transition that can synchronize two populations of atoms as in a quantum version of a Huygens clock with local and collective driving and dissipation \cite{Xu14}. 

For simplicity, here we perform a study of two TLS ensembles, with populations $N_1$ and $N_2$. 
In the bad-cavity limit, the cavity degree of freedom can be traced out. 
It has been shown in Refs.~\cite{Hama16,Hama18b} that if $N_1\neq N_2$, and both ensembles are driven and can dissipate only through a common channel, a peculiar exchange of spin inversion can be engineered in the system. 
We will consider a generalization of the collective dynamics considered in Ref.~\cite{Hama16}, to include the possibility of a single ensemble of TLS to experience a collective or local dissipative dynamics. 
We thus consider the master equation
\begin{eqnarray}
\label{ntemp0}
\dot{\rho}&=&-i \omega_0 \lbrack J_z^{(1)}+J_z^{(2)},\rho \rbrack
+\frac{\gamma_{\Downarrow}}{2}\mathcal{L}_{(J_{-}^{(1)}+J_{-}^{(2)})}[\rho]
\nonumber\\&&
+\frac{\gamma_{\Uparrow}}{2}\mathcal{L}_{(J_{+}^{(1)}+J_{+}^{(2)})}[\rho]
+\sum_{k=1}^2\frac{\gamma_{\Downarrow,k}}{2}\mathcal{L}_{J_{-}^{(k)}}[\rho]
\nonumber\\&&
+\sum_{n=1}^{N_k}\left(\frac{\gamma_{\downarrow}}{2}\mathcal{L}_{J_{-,n}^{(k)}}[\rho]
+\frac{\gamma_{\phi}}{2}\mathcal{L}_{J_{z,n}^{(k)}}[\rho]\right), 
\end{eqnarray}
where $\gamma_\Downarrow$ ($\gamma_{\Uparrow}$) is the rate of collective decay (pumping) of the two coupled ensembles, $\gamma_{\Downarrow,k}$ ($\gamma_{\Uparrow,k}$) is the rate of collective decay (pumping) of the single ensemble $k$, 
while $\gamma_{\downarrow}$ and $\gamma_{\phi}$ are the local emission and local dephasing rates, respectively. In \Eq{ntemp0}, the operators $J_{\alpha,n}^{(k)}$ and $J_{\alpha}^{(k)}$ are the local, and collective, operators of the $k$ ensemble, respectively.

\begin{figure*}[ht!]
\begin{center}
\includegraphics[width=8cm]{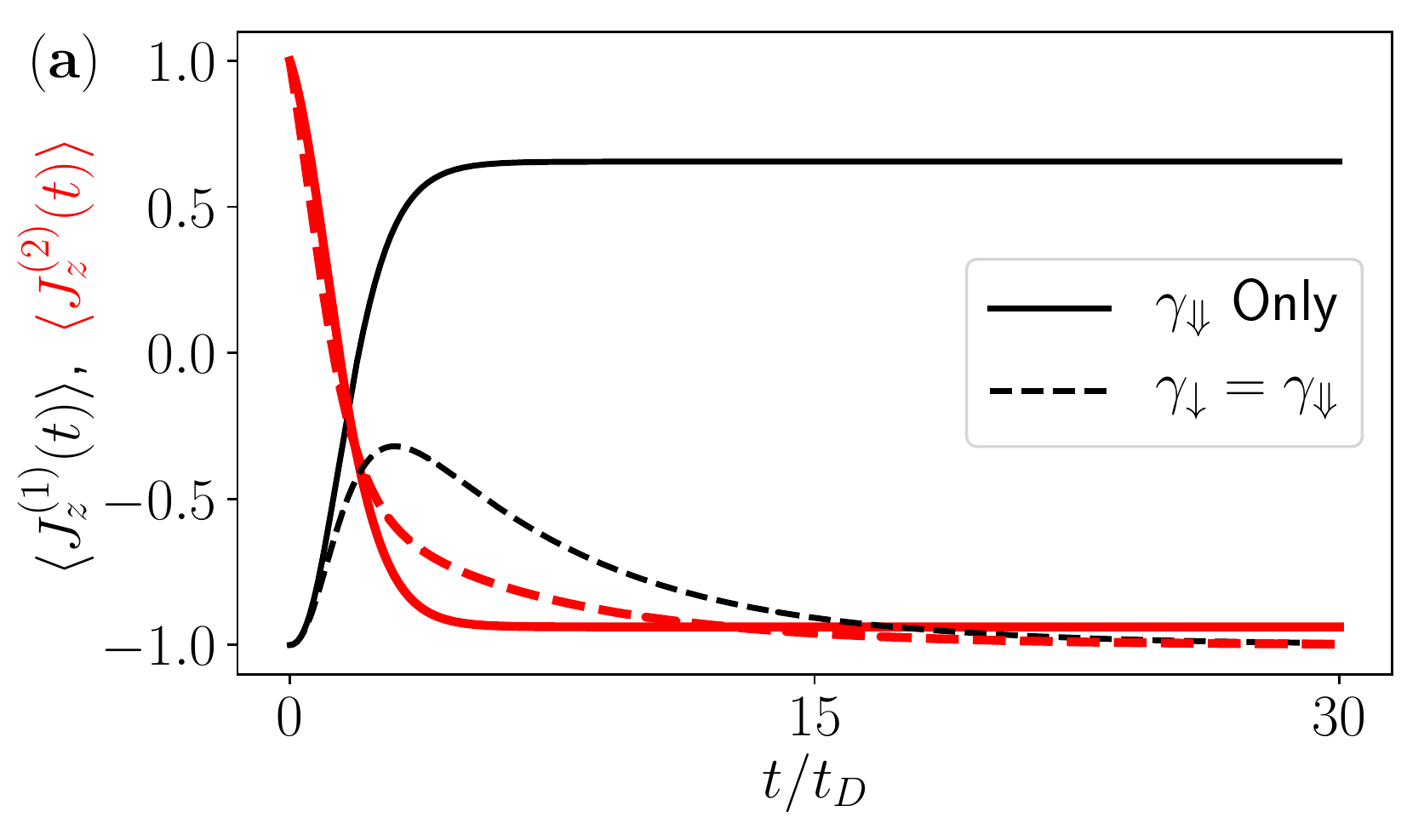}
\includegraphics[width=8cm]{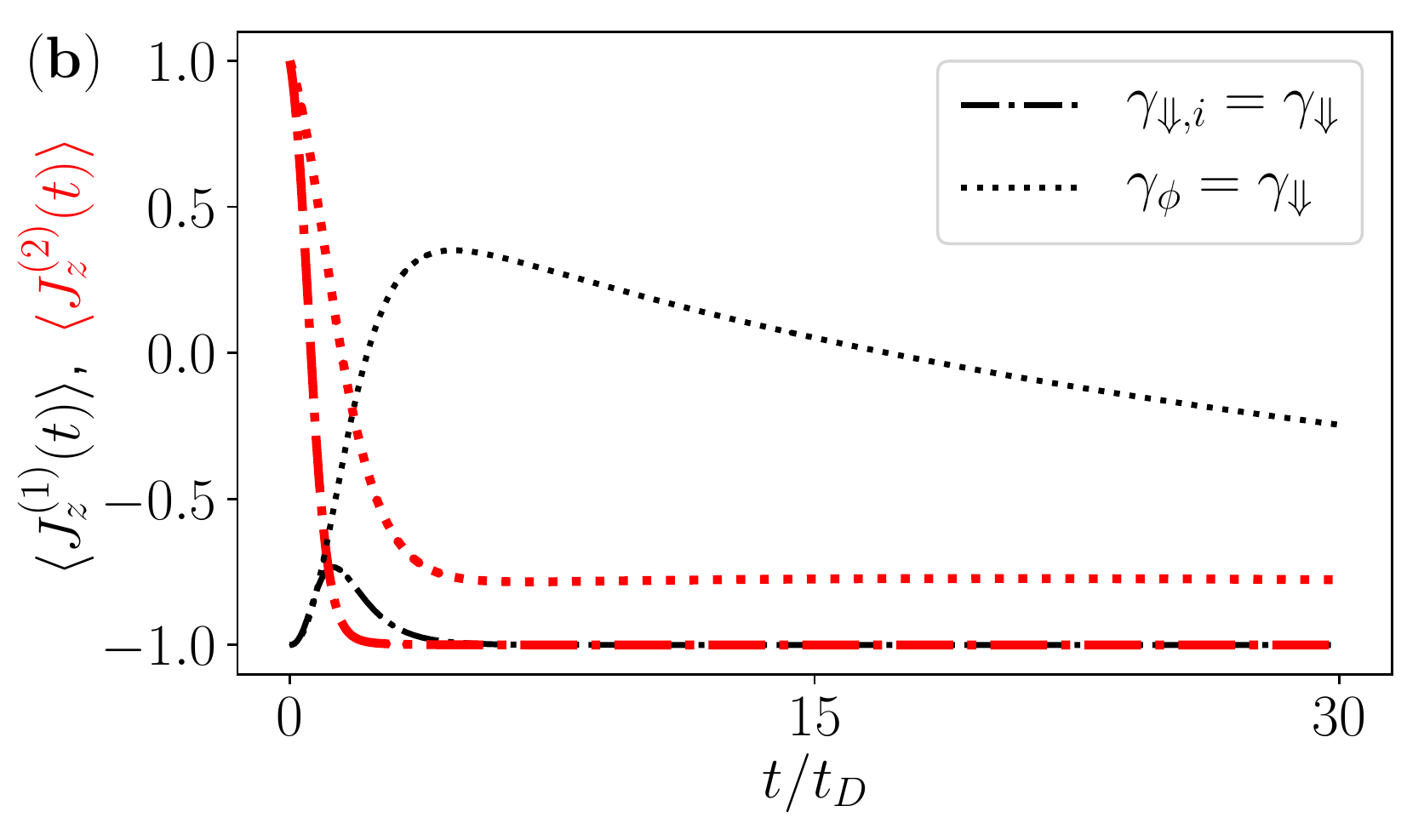}
\caption{\label{fignt} 
{\bf Spin-excitation exchange in multiple TLS ensembles}. 
The robustness of the spin-excitation exchange given by the dynamics of \Eq{ntemp0} is studied for two dissipatively-coupled ensembles of TLSs. 
Black and red curves display the time evolution of the total spin inversion of the first and second ensembles, respectively, normalized by $N_1/2$ and $N_2/2$, respectively. 
Here we set $N_1=5$, $N_2=15$, the system is initially in the state $\ket{\frac{N_1}{2},-\frac{N_1}{2}}\otimes\ket{\frac{N_2}{2},\frac{N_2}{2}}$, and $n_\text{T}\ll1$, thus neglecting collective pumping, $\gamma_\Uparrow$, in \Eq{ntemp0}.
(a) For $\gamma_{\Downarrow}=\omega_0$ (solid curves), the first ensemble, initially unexcited, starts at $\langle J_z^{(1)}(0)\rangle/(N_1/2)=-1$ (black thin solid curve) and ends up in an excited steady state, while the initially excited second ensemble, starting at $\langle J_z^{(2)}(0)\rangle/(N_2/2)=1$ (red thick solid curve) ends up de-excited;  these solid curves qualitatively reproduce the results of Ref.~\cite{Hama16}. 
We investigate the effect of local incoherent emission, $\gamma_{\downarrow}$, adding it to both ensembles, showing that the excitation exchange becomes only transient, as the excitation lost by the second ensemble (red thick dashed curve) is partly acquired by the first ensemble (black thin dashed curve) at short times; eventually also the second ensemble relaxes to the ground state.
(b) We show the effect of collective emission from each of the two ensembles, $\gamma_{\Downarrow,i}=\gamma_{\Downarrow}$ (dot-dashed curves), as well as that of local pure dephasing, for $\gamma_{\phi}=\gamma_{\Downarrow}$ (dotted curves). As in panel (a), the system is initialized in an antisymmetric state, with the first ensemble not excited at $t=0$ (black thin curves) and the second ensemble fully excited (red thick curves). 
The delay time is $t_\text{D}=\log(N_2)/N_2\gamma_{\Downarrow}$. }
\end{center}
\end{figure*}

For $N_1< N_2$, at $t=0$ all spins of the first ensemble are in the ground state, while the spins in the second ensemble are all excited, 
\begin{eqnarray}
\label{negt1}
\ket{\psi}&=&\ket{g\cdots g}_{N_1}\otimes\ket{e\cdots e}_{N_2}=\ket{\frac{N_1}{2},-\frac{N_1}{2}}\otimes\ket{\frac{N_2}{2},\frac{N_2}{2}},\nonumber\\
\end{eqnarray}
which on the right-hand side of \Eq{negt1} has been written as a tensor product of Dicke states, so that the system can be readily studied using PIQS formalism.
In Figure~\ref{fignt} we show that the total spin inversion of each ensemble, $\langle J_z^{(k)}(t)\rangle$ as a function of time for the first ensemble (black curves) and the second ensemble (red curves) for $N_1=5$ and $N_2=15$ for different dynamical conditions.
We investigate the low-temperature limit, setting the thermal occupation number to $n_\text{T}\ll1$.  
The time is normalized in terms of the superradiant dynamics of the second ensemble, using the standard definition of delay time as $t_\text{D}=\log(N_2)/N_2\gamma_\Downarrow$.

In Figure~\ref{fignt}(a), the solid curves in the plot shows the time evolution given by \Eq{ntemp0} for $\gamma_{\Downarrow,1}=\gamma_{\Downarrow,2}=\gamma_{\phi}=\gamma_{\downarrow}=0$ and $\gamma_{\Downarrow}>0$, with an exchange of collective spin excitation among the two ensembles and a negative-temperature effect in the steady-state for the spins of the first ensemble \cite{Hama16}. We note that this effect can be interpreted by resorting to the Dicke space picture, considering the total coupled ensembles as a single one, $N=N_1+N_2$. If only collective emission and pumping are allowed, the master equation can be rewritten simply as a unique system,
\begin{eqnarray}
\label{ntemp2}
\dot{\rho}&=&-i \omega_0 \lbrack J_z,\rho \rbrack
+\frac{\gamma_{\Downarrow}}{2}\mathcal{L}_{J_{-}}[\rho]
+\frac{\gamma_{\Uparrow}}{2}\mathcal{L}_{J_{+}}[\rho].
\end{eqnarray} 
The initial condition of \Eq{negt1} means that the system is prepared in a state with non-maximal cooperative number $j<\frac{N}{2}=\frac{N_1}{2}+\frac{N_2}{2}$. 
At zero temperature, \Eq{ntemp2} then reduces to the standard superradiant master equation. 
The persistence of excitation in the initially unexcited ensemble can be interpreted as the system being effectively confined in the ladder of dark Dicke states, as discussed in Refs.~\cite{Shammah17,Gegg17a}. We now generalize the dynamics: If each of the ensembles is allowed to dissipate incoherently, $\gamma_{\downarrow}$, the steady-state excitation of the first ensemble, initially unexcited, becomes only transient, and it eventually relaxes to the ground state, as shown by the dashed curves of Figure~\ref{fignt}(a).   
 
In Figure~\ref{fignt}(b), we assess the effect of dephasing, $\gamma_{\phi}=\gamma_{\Downarrow}$ (dot-dashed curves), and collective emission from each of the TLS ensembles, $\gamma_{\Downarrow,i}=\gamma_{\Downarrow,i}$, (dotted curves), for $i = 1,2$, which are both shown to deplete the population inversion of the first ensemble for $t\gg t_\text{D}$.
On the one hand, the introduction of these local and collective mechanisms is a detrimental effect for the preservation of the steady-state collective spin excitation of the $N_1$ TLSs in the first ensemble and prevents robust negative-temperature effects. On the other hand, these processes actually open the way to the investigation of excitation exchanges and delayed-light-emission in collections of TLS ensembles, e.g., coupled in series in an array of cavities. 
More complex experimental conditions than that of \Eq{ntemp0} can be simulated with PIQS, e.g., one or multiple ensembles of TLSs interacting with one or multiple bosonic environments, all able to dissipate. This could extend previous investigations to the dissipative regime or consider more complex connectivities \cite{Xu15,Gong16,Hama16,Kirton17,Saez17,Kiilerich17,Vaidya18,Saez18,Hama18b}.
%%%
% Figure 12
%%%
%%%
\subsection{Ultrastrong-coupling regime}
\label{sec:usc}
%%%
%%%
When the strength of the light-matter coupling becomes comparable to the bare excitation frequencies, the interaction becomes non-perturbative. The rich phenomenology which becomes then observable 
\cite{Kockum18b,Ciuti05,DeLiberato07,Ashhab10,Carusotto12,Auer12,Ashhab13,DeLiberato14,Ripoll15,Bamba15,Hwang15,Kockum17,Garziano17,Fedortchenko16,Jaako16,Cirio16,LeBoite16,Hagenmuller16,Armata17,Bamba14,DeLiberato17,Bamba17,Bamba17b,Flick18} has led to a remarkable interest in those non-perturbative regimes, which have 
been experimentally realized in a number of experimental implementations well described, at least in first approximation \cite{DeLiberato13a,Munoz18,DeBernardis18,Stokes18,DiStefano18}, by the Dicke model
\cite{Anappara09,Todorov10,Muravev11,Schwartz11,Geiser12,Scalari12,Benz13,Dietze13,Scalari13,KenaCohen13,Mazzeo14,Askenazi14,Gubbin14,Maissen14,George16,Zhang16,Bayer17}.
A number of works investigated the impact of losses in this regime, demonstrating in particular how the standard Lindblad form of the master equation can fail, leading to unphysical processes as emission of light from vacuum \cite{DeLiberato09,Beaudoin11,Cao11,Ridolfo12,Bamba13,Bamba14,DeLiberato14a,DeLiberato17}.

A commonly used approach to solve those problems is to write the master equation in the basis of the dressed states following, e.g., Ref.~\cite{Beaudoin11}. This allows us to avoid unphysical processes as the (ultrastrongly) coupled energies are used instead of the bare ones.
Here we show that PIQS, and more generally the permutational-invariant approach, is flexible enough to allow one to consider the correct ultrastrong-coupling (USC) master equation. 
For simplicity we neglect the Lamb shift in $H$ and assume that the TLSs interact with a white-noise bath and set all $\gamma_{i}=0$ besides $\gamma_\downarrow$. We thus consider the master equation
\begin{eqnarray}
\dot{\rho}&=&-\frac{i}{\hbar}\lbrack H,\rho\rbrack+\sum_{r,s>r}\left(\frac{\kappa}{2}|X^{r,s}|^{2} +\frac{\gamma_{\downarrow}}{2}\sum_{n=1}^N|J_{x,n}^{r,s}|^{2}\right)\mathcal{L}[\ket{r}\bra{s}](\rho),\nonumber\\
\label{eqlosss}
\end{eqnarray}
where $\ket{r},\ket{s}$ are the dressed light-matter eigenstates of $H$ in \Eq{hamdicke}, and the condition $s>r$ ensures that the jumps are from states of higher to lower energy only.
%% Figure 13
\begin{figure*}[ht!]
\begin{center}
\includegraphics[width=16cm]{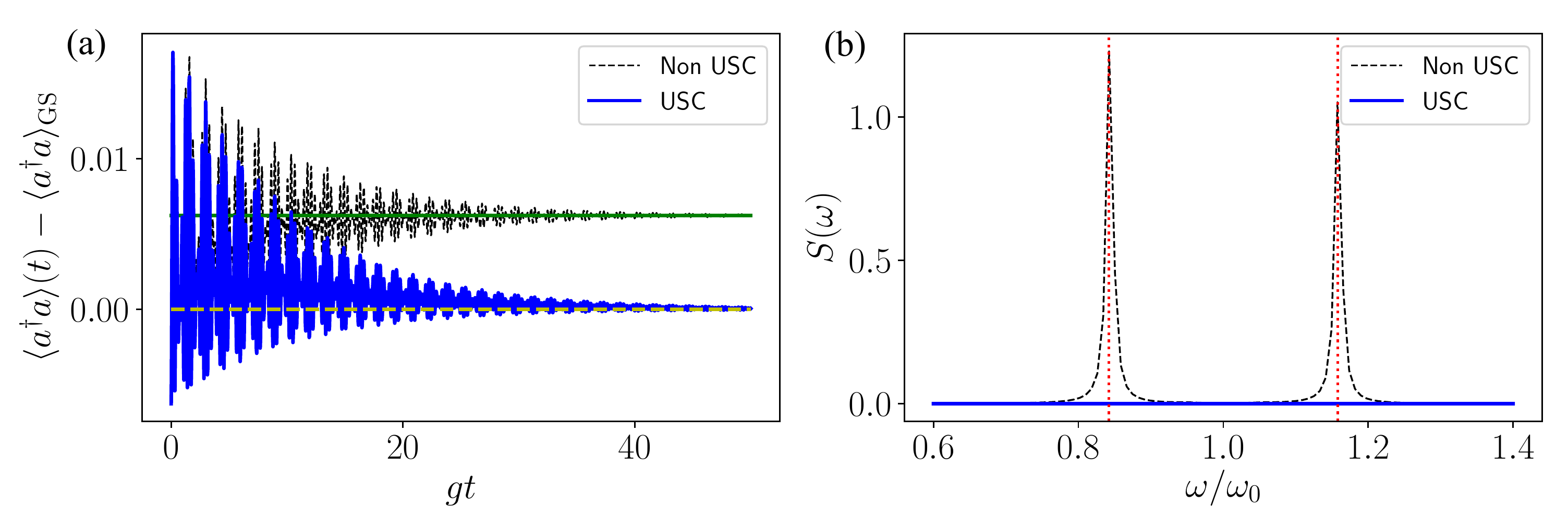}
\caption{\label{figusc}{\bf Ultrastrong-coupling regime}.  
We compare the solution of the master equation valid below the ultrastrong-coupling (USC) regime and for effective models (black dashed curve) with the correct USC master equation (blue solid curve) for $N=10$ TLSs in a lossy cavity. 
The parameters are $\omega_\text{cav}=\omega_0$, $g=0.1\omega_0$, $\gamma_\downarrow=\kappa=0.01\omega_0$, and the photonic Hilbert subspace has a cut-off of $n_\text{ph}=3$.
(a) The intra-cavity photon population, $\langle a^\dagger a \rangle (t)$, is calculated with two different Liouvillians, relative to the USC regime (solid blue curve) and to the non-USC regime (dashed black curve).
We subtract to these quantities the photon population in the steady state of the USC Liouvillian $\langle a^\dagger a \rangle_\text{GS}$. The initial state is the tensor state $\ket{\frac{N}{2},-\frac{N}{2}}\otimes\ket{0_\text{ph}}$. 
The different dynamics prompt different time evolutions tending to the USC steady-state (dashed yellow line) and non-USC steady state (solid green line). 
(b) The extra-cavity photon emission spectrum $S(\omega)$ is shown, according to the two calculations, here initializing the system in the ground state of the light-matter Hamiltonian. 
The dotted vertical red lines correspond to the polariton frequencies $\omega=\omega_0\pm\sqrt{N}\frac{g}{2}$.}
\end{center}
\end{figure*}
We have projected the spin operators onto the dressed basis, 
\begin{eqnarray}
\label{sms}
J_{\alpha,n}&=&\sum_{r,s}J_{\alpha,n}^{r,s}\ket{r}\bra{s},
\end{eqnarray}
with $\alpha=\{x,y,z,+,-\}$ and similarly $X^{r,s}=\bra{r}(a+a^\dagger)\ket{s}$. The problem of solving \Eq{eqlosss} for $N\gg1$ is that in order to explicitly write the brakets contained in \Eq{sms} for each of the $N$ TLSs, one seems to be forced to place them into the $2^N$ Hilbert space. 
Nevertheless, if we exploit the fact that with PIQS we can build the permutational-invariant Lindbladian superoperator for the undressed interactions, $\mathcal{L}_{J_{-,i}}$, in the Dicke basis $\ket{j,m}\bra{j,m'}$, we can explicitly write the USC master equation in the Liouvillian space given by the Dicke basis.  
By projecting this object onto the dressed USC basis, without requiring an explicit expression for the local operators, from \Eq{sms} we obtain
\begin{eqnarray}
\sum_{n=1}^N|J_{-,n}^{r,s}|^{2}_{s> r}&=&\frac{1}{2}\bra{s}\sum_{n=1}^N\mathcal{L}_{J_{-,n}}(\ket{r}\bra{r})\ket{s}_{s> r},
\label{eq7nm2a2}
\end{eqnarray}
and similarly for the other local spin operators to obtain the overlaps present in \Eq{eqlosss}. 
This approach can be generalized to the dynamics induced by the other local processes, dephasing and pumping.
In Figure~\ref{figusc}(a), we plot the time evolution of the excess intracavity photon population, with respect to the steady-state value, for a many-body system comprising $N=10$ TLSs, initialized in a tensor state with no cavity photons and the spins in the Dicke state $\ket{\frac{N}{2},-\frac{N}{2}}$. 
The Hamiltonian light-matter coupling is given by \Eq{hamdicke}, with $\omega_0=\omega_\text{cav}$ and $g=0.1\omega_0$.
We assume that no pure dephasing is present and we consider homogeneous local spin dissipation and photon loss, $\gamma_\downarrow=\kappa=0.01\omega_0$, with the bath temperatures zero, so that no pumping is included and the steady state is effectively the dressed ground state of the light-matter Hamiltonian of \Eq{hamdicke}. 

The solution of \Eq{eqlosss} is shown by the solid blue curve, with the steady-state photon population marked by the dashed yellow line.   
The solution of the master equation valid in the non-USC regime (dashed black curve) leads to the wrong steady state (solid green line), which overestimates the steady-state intra-cavity photon population. 

It is known that one of the problems of not employing the correct USC Liouvillian, as in \Eq{eqlosss}, is a unphysical estimate of the extra-cavity photon emission rate, which gives photon generation even when only dissipative processes are present \cite{DeLiberato09c,Beaudoin11}. 
In Figure~\ref{figusc}(b), we compare the steady-state photon emission spectrum, $S(\omega)$, which for positive frequencies and white reservoirs can be calculated from 
\begin{eqnarray}
\label{spectrumdw}
S(\omega>0) &\propto& \int_{-\infty}^{\infty}\langle a^\dagger(\tau) a(0) \rangle e^{-i\omega\tau}d\tau,
\end{eqnarray}
using the quantum regression theorem, implemented in QuTiP's \code{spectrum()} function.
The time evolution of the photon operators in \Eq{spectrumdw} is calculated according to the two master equations. 
In the first case, we use the non-USC master equation of \Eq{master} (dashed curve in Figure~\ref{figusc}(b)), which leads to the prediction of an unphysical photon spectrum with asymmetric-intensity peaks at the polariton frequencies $\omega=\omega_0\pm\sqrt{N}\frac{g}{2}$. The correct spectrum obtained using \Eq{eqlosss} shows no photon emission (solid curve in Figure~\ref{figusc}(b)). 

\section{Conclusions}
\label{conc}
We have provided a computational library for the investigation of the open quantum dynamics of many TLSs that leverages permutational invariance, PIQS \cite{Piqs}. 
We have shown how the Dicke states and the Dicke space are powerful tools to study the interplay of local and collective processes, giving a unifying analytical framework to visualize and quantify their effect, extending its application to the presence of collective pumping and collective dephasing. By coherently organizing an overview of existing works that use permutational-invariant methods, we could systematically highlight the rich physics that has been investigated in this setting. 
We have demonstrated how PIQS can be used to investigate a range of physical phenomena in the context of driven-dissipative open quantum systems, and we have shown how they are influenced by local dissipation.

We have provided original results in all subsections of Section~\ref{results}. Since we can resolve the time evolution of the collective density matrix, we could study how the same dynamics leads to very different time evolutions, depending on the initial preparation of the system. 
We began by studying the dynamics governed by superradiant decay in the bad-cavity regime and with local dephasing. 
We could pinpoint the different evolutions of the maximally-symmetric Dicke state $\ket{\frac{N}{2},0}$ and the GHZ state, which cannot be captured by second-order approximate methods relying on the factorization of collective spin moments. On the other hand, we verified that there is no significant difference between the evolution of the entangled $\ket{\frac{N}{2},0}$ state and the symmetric and antisymmetric CSS, separable states that are easier to initialize in experiments. In the same setting, we have also pointed out that local dephasing can be beneficial to light emission for a state initialized in a dark state, $\ket{0,0}$, as long as collective emission and dephasing mechanisms are faster than local emission. 

Turning to the case of steady-state superradiance in the bad-cavity limit and under incoherent local pumping and local losses, we have found that a system at detailed balance does not display a threshold pump with regard to the nonlinear enhancement of emitted light. 
We have then investigated spin squeezing in the two-axis twisting model in the presence of dissipative local or collective spin flips. We have used PIQS for state engineering, finding that the initial state with longest spin squeezing time does not belong to the Dicke symmetric ladder.

In the context of non-equilibrium phase transitions, we have studied the open Dicke model in the presence of local dephasing and local pumping. By assessing the effect of collective dissipation and incoherent driving to a system experiencing local dephasing, we found that, for a fixed resonance frequency, collective pumping is more detrimental than collective emission to the occurrence of the superradiant phase. We then studied time crystallization in a driven-dissipative open quantum system out of equilibrium, verifying that the related collective spin oscillations are affected by the presence of local dephasing. The reduction in their visibility is marked by a decrease of the system's cooperation number.

Generalizing our study to the dissipative dynamics of multiple ensembles of TLSs, we have investigated how the exchange of collective spin excitations among two ensembles of TLSs is affected by local dephasing, local losses or collective losses in each of the ensembles. On the one hand, we have found that such processes are detrimental for the observation of negative-temperature effects in the steady state. On the other hand, we proposed to use the transient dynamics arising under such conditions to exchange excitations in arrays of coupled ensemble of TLSs. 

Finally, we have shown that the permutational invariance of the TLS Lindblad superoperators can be used to analytically derive the correct Lindblad master equation for the USC regime in terms of dressed light-matter superoperators, thus unlocking the study of local dissipative processes in the USC regime for $N\gg 1$. We have thus used PIQS to investigate time-dependent and steady-state properties of the open Dicke model in the USC regime in the presence of cavity and local TLS losses. We calculated the relaxation of the system to the correct steady state, and showed that this model correctly predicts no photon emission from the steady state.

%%%%
There are multiple opportunities for future research directions involving the interplay of macroscopic cooperative effects and noise both for fundamental aspects and for applications to quantum technology \cite{Lambert16a,Campaioli17,Cabot17,Ferraro18,Quijandria18,Venkatesh18,Sinha18,Roulet18,Niedenzu18}. 
Effective spin models relevant to photon-mediated long-range interactions \cite{Tsomokos08,Lee11,Chaves13,Wilson16,FossFeig17,Russomanno17,Bermudez17b,Safavi17,Owen18} can be studied, especially as PIQS allows us to explore the range of qubit systems engineered in current and near-term quantum simulators \cite{Martin13,Bohnet16,Kakuyanagi16,Garttner17,Bernien17,Zhang17b}.
With regard to the permutational-invariance numerical tool employed here, it could be further extended to include processes of other and more general Lindblad superoperators, e.g., terms $\mathcal{L}_{J_x}$, $\mathcal{L}_{J_y}$, $\mathcal{L}_{J_{x,n}}$, and $\mathcal{L}_{J_{y,n}}$ \cite{Xu13,Gegg16}. Another interesting open question is the extension of permutational invariant approaches to non-Markovian baths \cite{Iles16,Zeb17}, Floquet driving \cite{Gambetta18}, stochastic processes and continuous-measurement protocols \cite{Baragiola09b,Albarelli17,Albarelli18}, and out-of-time-ordered-correlators \cite{Seshadri18}.   

Finally, the USC regime seems a promising field in which to investigate the effect of local dissipation for ensembles with $N\gg1$, as using PIQS one can retain the full nonlinearity of the TLSs beyond the usually explored dilute-excitation regime \cite{Garziano17,Garbe17}.

%%%
\section*{Acknowledgements}
%%%
We thank Peter Kirton, Jonathan Keeling, Michael Gegg, Anton Frisk Kockum, Marcello Dalmonte, Yu-Ran Zhang, Jiabao Chen, Michael Foss-Feig, and Peter Groszkowski for useful discussions and comments. 
N.L. and F.N. acknowledge support from the RIKEN-AIST Challenge Research Fund, and the John Templeton Foundation.
S.D.L. acknowledges support from a Royal Society research fellowship.
N.L. acknowledges partial support from Japan Science and Technology Agency (JST) (JST PRESTO Grant No. JPMJPR18GC).
F.N. is partly supported by the MURI Center for Dynamic Magneto-Optics via the Air Force Office of Scientific Research (AFOSR) (FA9550-14-1-0040), Army Research Office (ARO) (Grant No. W911NF-18-1-0358), Asian Office of Aerospace Research and Development (AOARD) (Grant No. FA2386-18-1-4045), Japan Science and Technology Agency (JST) (the Q-LEAP program, the ImPACT program and CREST Grant No. JPMJCR1676), Japan Society for the Promotion of Science (JSPS) (JSPS-RFBR Grant No. 17-52-50023, JSPS-FWO Grant No. VS.059.18N).
\bibliography{references11}
\onecolumngrid
\appendix
%%%
%%%
\section{Coefficients of the permutational-invariant dynamics}
\label{app1}
%%%
In \Eq{master}, which we rewrite here,
\begin{eqnarray}
\label{masterapp}
\dot{\rho} &=& 
- \frac{i}{\hbar}\lbrack H,\rho \rbrack 
 +\frac{\gamma_{\Downarrow}}{2}\mathcal{L}_{J_{-}}[\rho] +\frac{\gamma_{\Phi}}{2}\mathcal{L}_{J_{z}}[\rho]  +\frac{\gamma_{\Uparrow}}{2}\mathcal{L}_{J_{+}}[\rho] 
 +\sum_{n=1}^{N}\left(\frac{\gamma_{\downarrow}}{2}\mathcal{L}_{J_{-,n}}[\rho] +\frac{\gamma_{\phi}}{2}\mathcal{L}_{J_{z,n}}[\rho]
+\frac{\gamma_{\uparrow}}{2}\mathcal{L}_{J_{+,n}}[\rho]\right), \nonumber\\
\end{eqnarray}
the problematic terms with regard to the exponential increase of the Liouvillian space size are the jump terms relative to the local Lindbladians. We can rewrite them explicitly using the relations of the SU(2) algebra 
\begin{subequations}
\label{eq3}
\begin{eqnarray}
\frac{\gamma_{\downarrow}}{2}\sum_{n=1}^{N}\mathcal{L}_{J_{-,n}}[\rho]&=&\frac{\gamma_{\downarrow}}{2}\lbrack2\left(\sum_{n=1}^{N}J_{-,n}\rho J_{+,n}\right)-J_z\rho-\rho J_z-N\rho\rbrack,\\
\frac{\gamma_{\phi}}{2}\sum_{n=1}^{N}\mathcal{L}_{J_{z,n}}[\rho]&=&\frac{\gamma_{\phi}}{2}\lbrack2\left(\sum_{n=1}^{N}J_{z,n}\rho J_{z,n}\right)-\frac{N}{2}\rho\rbrack,\\
\frac{\gamma_{\uparrow}}{2}\sum_{n=1}^{N}\mathcal{L}_{J_{+,n}}[\rho]&=&\frac{\gamma_{\uparrow}}{2}\lbrack2\left(\sum_{n=1}^{N}J_{+,n}\rho J_{-,n}\right)+J_z\rho+\rho J_z-N\rho\rbrack.
\end{eqnarray}
\end{subequations}
In Refs.~\cite{Chase08,Baragiola10}, it is shown that the first term can be rewritten in terms of the Dicke states
\begin{subequations}
\begin{eqnarray}
\label{eq4a}
\sum_{n=1}^{N}J_{r,n} \ket{j,m}\bra{j,m'} J_{q,n}^\dagger&=&
a_{qr}^N(j,m,m')\ket{j,m+\tilde{q}}\bra{j,m'+\tilde{r}}\\
\label{eqb}
&&+b_{qr}^N(j,m,m')\ket{j-1,m+\tilde{q}}\bra{j-1,m'+\tilde{r}}\\
&&+c_{qr}^N(j,m,m')\ket{j+1,m+\tilde{q}}\bra{j+1,m'+\tilde{r}}
\label{eq4c}
\end{eqnarray}
\label{eq4}
\end{subequations}
where $q,r=\{+,-,z\}$ so that 
$J_{-,n}=\sigma_{-,n}$, $J_{+,n}=\sigma_{+,n}$, $J_{z,n}=\frac{1}{2}\sigma_{z,n}$ and $\tilde{q},\tilde{r}=\{+1,-1,0\}$, 
respectively, and
\begin{subequations}
\label{eq5}
\begin{eqnarray}
a_{qr}^N(j,m,m')&=&A^{j,m}_{q}A^{j,m'}_{r}\frac{x_{N,j,a}}{2}=A^{j,m}_{q}A^{j,m'}_{r}\frac{1}{2j}\left(1+\frac{\alpha_N^{j+1}(2j+1)}{d_N^{j}(j+1)}\right),\\
b_{qr}^N(j,m,m')&=&B^{j,m}_{q}B^{j,m'}_{r}\frac{x_{N,j,b}}{2}=B^{j,m}_{q}B^{j,m'}_{r}\frac{\alpha_N^{j}}{2jd^j_N},\\
c_{qr}^N(j,m,m')&=&D^{j,m}_{q}D^{j,m'}_{r}\frac{x_{N,j,c}}{2}=D^{j,m}_{q}D^{j,m'}_{r}\frac{\alpha_N^{j+1}}{2(j+1)d^j_N},
\end{eqnarray}
\end{subequations}
where
\begin{eqnarray}
\label{eq6}
&A^{j,m}_{\pm}=\sqrt{(j\mp m)(j\pm m +1)},\ \ \ \ \ \ &A^{j,m}_{z}=m,\\
&B^{j,m}_{\pm}=\pm\sqrt{(j\mp m)(j\mp m -1)},\ \ \ \ \ \  &B^{j,m}_{z}=\sqrt{(j + m)(j - m )},\\
&D^{j,m}_{\pm}=\mp\sqrt{(j\pm m+1)(j\pm m +2)}, \ \ \ \ \ \  &D^{j,m}_{z}=\sqrt{(j + m+1)(j - m+1 )},
\end{eqnarray}
and
\begin{eqnarray}
\label{eq8}
\alpha_N^{j}&=&\sum_{j'=j}^{N/2}d_N^{j'}=\frac{N!}{(N/2-j)!(N/2+j)!},
\end{eqnarray}
with $d_N^j$ given by \Eq{eq8d}. By definition, $\alpha^{j+1}_{N}=0$ for $j=\frac{N}{2}$. 
Here $\alpha_N^{j}$ is not the symmetric quantum number of the Dicke state $\ket{j,m,\alpha_j}$, but we keep the notation to be consistent with previous works. 
We have also introduced the coefficients $x_{N,j,a}$, $x_{N,j,b}$, and $x_{N,j,c}$, defined by the right-hand side of \Eq{eq5} as they will be convenient to write the rates in a more compact form. 
In \Eq{eq4} there is mixing only between $j-$blocks of the density matrix with $\Delta j=\pm 1$, while within each block $j$, the change in $m$ is $\Delta m=\pm 1, 0$.  
Using the identity for density matrix, \Eq{eq1b}, we can rewrite the master equation of \Eq{master} in terms of the Dicke states $\ket{j,m}$
\begin{eqnarray}
\frac{d}{dt}p_{jmm'}(t)\ket{j,m}\bra{j,m'}&=&
p_{jmm'}\Bigg\{
\frac{\gamma_{\Downarrow}}{2}\left(2A^{j,m}_{-}A^{j,m'}_{-}\ket{j,m-1}\bra{j,m'-1}\right) -\frac{\gamma_{\Downarrow}}{2}\lbrack\left((A^{j,m}_{-})^{2}+(A^{j,m'}_{-})^{2}\right)\ket{j,m}\bra{j,m'}\rbrack\nonumber\\
&&+\frac{\gamma_{\Phi}}{2}\lbrack2mm' -\left(m^{2}+m'^{2}\right)\rbrack\ket{j,m}\bra{j,m'}\nonumber\\
&&+\frac{\gamma_{\Uparrow}}{2}\{2A^{j,m}_{+}A^{j,m'}_{+}\ket{j,m+1}\bra{j,m'+1} -\lbrack(A^{j,m}_{+})^{2}+(A^{j,m'}_{+})^{2}\rbrack\ket{j,m}\bra{j,m'}\}\nonumber\\
&&+\frac{\gamma_{\downarrow}}{2}\lbrack2\left(\sum_{n=1}^{N} J_{-,n}\ket{j,m}\bra{j,m'} J_{+,n}\right)-(N+m+m')\ket{j,m}\bra{j,m'}\rbrack\nonumber\\
&&+\frac{\gamma_{\phi}}{2}\lbrack2\left(\sum_{n=1}^{N}J_{z,n}\ket{j,m}\bra{j,m'}J_{z,n}\right)-\frac{N}{2}\ket{j,m}\bra{j,m'}\rbrack\nonumber\\
&&+\frac{\gamma_{\uparrow}}{2}\lbrack2\left(\sum_{n=1}^{N} J_{+,n}\ket{j,m}\bra{j,m'}J_{-,n}\right)-(N-m-m')\ket{j,m}\bra{j,m'}\rbrack
\Bigg\}.
\label{eq10}
\end{eqnarray} 

We have from \Eq{eq4} that dephasing gives
\begin{eqnarray}
\label{eq11a}
\sum_{n=1}^{N}J_{z,n} \ket{j,m}\bra{j,m'} J_{z,n}&=&
mm'
\frac{x_{N,j,a}}{2}
\ket{j,m}\bra{j,m'}
+B^{j,m}_{z}B^{j,m'}_{z}
\frac{x_{N,j,b}}{2}
\ket{j-1,m}\bra{j-1,m'}\nonumber\\&&+D^{j,m}_{z}D^{j,m'}_{z}
\frac{x_{N,j,c}}{2}
\ket{j+1,m}\bra{j+1,m'}.
\label{eq11}
\end{eqnarray}
Note that in the relative definition of Refs.~\cite{Chase08,Baragiola10} there is a factor $\frac{1}{2}$ missing for local dephasing and that the result here is correct.  
The jumps from losses are
\begin{eqnarray}
\label{eq12a}
\sum_{n=1}^{N} J_{-,n} \ket{j,m}\bra{j,m'} J_{+,n}&=&
A^{j,m}_{-}A^{j,m'}_{-}
\frac{x_{N,j,a}}{2}
\ket{j,m-1}\bra{j,m'-1}
+B^{j,m}_{-}B^{j,m'}_{-}\nonumber
\frac{x_{N,j,b}}{2}
\ket{j-1,m-1}\bra{j-1,m'-1}
\\ \label{eq11b}&&
+D^{j,m}_{-}D^{j,m'}_{-}
\frac{x_{N,j,c}}{2}
\ket{j+1,m-1}\bra{j+1,m'-1},
\label{eq12}
\end{eqnarray}
and the jumps from the pump
\begin{eqnarray}
\label{eq13a}
\sum_{n=1}^{N} J_{+,n} \ket{j,m}\bra{j,m'} J_{-,n}&=&
A^{j,m}_{+}A^{j,m'}_{+}
\frac{x_{N,j,a}}{2}
\ket{j,m+1}\bra{j,m'+1}+B^{j,m}_{+}B^{j,m'}_{+}
\frac{x_{N,j,b}}{2}
\ket{j-1,m+1}\bra{j-1,m'+1}\nonumber\\&&
+D^{j,m}_{+}D^{j,m'}_{+}
\frac{x_{N,j,c}}{2}
\ket{j+1,m+1}\bra{j+1,m'+1}.
\label{eq13}
\end{eqnarray}
It is possible to perform an analytical check on the probability density current flow in the dynamics equations of \Eq{eq10}. 
If we consider \Eq{master}, since $\text{Tr}[\rho]=1$ we have that $\sum_{j=j_\text{min}}^{N/2}\sum_{m=-j}^{j}\frac{d}{dt}p_{jmm}(t)=0$. 
Since this equation is valid at any time $t$ and in principle at a time $t=0$ it is possible to initialize the system in a specific Dicke state $\ket{j,m}$, then for each population probability the sum of the relative rates must be $0=-\Gamma^{(1)}_{j,m,m}+\sum_{i>1}\Gamma^{(i)}_{j,m,m}$. 

We can express the functions as $\Gamma^{(i)}_{j,m,m'}$ explictly as
\begin{subequations}
\begin{eqnarray}
\label{eq15bX}
\Gamma^{(1)}_{j,m,m'}&=&
\frac{\gamma_{\Downarrow}}{2}\lbrack(A^{j,m}_{-})^2+(A^{j,m'}_{-})^2\rbrack
+\frac{\gamma_{\Uparrow}}{2}\left((A^{j,m}_{+})^2+(A^{j,m'}_{+})^2\right)
+\frac{\gamma_{\Phi}}{2}\left(m-m'\right)^2\nonumber\\&&
+\frac{\gamma_{\downarrow}}{2}\left(N+m+m'\right)
+\frac{\gamma_{\uparrow}}{2}\left(N-m-m'\right)
+\frac{\gamma_{\phi}}{2}\left( \frac{N}{2}
-mm'\frac{\left(\frac{N}{2} + 1\right)
}{j(j+1)}\right), 
\\
\Gamma^{(2)}_{j,m,m'}&=&\gamma_{\Downarrow}A^{j,m}_{-}A^{j,m'}_{-}+\frac{\gamma_{\downarrow}}{2}A^{j,m}_{-}A^{j,m'}_{-}\frac{\left(\frac{N}{2} + 1\right)}{j (j+1)},
\\
\Gamma^{(3)}_{j,m,m'}&=&\frac{\gamma_{\downarrow}}{2}B^{j,m}_{-}B^{j,m'}_{-}\frac{\left(\frac{N}{2} + j + 1\right)}{j(2j+1)},
\\
\Gamma^{(4)}_{j,m,m'}&=&\frac{\gamma_{\downarrow}}{2}D^{j,m}_{-}D^{j,m'}_{-}\frac{\left(\frac{N}{2} - j \right)}{(j+1)(2j+1)},\\
\Gamma^{(5)}_{j,m,m'}&=&\frac{\gamma_{\phi}}{2}B^{j,m}_{z}B^{j,m'}_{z}\frac{\left(\frac{N}{2} + j + 1\right)}{j(2j+1)},\\
\Gamma^{(6)}_{j,m,m'}&=&\frac{\gamma_{\phi}}{2}D^{j,m}_{z}D^{j,m'}_{z}\frac{\left(\frac{N}{2} - j \right)}{(j+1)(2j+1)},
\\
\Gamma^{(7)}_{j,m,m'}&=&\frac{\gamma_{\uparrow}}{2}B^{j,m}_{+}B^{j,m'}_{+}\frac{\left(\frac{N}{2} + j + 1\right)}{j(2j+1)},\\
\Gamma^{(8)}_{j,m,m'}&=&\gamma_{\Uparrow}A^{j,m}_{+}A^{j,m'}_{+}
+\frac{\gamma_{\uparrow}}{2}A^{j,m}_{+}A^{j,m'}_{+}\frac{\left(\frac{N}{2} + 1\right)}{j (j+1)},
\\
\Gamma^{(9)}_{j,m,m'}&=&\frac{\gamma_{\uparrow}}{2}D^{j,m}_{+}D^{j,m'}_{+}\frac{\left(\frac{N}{2} - j \right)}{(j+1)(2j+1)}.
\end{eqnarray}
\label{eq15bbX}
\end{subequations}
The terms of the coefficients proportional to $\gamma_{\Downarrow}$, $\gamma_{\Phi}$ and $\gamma_{\Uparrow}$, differently from the local phase-breaking mechanisms in $\gamma_{\downarrow}$, $\gamma_{\phi}$ and $\gamma_{\uparrow}$, do not depend explicitly on $N$. The $\code{lindbladian}$ function at the core of the PIQS library 
uses \Eq{eq15bbX} to build the matrix corresponding to the total Liouvillian superoperator.

Hereafter we consider the special case of a diagonal problem. For the terms on the main diagonal for which $m'=m$, we can simplify the notation $\Gamma^{(i)}_{j,m,m}=\Gamma^{(i)}_{j,m}$ and write in explicit form the compact expressions
\begin{subequations}
\begin{eqnarray}
\Gamma^{(1)}_{j,m}&=&\gamma_{\Downarrow}(1 + j - m) (j + m)+\gamma_{\Uparrow}(1 + j + m) (j - m)+\gamma_{\downarrow}\left(\frac{N}{2} + m\right)+\gamma_{\uparrow}\left(\frac{N}{2}-m\right)+\gamma_{\phi}\left(\frac{N}{4}- m^2\frac{1+{N}/{2}}{2j(j+1)}\right),
\nonumber\\
\\
\Gamma^{(2)}_{j,m}&=&\gamma_{\Downarrow}(1 + j - m) (j + m)+\gamma_{\downarrow}\frac{(N+2) (j-m+1) (j+m)}{4 j (j+1)},\\
\Gamma^{(3)}_{j,m}&=&\gamma_{\downarrow}\frac{ (j+m-1) (j+m) (j+1+N/2)}{2 j (2 j+1)},\\
\Gamma^{(4)}_{j,m}&=&\gamma_{\downarrow} \frac{(j-m+1) (j-m+2) (N/2- j)}{2 (j+1) (2 j+1)},\\
\Gamma^{(5)}_{j,m}&=&\gamma_{\phi}\frac{(j-m) (j+m) (j+1+N/2)}{2 j (2 j+1)},\\
\Gamma^{(6)}_{j,m}&=&\gamma_{\phi}\frac{ (j-m+1) (j+m+1) (N/2- j)}{2 (j+1) (2 j+1)},\\
\Gamma^{(7)}_{j,m}&=&\gamma_{\uparrow}\frac{(j-m-1) (j-m) (j+1+N/2)}{2 j (2 j+1)},\\
\Gamma^{(8)}_{j,m}&=&\gamma_{\Uparrow}(1 + j + m) (j - m)+\gamma_{\uparrow}\frac{(1+N/2) (j-m) (j+m+1)}{2 j (j+1)},\\
\Gamma^{(9)}_{j,m}&=&\gamma_{\uparrow}\frac{(j+m+1) (j+m+2) (N/2- j)}{2 (j+1) (2 j+1)}.
\end{eqnarray}
\label{eq15xx}
\end{subequations}
\Eq{eq15xx} fully determines the Lindbladian part of the dynamics of \Eq{master} for problems for which $(i)$ the Hamiltonian is diagonal in the $\ket{j,m}\bra{j,m'}$ basis and $(ii)$ the system is initialized in a state that is diagonal in this basis. 
In this special case, we can write the master equation \Eq{master} simply as a rate equation in matrix form, see \Eq{mp}, 
\begin{eqnarray}
\label{MN2}
\frac{d}{dt}p&=&M p,
\end{eqnarray}
with $p_{j,m}\equiv p_{jmm}$ that can be ordered as
\begin{eqnarray}
\frac{d}{dt}\left(\begin{tabular}{c} $p_{N/2,N/2}$\\$p_{N/2,N/2-1}$ \\ $\dots$ \\ $p_{N/2,-N/2}$\\ $p_{N/2-1,N/2-1}$\\ $\dots$\\ $p_{j_\text{min},-j_\text{min}}$ \end{tabular}\right)&=&\text{M} \left(\begin{tabular}{c} $p_{N/2,N/2}$\\$p_{N/2,N/2-1}$ \\ $\dots$ \\ $p_{N/2,-N/2}$\\ $p_{N/2-1,N/2-1}$\\ $\dots$\\ $p_{j_\text{min},-j_\text{min}}$ \end{tabular}\right),
\end{eqnarray}
$p$ is the $n_\text{DS}$-dimensional vector of the diagonal matrix elements of the general density matrix in the Dicke basis. 
$M$ is the $n_\text{DS}\cdot n_\text{DS}$-dimensional matrix of coefficients. 
Since, for each problem, the number of two-level systems can be fixed, we just write $\Gamma_{N,j,m,m}^{(i)}$ as $\Gamma_{j,m}^{(i)}$ for simplicity.
Let us now write the matrix $M$ explicitly for $N=4$, setting only $\gamma_\downarrow$ and $\gamma_\Downarrow$ different from zero, 
\begin{eqnarray}
\label{MN4}
M=\left(\begin{tabular}{lllllllll}
$-\Gamma_{2,2}^{(1)}$&0&0&0&0&0&0&0&0\\
$\Gamma_{2,2}^{(2)}$&$-\Gamma_{2,1}^{(1)}$&0&0&0&$\Gamma_{1,1}^{(6)}$&0&0&0\\
0&$\Gamma_{2,1}^{(2)}$&$-\Gamma_{2,0}^{(1)}$&0&0&$\Gamma_{1,1}^{(4)}$&$\Gamma_{1,0}^{(6)}$&0&0\\
0&0&$\Gamma_{2,0}^{(2)}$&$-\Gamma_{2,-1}^{(1)}$&0&0&$\Gamma_{1,0}^{(4)}$&$\Gamma_{1,-1}^{(6)}$&0\\
0&0&0&$\Gamma_{2,-1}^{(2)}$&$-\Gamma_{2,-2}^{(1)}$&0&0&$\Gamma_{1,-1}^{(4)}$&0\\
$\Gamma_{2,2}^{(3)}$&$\Gamma_{2,1}^{(5)}$&0&0&0&$-\Gamma_{1,1}^{(1)}$&0&0&0\\
0&$\Gamma_{2,1}^{(3)}$&$\Gamma_{2,0}^{(5)}$&0&0&$\Gamma_{1,1}^{(2)}$&$-\Gamma_{1,0}^{(1)}$&0&$\Gamma_{0,0}^{(6)}$\\
0&0&$\Gamma_{2,0}^{(3)}$&$\Gamma_{2,-1}^{(5)}$&0&0&$\Gamma_{1,0}^{(2)}$&$-\Gamma_{1,-1}^{(1)}$&$\Gamma_{0,0}^{(4)}$\\
0&0&0&0&0&$\Gamma_{1,1}^{(3)}$&$\Gamma_{1,0}^{(5)}$&0&$-\Gamma_{0,0}^{(1)}$
\end{tabular}\right).\nonumber\\
\end{eqnarray}
The 4-dimensional block in the lower-right corner of $M$ in \Eq{MN4} is given by $M$ for $N=2$, and the structure of the matrix $M$ iterates similarly for greater $N$. For different $N$, the matrix elements change, as all the rates $\Gamma_{j,m}^{(i)}$ do change with $N$, see \Eq{eq15xx}. Only in the case of collective phenomena, the values of the matrix $M$ are independent of the number of TLSs. 
The sum of the elements of each of the columns of the matrix $M$ must add to zero.
Using \Eq{eq15xx} we obtain that for a row $k$ of the matrix $M$, in the middle of the Dicke space (to avoid special boundary conditions),
\begin{eqnarray}
\text{M}_{k}&=&\left(\dots\stackrel{{\color{red}k-(2j+3)}}{\Gamma_{N,j+1,m+1}^{(3)}},\stackrel{{\color{red}k-(2j+2)}}{\Gamma_{N,j+1,m}^{(5)}},\stackrel{{\color{red}k-(2j+1)}}{\Gamma_{N,j+1,m-1}^{(7)}}\dots\stackrel{{\color{red}k-1}}{\Gamma_{N,j,m+1}^{(2)}},\stackrel{{\color{red}k}}{-\Gamma_{N,j,m}^{(1)}},\stackrel{{\color{red}k+1}}{\Gamma_{N,j,m-1}^{(8)}},\dots\stackrel{{\color{red}k+(2j-1)}}{\Gamma_{N,j-1,m+1}^{(4)}},\stackrel{{\color{red}k+2j}}{\Gamma_{N,j-1,m}^{(6)}},\stackrel{{\color{red}k+(2j+1)}}{\Gamma_{N,j-1,m-1}^{(9)}}\right),\nonumber\\
\label{Mk}
\end{eqnarray}
where the red superscripts show the column corresponding to each row element.
The matrix $M$ is a square matrix of side $n_\text{DS}$, which is extremely sparse, with at most nine non-zero elements per row. 

\end{document}